\documentclass[fleqn,usenatbib]{mnras}

\usepackage{newtxtext,newtxmath}
\usepackage[T1]{fontenc}
\usepackage{ae,aecompl}
\usepackage{placeins}

\usepackage{graphicx}	 
\usepackage{amsmath}	
\usepackage{float}	
\usepackage{color}
\usepackage{gensymb}
\usepackage{longtable}
\usepackage{adjustbox}

\usepackage{scalerel,tikz}
\usetikzlibrary{svg.path}
\definecolor{orcidlogocol}{HTML}{A6CE39}
\tikzset{orcidlogo/.pic={
 \fill[orcidlogocol] svg{M256,128c0,70.7-57.3,128-128,128C57.3,256,0,198.7,0,128C0,57.3,57.3,0,128,0C198.7,0,256,57.3,256,128z};
 \fill[white] svg{M86.3,186.2H70.9V79.1h15.4v48.4V186.2z}
 svg{M108.9,79.1h41.6c39.6,0,57,28.3,57,53.6c0,27.5-21.5,53.6-56.8,53.6h-41.8V79.1z M124.3,172.4h24.5c34.9,0,42.9-26.5,42.9-39.7c0-21.5-13.7-39.7-43.7-39.7h-23.7V172.4z}
 svg{M88.7,56.8c0,5.5-4.5,10.1-10.1,10.1c-5.6,0-10.1-4.6-10.1-10.1c0-5.6,4.5-10.1,10.1-10.1C84.2,46.7,88.7,51.3,88.7,56.8z};
}}
\newcommand\orcidicon[1]{\href{https://orcid.org/#1}{\mbox{\scalerel*{
\begin{tikzpicture}[yscale=-1,transform shape]
\pic{orcidlogo};
\end{tikzpicture}
}{|}}}}

\newcommand{\aref}[1]{\hyperref[#1]{Appendix~\ref{#1}}}

\usepackage[normalem]{ulem}

\definecolor{darkgreen}{rgb}{0.13, 0.55, 0.13}
\definecolor{brown}{rgb}{0.65, 0.16, 0.16}

\title[IMF and non Solar-scaled C/O]{The impact of carbon and oxygen abundances on the metal-poor initial mass function}

\author[P. Sharda et al.]{Piyush Sharda$^{\orcidicon{0000-0003-3347-7094}\,1,2,3}$\thanks{sharda@strw.leidenuniv.nl (PS)},
Anish M. Amarsi$^{\orcidicon{0000-0002-3181-3413}\,4}$\thanks{anish.amarsi@physics.uu.se (AMA)},
Kathryn Grasha$^{\orcidicon{0000-0002-3247-5321}\,1,2}$\thanks{kathryn.grasha@anu.edu.au (KG)},
Mark R. Krumholz$^{\orcidicon{0000-0003-3893-854X}\,1,2}$,
\newauthor
David Yong$^{\orcidicon{0000-0002-6502-1406}\,1,2}$,
Gen Chiaki$^{\orcidicon{0000-0001-6246-2866}\,5,6}$,
Arpita Roy$^{\orcidicon{0000-0002-5021-6737}\,7}$, and
Thomas Nordlander$^{\orcidicon{0000-0001-5344-8069}\,1,2}$
\\
$^{1}$Research School of Astronomy and Astrophysics, Australian National University, Canberra, ACT 2611, Australia\\
$^{2}$Australian Research Council Centre of Excellence for All Sky Astrophysics in 3 Dimensions (ASTRO 3D), Australia\\
$^{3}$Leiden Observatory, Universiteit Leiden, NL-2300 RA Leiden, The Netherlands\\
$^{4}$Theoretical Astrophysics, Department of Physics and Astronomy, Uppsala University, SE-751 20 Uppsala, Sweden\\
$^{5}$Center for Relativistic Astrophysics, School of Physics, Georgia Institute of Technology, Atlanta, GA 30332, USA\\
$^{6}$Astronomical Institute, Graduate School of Science, Tohoku University, Aoba, Sendai 980-8578, Japan\\
$^{7}$Cosmology Research Group, Scuola Normale Superiore, 56126 Pisa, Italy
}

\date{Accepted 2022 November 10. Received 2022 November 10; in original form 2022 September 30}

\pubyear{2022}

\begin{document}
\label{firstpage}
\pagerange{\pageref{firstpage}--\pageref{lastpage}}
\maketitle

\begin{abstract}
Star formation models predict that the metal-poor initial mass function (IMF) can be substantially different from that observed in the metal-rich Milky Way. This changeover occurs because metal-poor gas clouds cool inefficiently due to their lower abundance of metals and dust. However, predictions for the metal-poor IMF to date rely on assuming Solar-scaled abundances, that is, [X/O] = 0 at all [O/H]. There is now growing evidence that elements such as C and O that dominate metal line cooling in the ISM do not follow Solar scaling at low metallicities. In this work, we extend models that predict the variation in the characteristic (or, the peak) IMF mass as a function of metallicity using [C/O] ratios derived from observations of metal-poor Galactic stars and of \ion{H}{ii} regions in dwarf galaxies. These data show [C/O] < 0 at sub-Solar [O/H], which leads to a substantially different metal-poor IMF in the metallicity range where \ion{C}{i} and \ion{C}{ii} cooling dominate ISM thermodynamics, resulting in an increase in the characteristic mass by a factor as large as 7. An important consequence of this difference is a shift in the location of the transition from a top- to a bottom-heavy IMF upwards by 0.5 -- 1 dex in metallicity. Our findings indicate that the IMF is very sensitive to the assumptions around Solar-scaled ISM compositions in metal-poor systems (e.g., dwarf galaxies, the Galactic halo and metal-poor stars) that are a key focus of JWST.
\end{abstract}




\section{Introduction}
\label{s:introduction}

Understanding the impact of metallicity on the stellar initial mass function (IMF) is of utmost importance for several key areas of galaxy evolution. It is now becoming clear that although the IMF seems to be invariant in the metal-rich components of the Milky Way \citep{2001MNRAS.322..231K,2003PASP..115..763C,2010ARA&A..48..339B}, it could have been different in other environments with metallicities that are sub- \citep[e.g.,][]{2007ApJ...658..367K,2010A&A...515A..68M,2011MNRAS.412..843S,2013MNRAS.432L..46S,2018ApJ...855...20G,2021MNRAS.503.6026R,2022arXiv220910461F} or super-Solar \citep[e.g.,][]{2010Natur.468..940V,2012ApJ...760...71C,2015ApJ...806L..31M,2020ARA&A..58..577S,2022ApJ...932..103G}. It is not surprising that the IMF should be sensitive to the metallicity, since the amount of metals present in the interstellar medium (ISM) directly sets the thermodynamics of the collapsing gas that ultimately forms stars \citep{2000ApJ...534..809O,2001MNRAS.328..969B,2005ApJ...626..627O,2010ApJ...722.1793O,2015MNRAS.446.2659C,2021MNRAS.508.4175C,2022MNRAS.509.1959S}.

The number of newly discovered metal-poor stars in the Milky Way and dwarf galaxies has grown exponentially over the past few decades \citep{2015ARA&A..53..631F}. From a theoretical perspective, several models and simulations have been developed to investigate star formation in low metallicity environments. However, most of these simulations do not extend down to metallicities low enough ($\rm{[O/H]} < -1.5$)\footnote{$\rm{[O/H]} = \log_{10}(\rm{O/H}) - \log_{10}(\rm{O/H})_{\odot}$} that we expect major variations in the IMF \citep[e.g.,][]{2011ApJ...735...49M,2014MNRAS.442..285B,2015MNRAS.449.2643B,2019MNRAS.484.2341B}. 

Only a handful of simulations exist that self-consistently evolve the abundances of different elements in the metal-poor ISM by solving for chemistry on the fly with hydrodynamics \citep[e.g.,][]{2016MNRAS.463.2781C,2021MNRAS.508.4175C,2022MNRAS.510.5199C}. While these simulations provide a realistic picture of how the abundances of C, N and O ultimately shape the metal-poor IMF, they cannot disentangle the effects of one element versus another because the evolution of collapsing gas clouds is highly non-linear. Since these simulations do not sweep across the possible range of ISM properties and exclude protostellar radiation feedback, they cannot quantify the relative importance of metallicity as compared to physical properties like density and pressure for the IMF \citep[e.g.,][]{2014ApJ...781L..14M,2022arXiv220604999T}. This is where analytical models have proved useful owing to the level of control and range one can achieve. \cite{2022MNRAS.509.1959S} present an analysis of the thermodynamic budget in collapsing dusty gas clouds across a wide range of metallicities ($10^{-6}$ to $3$ times Solar) to explore variations in the characteristic mass (or, the peak mass, depending on the functional form of the IMF; see Section~2.2 of \citealt{2018PASA...35...39H}) of the IMF. They identify a distinct low-metallicity regime where gaseous metals that are not bound into molecules are the primary contributors to cooling in the ISM, and subsequently set the characteristic mass of the IMF. 

However, \cite{2022MNRAS.509.1959S} assume Solar-scaled abundances to calculate metal line cooling in their models. While this is a reasonable approach to adopt in the first instance, with increasing observational evidence it is becoming clear that several key elements like C and N do not scale with their Solar values at low metallicity \citep[see the recent review by][]{2022arXiv221004350R}. This is because N production changes from primary (not dependent on [O/H]) to secondary (dependent on [O/H]) as the ISM metallicity increases beyond [O/H] > -0.4 (e.g., \citealt{1978MNRAS.185P..77E,2010ApJ...720..226P,2022arXiv220204666J}; however, see \citealt{2021MNRAS.502.4359R} for a somewhat more complex picture). Similarly, at low [O/H], C is produced as a primary element, but its production pathway switches to \textit{pseudo}-secondary at high [O/H] due to enhanced mass loss in the form of stellar winds \citep[e.g.,][]{2000ApJ...541..660H,2014MNRAS.443..624E,2016ApJ...827..126B}. At the lowest metallicities, the individual stellar C, N and O abundances vary by several orders of magnitude \citep[figure 2]{2013ApJ...762...28N}. The primary versus secondary production also impacts the observed metallicity plateau in the gas phase in local galaxies at large galactocentric distances \citep{2022ApJ...929..118G}, and is important for photoionization modeling of \ion{H}{ii} regions \citep{2017MNRAS.466.4403N,2021ApJ...908..241G}. .

The consequence of multiple production pathways is that the [C/O] and [N/O] ratios measured in metal-poor stars in the Galaxy significantly deviate from 0 at low metallicities \citep{1999A&A...342..426G,2004A&A...414..931A,2004A&A...421..649I,2005A&A...430..655S,2009A&A...500.1143F,2014A&A...568A..25N,2019A&A...622L...4A,2019A&A...630A.104A}. Similarly, the [C/O] and [N/O] ratios in low metallicity galaxies also deviate from 0 \citep{1995ApJ...443...64G,1996ApJ...471..211K,2006ApJ...636..214V,2006ApJ...652..257L,2009MNRAS.398..949P,2016ApJ...827..126B,2019ApJ...874...93B,2017MNRAS.466.4403N,2021MNRAS.502.4359R,2022arXiv220802562A}. Both these groups of works also find a scatter as high as 0.5 dex in [C/O] at fixed [O/H]. 

It therefore seems important to take these non-linearities in the [C/O] ratio into account when making predictions for the metal-poor IMF, because C and O are the main metal ISM coolants \citep{2003Natur.425..812B,2007MNRAS.380L..40F}. The vast majority of available research in this area has focused on understanding how a non-universal IMF could give rise to the observed trends in [C/O] at low metallicities, but not the other way around \citep[e.g.,][]{1995ApJ...443...64G,2005ApJ...623..213C,2008MNRAS.390..582C,2010A&A...515A..68M,2011A&A...530A..78T,2019MNRAS.490.2838R,2020A&A...639A..37R,2020MNRAS.495.3276L,2020MNRAS.494.2355P}. The few works that do look at the influence of C and O cooling on the IMF assume Solar-scaled abundances \citep[e.g.,][]{2000ApJ...534..809O,2021MNRAS.508.4175C,2022MNRAS.514.4639C} or exclude protostellar radiation feedback \citep[e.g.,][]{2006MNRAS.369.1437S,2022MNRAS.510.5199C}. In reality, the IMF and [C/O] ratio influence each other because the IMF sets the yield of C and O from massive and intermediate-mass stars \citep[e.g.,][]{1992A&A...264..105M,2005Sci...309..451I,2020ApJ...900..179K}, and the ISM C and O abundance directly set the cooling rate of collapsing molecular gas that dictates the IMF \citep[e.g.,][]{2012MNRAS.419.1566S,2021MNRAS.508.4175C,2022MNRAS.509.1959S}. Additionally, preferential ejection of O through outflows \citep{1999ApJ...513..142M,2018MNRAS.481.1690C,2021MNRAS.502.5935S} can also alter the [C/O] ratio in low metallicity systems \citep{2011A&A...531A.136Y,2019ApJ...874...93B,2022arXiv220909345Y}. Thus, it is crucial to understand the dependence of the IMF on the [C/O] ratio in the ISM.

In this work, we extend the calculations of \cite{2022MNRAS.509.1959S} by accounting for realistic variations in [C/O] to explore how non-Solar-scaled C and O abundances impact the thermodynamics of the metal-poor ISM and thence characteristic stellar mass of the metal-poor IMF. We arrange the rest of the paper as follows: \autoref{s:theory} summarizes the theoretical framework of \cite{2022MNRAS.509.1959S}, which we also adopt in the current work, \autoref{s:results} describes the resulting ISM thermodynamic budget and the IMF when we use a non-zero [C/O], and \autoref{s:impact_co} discusses the impact of varying [C/O] as found in observations of metal-poor stars and of \ion{H}{ii} regions in dwarf galaxies on the characteristic mass of the IMF. Finally, we conclude and provide a future outlook in \autoref{s:conclusions}.

\section{Model Summary}
\label{s:theory}

The basic premise of the \cite{2022MNRAS.509.1959S} model is that radiation feedback from an existing protostar plays a key role in setting the characteristic mass, $M_{\rm{ch}}$, of the IMF \citep{2011ApJ...743..110K,2016MNRAS.460.3272K,2016MNRAS.458..673G}. To explore the implications of this $\textit{ansatz}$, the authors consider a spherically-symmmetric collapsing dusty gas cloud at a range of densities, pressures and metallicities. At the centre of this cloud is a protostar, the radiation feedback from which heats the surrounding gas. Following simulations \citep[e.g.,][]{2011ApJ...740...74K,2013ApJ...763...51F,2016MNRAS.463.2781C,2019MNRAS.490..513S,2021MNRAS.503.2014S} as well as observations \citep[e.g.,][]{1995ApJ...446..665C,2000ApJ...537..283V,2015A&A...578A..29S,2021A&A...648A..66G}, the authors adopt a power-law for the density profile of the cloud. With such a profile, the dust temperature profile in the presence of radiation feedback is also given by a power-law \citep{2005ApJ...631..792C,2011ApJ...743..110K,2013ApJ...773..113C}. Once the dust temperature is determined, the authors include a variety of thermodynamic processes, such as dust-gas energy exchange, cooling due to metal lines, $\rm{H_2}$, $\rm{HD}$, cosmic-ray heating, compressional heating, and H$_2$ formation heating, to find the equilibrium gas temperature. Using the gas temperature profile, the authors then quantify how far away from the existing protostar the cloud becomes Jeans unstable such that it can fragment and collapse to form a new star instead of accreting onto the existing star. The mass that is Jeans unstable is then classified as $M_{\rm{ch}}$. We refer the reader to \cite{2022MNRAS.509.1959S} for additional information on the model.

\cite{2022MNRAS.509.1959S} parameterize metallicity using the generic notation $\mathcal{Z}$ that corresponds to the logarithmic abundance of all metals with respect to Solar abundances \citep{2009ARA&A..47..481A}. Since we are interested in studying non-Solar-scaled abundances for C and O, we fix our metallicity scale to represent the O abundance ($\textit{i.e.,}$ $\mathcal{Z} \equiv \rm{[O/H]}$). We will therefore study how variations in $\rm{[C/O]}$ as a function of $\rm{[O/H]}$ impact $M_{\rm{ch}}$. The abundances of neutral C and O atoms per H nucleus \textit{in the gas phase} for the Solar case that we adopt are $x_{\rm (C,MW)} = 1.4\times 10^{-4}$ and $x_{\rm (O,MW)} = 3.0\times 10^{-4}$, following \citet[table 23.1]{2011piim.book.....D}. Our fiducial model uses a Solar-scaled dust to gas ratio $\delta = \delta_{\rm{MW}}\times 10^{[\rm{O/H}]}$, where $\delta_{\rm{MW}} = 1/162$ \citep{2004ApJS..152..211Z}. We also consider an alternative scaling of $\delta$ from \cite{2014A&A...563A..31R} below. We do not differentiate between $\delta$ for carbonaecous and silicate grains to calculate the gas-dust energy exchange. This distinction is more important for heating due to H$_2$ formation on dust \citep[e.g.,][]{2009A&A...496..365C} that we do not include in this work. This is because it has a negligible contribution to gas thermodynamics at the location that sets $M_{\rm{ch}}$ \citep[section 4.1]{2022MNRAS.509.1959S}. We import variations in the chemical state of C and O (\ion{C}{ii}, \ion{C}{i}, \ion{O}{i}, and CO) as a function of $\rm{[O/H]}$ from \citet[figure 1]{2022MNRAS.509.1959S}, but we also investigate how variations in the chemical state of C and O impact $M_{\rm{ch}}$ in \autoref{s:results_chemicalcompositions}.

As in \cite{2022MNRAS.509.1959S}, we study a wide variety of ISM physical conditions by varying the pressure $P$ between $10^4\,k_{\rm{B}}\,\rm{K\,cm^{-3}}$ (representing molecular clouds in main sequence star-forming galaxies $-$ \citealt{2001ApJ...547..792D,2016ApJ...821..118W,2017ApJ...834...57M,2019ApJ...880...16K}) and $10^8\,k_{\rm{B}}\,\rm{K\,cm^{-3}}$ (representing molecular clouds in starburst galaxies $-$ \citealt{2000ApJ...532L.109T,2008ApJ...677L...5V,2008ApJ...686..948B}, or progenitors of super star clusters and globular clusters $-$ \citealt{1997ApJ...480..235E,2006A&A...448..881B,2015ApJ...806...35J,2019ApJ...874..120F}). We also effectively vary the cloud density by sweeping across a range of values for the cloud velocity dispersion $\sigma_{\rm{v}}$, from $0.5-5\,\rm{km\,s^{-1}}$ \citep{1997MNRAS.288..145P,2011ApJ...735...49M,2014ApJ...796...75C,2016ApJ...828...50K,2022MNRAS.509.2180S}. Since the density is given by the ratio of $P$ and $\sigma^2_{\rm{v}}$, models with low $\sigma_{\rm{v}}$ correspond to the case of high density, and vice-versa. We use density and effective velocity dispersion interchangeably. Thus, our model grid is 3D, sweeping across $P$, $\sigma_{\rm{v}}$ and $\rm{[O/H]}$.

\begin{figure*}
\includegraphics[width=1.0\linewidth]{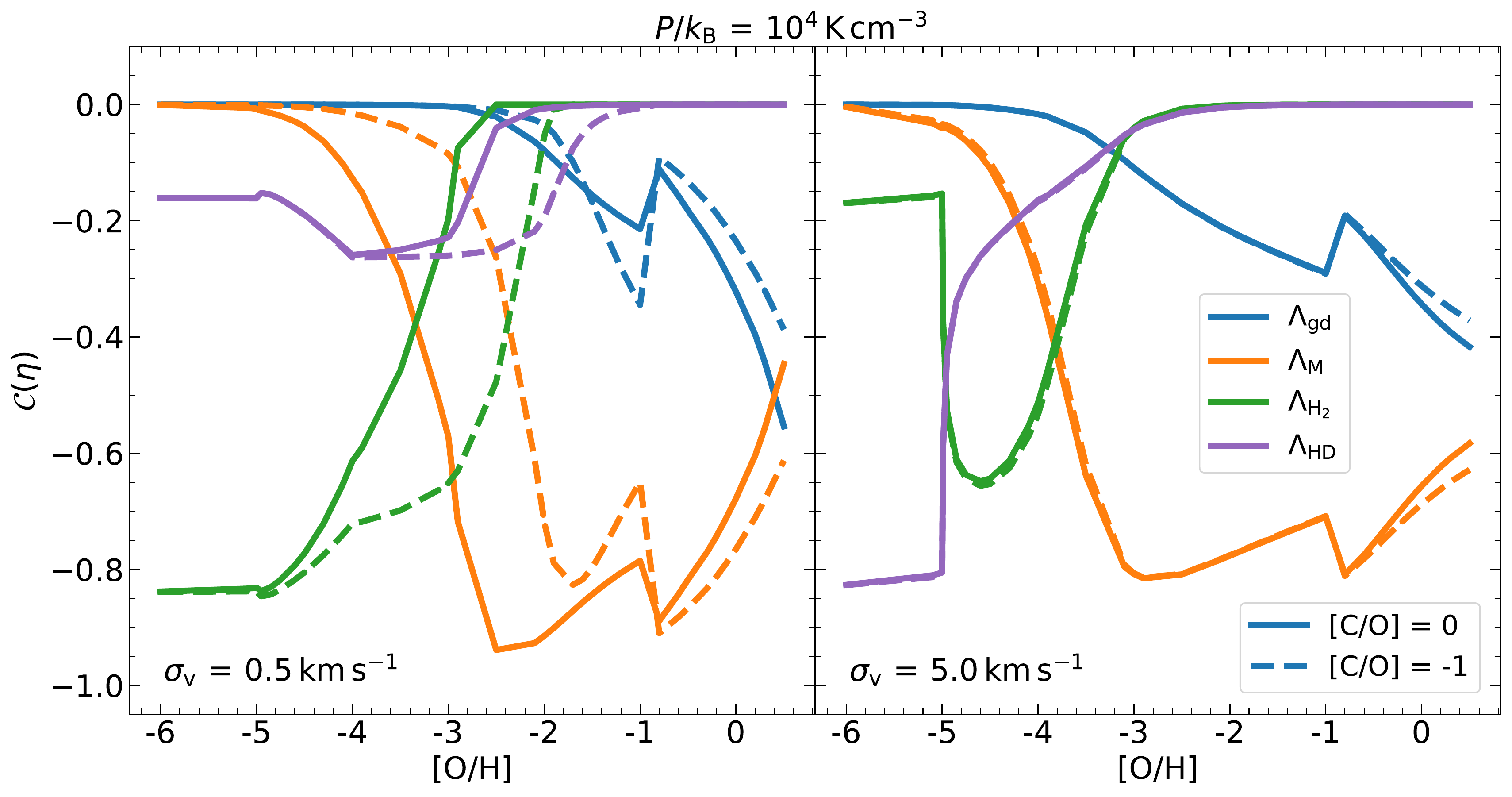}
\caption{Importance of different cooling processes under thermal balance at the location in the cloud where the characteristic stellar mass of the IMF is defined, $\mathcal{C}(\eta)$, as a function of the oxygen abundance, [O/H]. The results are plotted for pressure $P/k_{\rm{B}}=10^4\,\rm{K\,cm^{-3}}$ and effective velocity dispersion $\sigma_{\rm{v}}=0.5\,\rm{km\,s^{-1}}$ (left panel) and $5\,\rm{km\,s^{-1}}$ (right panel). The solid and dashed curves correspond to models where carbon is Solar-scaled ([C/O] = 0) and not Solar-scaled ([C/O] = $-$1), respectively. The cooling processes we study are cooling due to gas-dust energy exchange ($\Lambda_{\rm{gd}}$), metals like C and O ($\Lambda_{\rm{M}}$), $\rm{H_2}$ ($\Lambda_{\rm{H_2}}$), and HD ($\Lambda_{\rm{HD}}$). If $\mathcal{C}(\eta)$ is zero for a process $\eta$, it is unimportant for cooling the gas. If $\mathcal{C}(\eta)$ is $-$1, it dominates gas cooling. Non-Solar-scaled [C/O] significantly impacts the gas cooling budget for $\sigma_{\rm{v}}=0.5\,\rm{km\,s^{-1}}$ but not for $\sigma_{\rm{v}}=5\,\rm{km\,s^{-1}}$.}
\label{fig:gtb_lowpres}
\end{figure*}

\begin{figure*}
\includegraphics[width=1.0\linewidth]{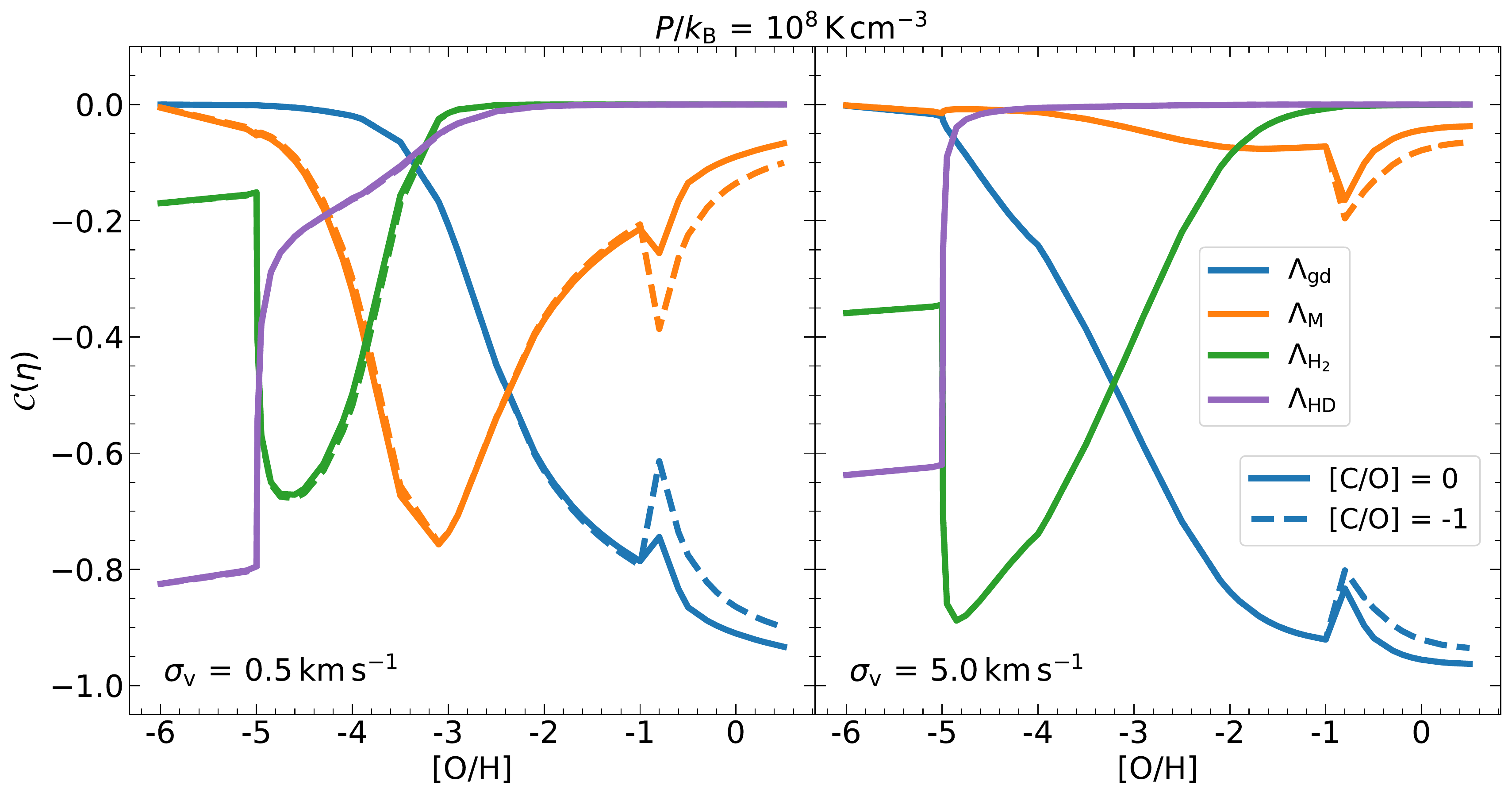}
\caption{Same as \autoref{fig:gtb_lowpres}, but for high cloud pressure $P/k_{\rm{B}}=10^8\,\rm{K\,cm^{-3}}$. Non-Solar-scaled [C/O] has a negligible impact on the gas cooling budget at high pressure.}
\label{fig:gtb_highpres}
\end{figure*}

\section{Results}
\label{s:results}

\subsection{Gas thermal balance}
\label{s:results_gasthermalbalance}
We first discuss how the overall cooling budget of the gas at the location in the cloud where $M_{\rm{ch}}$ is defined changes when we fix $\rm{[C/O]}$ to a value other than 0, implying a non-Solar scaling of C with respect to O. We remind the reader that $M_{\rm{ch}}$ is set where the gas mass around an existing protostar in the cloud is sufficient to collapse by itself (\textit{i.e.,} where the enclosed mass equals the Bonnor-Ebert mass; see \citealt{1955ZA.....37..217E,1957MNRAS.117..104B}). For the purpose of this demonstration, we consider [C/O] = 0 (fiducial model) and [C/O] = $-1$, and investigate the corners of our grid in $P$ and $\sigma_{\rm{v}}$. As we will discuss later in \autoref{s:impact_co}, [C/O] = $-$1 represents the lowest [C/O] observed in dwarf galaxies \citep{2019ApJ...874...93B}, and a loose lower bound of that measured in metal-poor stars after factoring in uncertainties \citep{2019A&A...630A.104A}.

As in \citet[equation 31]{2022MNRAS.509.1959S}, we quantify the relative contribution of a process $\eta$ to gas thermodynamics as follows
\begin{equation}
\mathcal{C}(\eta) = \frac{2 \eta}{|\Gamma_{\rm{c}}| + |\Gamma_{\rm{gd}}| + |\Lambda_{\rm gd}| + |\Lambda_{\rm{M}}| + |\Lambda_{\rm{H_2}}| + |\Lambda_{\rm{HD}}|}\,,
\label{eq:term_contribution}
\end{equation}
where the $\Gamma$ terms represent heating processes: $\Gamma_{\rm{c}} -$ compressional heating, and $\Gamma_{\rm{gd}} -$ heating due to dust-gas energy exchange, and the $\Lambda$ terms represent cooling processes: $\Lambda_{\rm{gd}} -$ cooling due to dust-gas energy exchange, $\Lambda_{\rm{M}} -$ cooling due to fine structure lines of C and O (\ion{C}{i}, \ion{C}{ii}, \ion{O}{i}) as well as cooling due to low $J$ transitions of CO, $\Lambda_{\rm{H_2}} -$ cooling due to lines of $\rm{H_2}$ and collisional excitation of $\rm{H_2}$, and $\Lambda_{\rm{HD}} -$ cooling due to lines of HD and collisional excitation of HD\footnote{As in \cite{2022MNRAS.509.1959S}, we use the maximum possible cooling due to HD.}. When written in this manner, $C(\eta)$ varies from $-$1 to $+$1; it is positive if $\eta$ is a heating processes and negative if $\eta$ is a cooling processes. Additionally, absolute values of $C(\eta)$ close to unity imply that the corresponding process $\eta$ is important for gas thermodynamics. The sum of all heating (cooling) processes is $+$1 ($-$1). We also ensure that the luminosity due to all the cooling radiation never exceeds that due to blackbody cooling for all variations of [C/O] we present in this work.

\autoref{fig:gtb_lowpres} shows the results for the low pressure case $P/k_{\rm{B}}=10^4\,\rm{K\,cm^{-3}}$ with effective velocity dispersion $\sigma_{\rm{v}}=0.5$ (left panel) and $5\,\rm{km\,s^{-1}}$ (right panel). We observe cooling due to $\rm{H_2}$ and HD dominates at extremely low metallicities ($\rm{[O/H]} \lesssim -5$) as expected, and cooling due do dust starts to become significant at high metallicities ($\rm{[O/H]} \gtrsim -2$).\footnote{Several authors find that dust already becomes a significant gas coolant at [O/H] > $-$5 \citep[e.g.,][]{2006MNRAS.369.1437S,2010MNRAS.402..429S,2014ApJ...783...75M,2021arXiv210206312S,2022MNRAS.510.5199C}. The key difference between their work and ours is that we include radiation feedback that heats the dust around the existing protostar.} We also find that cooling due to C and O dominates at intermediate metallicities, as found in \cite{2022MNRAS.509.1959S}. 

We can compare the left and right panels of \autoref{fig:gtb_lowpres} to see the impact of using non-Solar-scaled [C/O]. Interestingly, we see that changing the [C/O] ratio has a negligible impact on the gas cooling budget in the model with $\sigma_{\rm{v}} = 5\,\rm{km\,s^{-1}}$. This is because the location of $M_{\rm{ch}}$ in models with $\sigma_{\rm{v}} = 5\,\rm{km\,s^{-1}}$ is much closer to the existing protostar as compared to the models with $\sigma_{\rm{v}} = 0.5\,\rm{km\,s^{-1}}$. Since the gas is denser closer to the protostar, O provides much more cooling as compared to C. However, changing [C/O] to $-$1 considerably impacts the model with $\sigma_{\rm{v}} = 0.5\,\rm{km\,s^{-1}}$ where the location of $M_{\rm{ch}}$ is farther away from the protostar. Far away from the protostar, the low density gas is more efficiently cooled by C than O. Thus, the low amount of C as compared to O in the latter case leads to an overall reduction in the cooling provided by metals. As a consequence, the transition where metals start dominating the cooling budget shifts up by more than a dex in metallicity as compared to the model where [C/O] = 0.

\begin{figure}
\includegraphics[width=1.0\columnwidth]{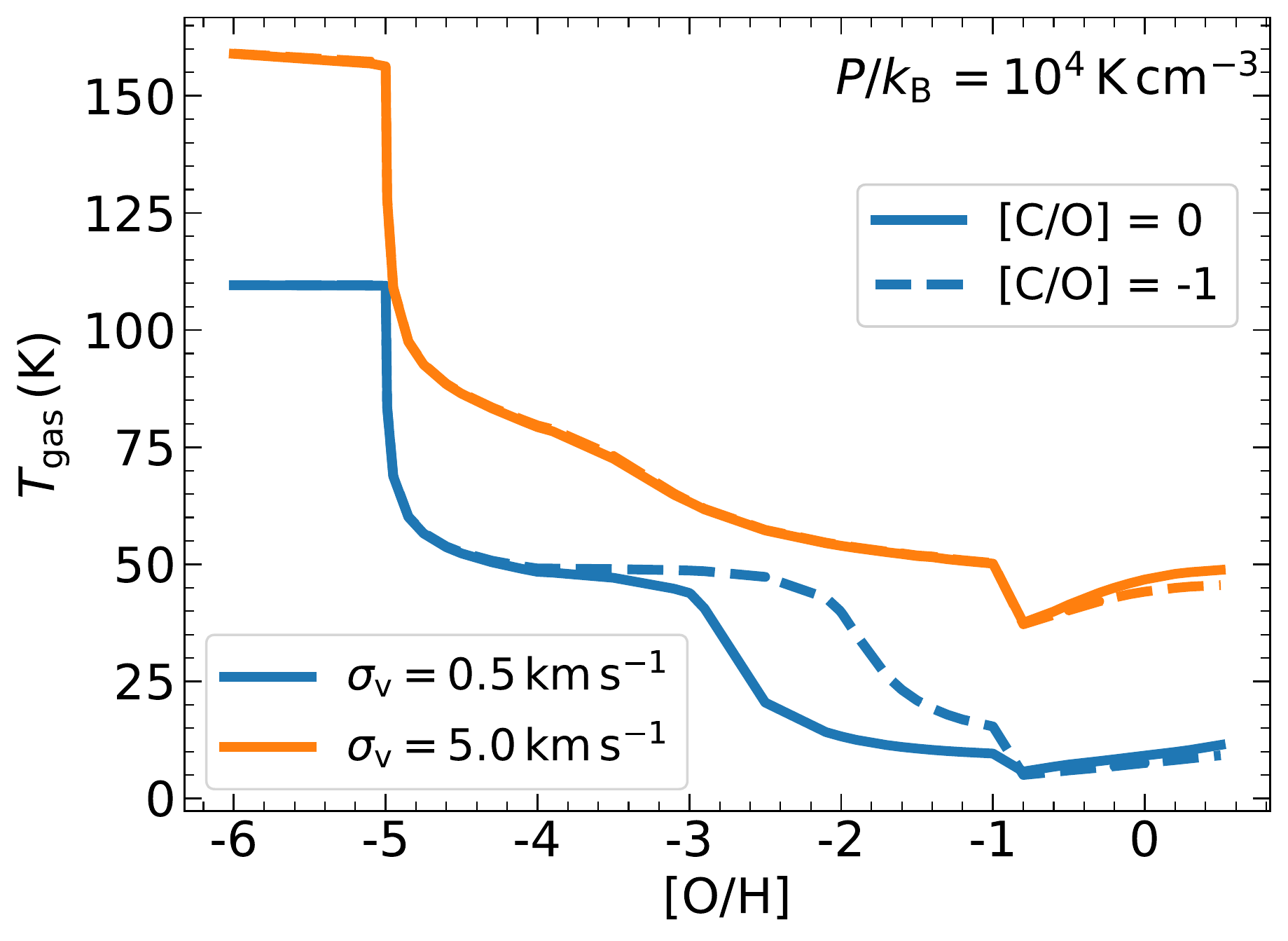}
\includegraphics[width=1.0\columnwidth]{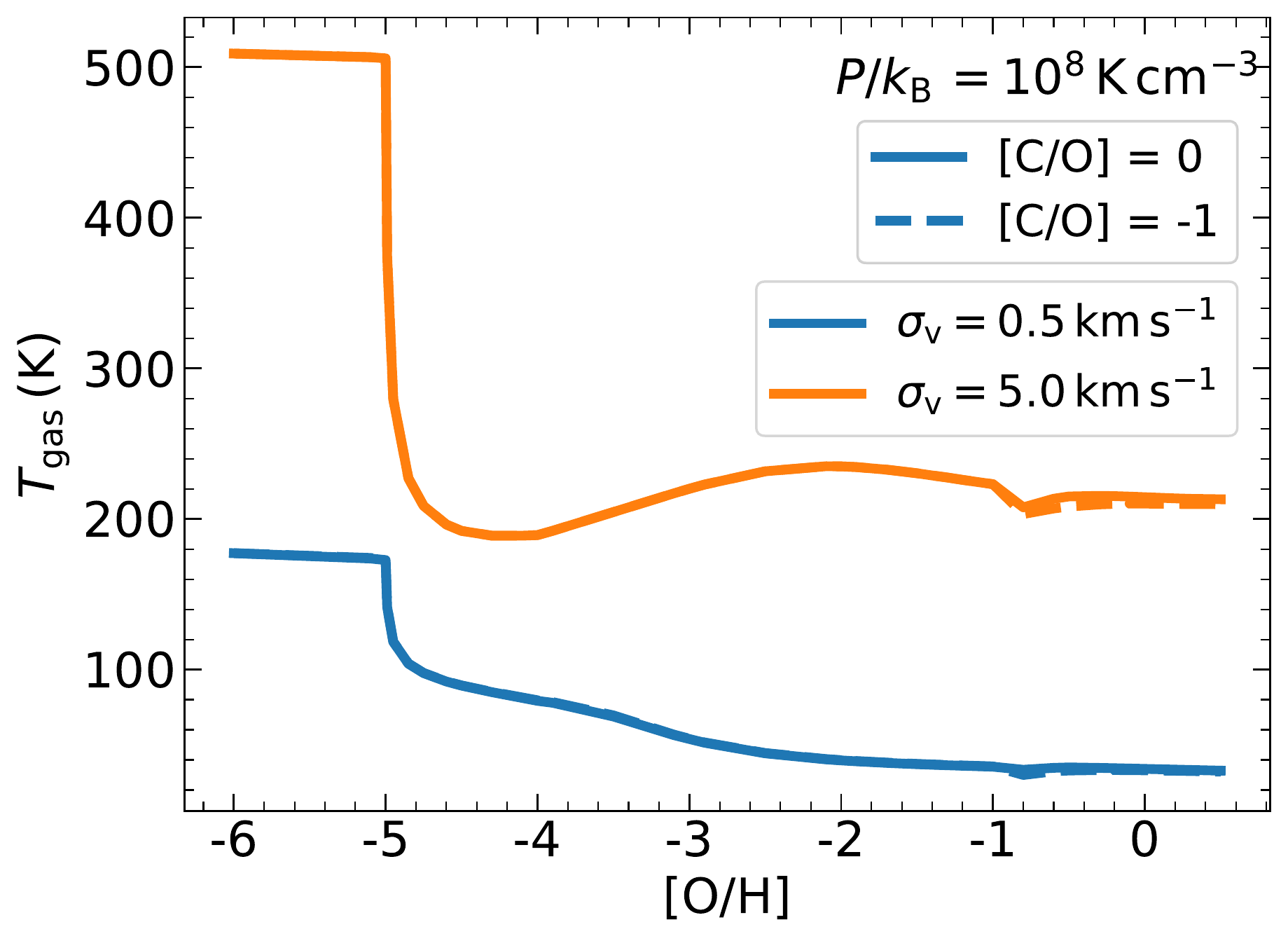}
\caption{\textit{Top panel:} Gas temperature, $T_{\rm{gas}}$, at the location that sets the characteristic stellar mass as a function of oxygen abundance, [O/H], for a fixed cloud pressure $P/k_{\rm{B}}=10^4\,\rm{K\,cm^{-3}}$ at different effective velocity dispersions $\sigma_{\rm{v}}=0.5\,\rm{km\,s^{-1}}$ (blue curves) and $\sigma_{\rm{v}}=5\,\rm{km\,s^{-1}}$ (orange curves). The solid and dashed curves correspond to Solar-scaled ([C/O] = 0) and non-Solar-scaled ([C/O] = -1) carbon abundances, respectively. \textit{Bottom panel:} Same as the top panel, but for high cloud pressure $P/k_{\rm{B}}=10^8\,\rm{K\,cm^{-3}}$. Non-Solar-scaled [C/O] changes the gas temperature a $-4 \leq \rm{[O/H]} \leq -1$ at low $P$ and low $\sigma_{\rm{v}}$.}
\label{fig:Tgas_pres}
\end{figure}

\autoref{fig:gtb_highpres} shows the results in the high pressure case ($P/k_{\rm{B}}=10^8\,\rm{K\,cm^{-3}}$). In contrast to \autoref{fig:gtb_lowpres}, we see that there is no appreciable reduction in metal cooling when the abundance of C is reduced by a dex. This is due to O being the primary metal coolant in high pressure environments, and the fact that dust anyway controls gas thermodynamics at high pressures. Although dust abundance also changes with metallicity, and presumably would change at least somewhat in response to changes in [C/O] (see \autoref{s:discussions_metalscaling} for a detailed discussion), at high densities (equivalently, high $P$) the gas temperature is quite insensitive to the total dust abundance \citep[e.g.,][]{2011ApJ...735...49M, 2019MNRAS.484.2341B}. This is because the dust temperature itself is not altered by the dust abundance as long as there is enough dust to render the circumstellar environment optically thick (true even at very sub-Solar metallicities due to the high optical/UV opacity of dust grains), and because at high densities dust and gas become well-coupled even if the dust abundance is sub-Solar.

Overall, we learn that a non-Solar scaled [C/O] only impacts the thermodynamics of collapsing dusty molecular clouds at low pressures and low effective velocity dispersions. In this particular case, the transition from a $\rm{H_2}$-dominated cooling to metal-dominated cooling occurs at a higher metallicity. The reduced amount of cooling due to a negative [C/O] would lead to higher gas temperatures (since $\rm{H_2}$ is a poor gas coolant; e.g., \citealt{1998A&A...335..403G,2005ApJ...626..627O,2012MNRAS.426..377G}), which is what we see from the top panel of \autoref{fig:Tgas_pres}. Since a negative [C/O] does not impact the gas thermal balance in high pressure environments, the gas temperature does not deviate from the model with [C/O] = 0 (see the bottom panel of \autoref{fig:Tgas_pres}).

\subsection{Characteristic stellar mass of the IMF}
\label{s:results_IMF}
We characterize the IMF as bottom-heavy if the characteristic stellar mass $M_{\rm{ch}} < 1\,\rm{M_{\odot}}$, and top-heavy otherwise. The variations in the gas temperatures effected by reducing [C/O] discussed in \autoref{s:results_gasthermalbalance} directly impact $M_{\rm{ch}}$ because the temperature structure of the gas determines the Bonnor-Ebert mass. 

\begin{figure}
\includegraphics[width=1.0\columnwidth]{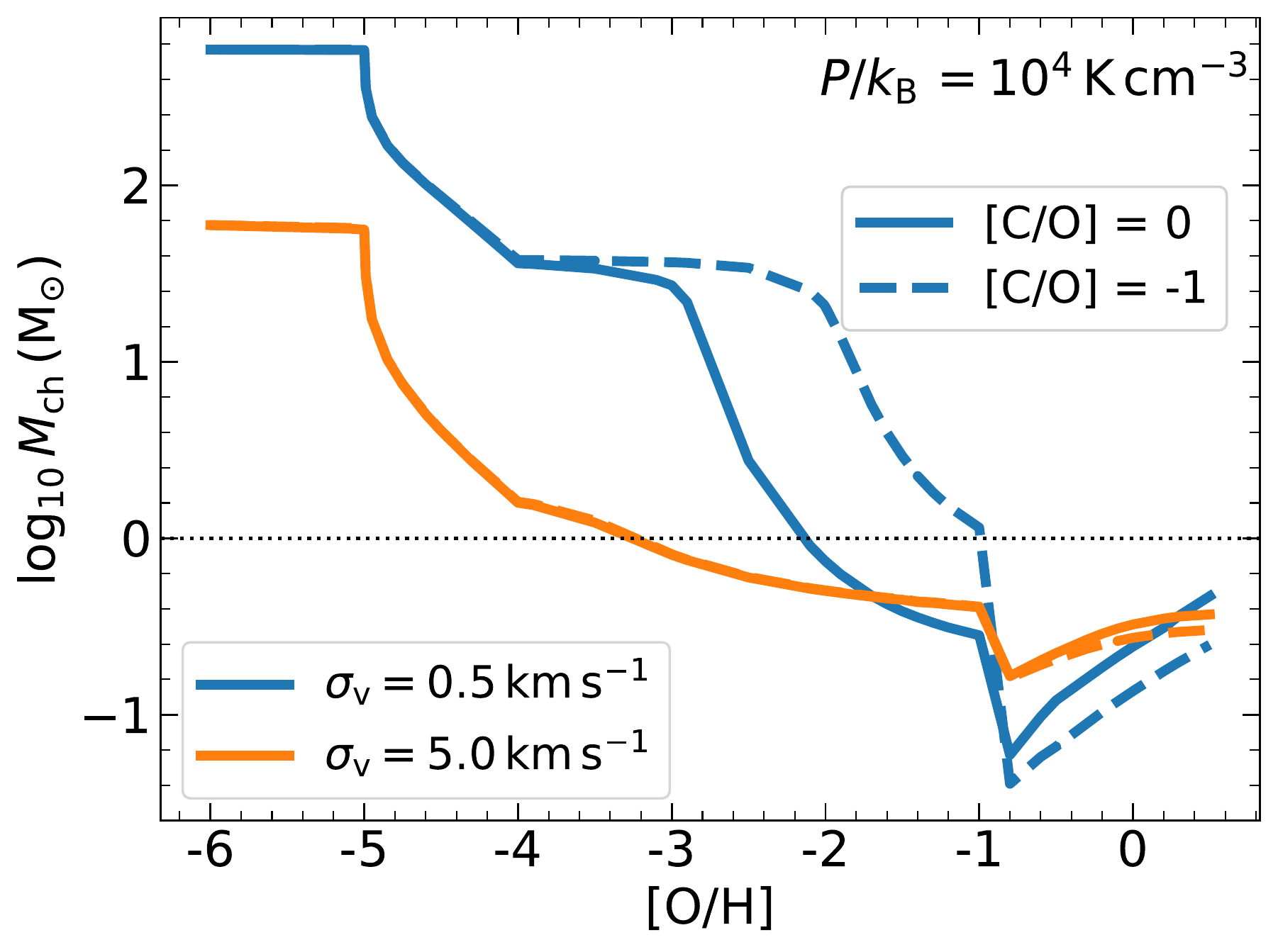}
\includegraphics[width=1.0\columnwidth]{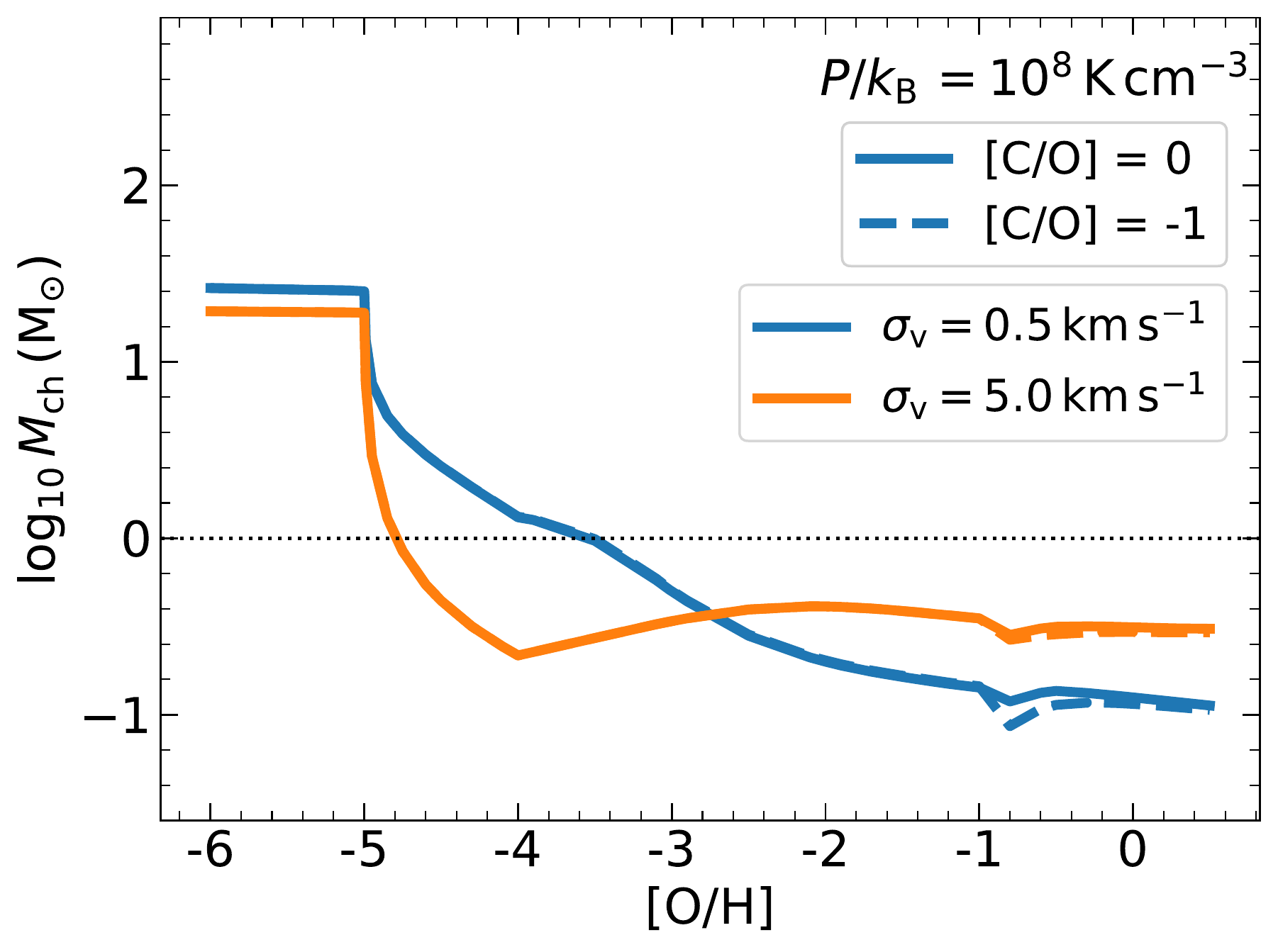}
\caption{\textit{Top panel:} Characteristic stellar mass, $M_{\rm{ch}}$, as a function of oxygen abundance, [O/H], for a fixed cloud pressure $P/k_{\rm{B}}=10^4\,\rm{K\,cm^{-3}}$ at different effective velocity dispersions $\sigma_{\rm{v}}$ as shown in the legend. The solid and dashed curves correspond to Solar-scaled ([C/O] = 0) and non-Solar-scaled ([C/O] = $-$1) carbon abundances, respectively. $M_{\rm{ch}}$ increases by a factor as high as 7 at $-4 \leq \rm{[O/H]} \leq -1$ when [C/O] = $-$1. \textit{Bottom panel:} Same as the top panel, but for high cloud pressure $P/k_{\rm{B}}=10^8\,\rm{K\,cm^{-3}}$. $M_{\rm{ch}}$ is not impacted by non-Solar-scaled [C/O] at high $P$.}
\label{fig:Mch_pres}
\end{figure}

We see from \autoref{fig:Mch_pres} that the high $\sigma_{\rm{v}}$ or high $P$ models, for which the gas temperature is insensitive to changes in [C/O], also do not show any variations in $M_{\rm{ch}}$ as a function of [O/H]. However, $M_{\rm{ch}}$ significantly changes for the model with low $\sigma_{\rm{v}}$ and low $P$, where metal cooling is dominated by C. The higher gas temperatures due to the lower amount of C in the case with [C/O] = $-$1 result in higher $M_{\rm{ch}}$ because the location where the enclosed mass equals the Bonnor-Ebert mass shifts further away from the existing protostar in the cloud. For example, $M_{\rm{ch}}$ increases from $\sim\,3\,\rm{M_{\odot}}$ to $34\,\rm{M_{\odot}}$ around [O/H] = $-$2.5. Additionally, the transition where the IMF is expected to become bottom-heavy now occurs around $\rm{[O/H]} \sim -1$ instead of $\rm{[O/H]} \sim -2$ as in the fiducial model with [C/O] = 0. Thus, a non-zero [C/O] matters for $M_{\rm{ch}}$ (and consequently, for the IMF) at low pressures and high densities.

Using [C/O] = $-$1 at [O/H] > $-$1 also has a minor but noticeable effect on $M_{\rm{ch}}$. In this case, we observe from \autoref{fig:Mch_pres} that [C/O] = $-$1 produces a lower $M_{\rm{ch}}$ as compared to the fiducial model with [C/O] = 0. This is opposite of the influence of a lower [C/O] at [O/H] < $-$1, and occurs because of enhanced CO cooling due to higher gas density as the equilibrium location where the enclosed mass equals the Bonnor-Ebert mass shifts further away from the existing protostar. Nonetheless, this effect is likely not realistic since we expect [C/O] to be closer to zero near [O/H] $\sim 0$.

\subsection{Effects of different chemical states of C and O}
\label{s:results_chemicalcompositions}
So far, we have assumed a particular chemical state of C and O as a function of [O/H] \citep[figure 1]{2022MNRAS.509.1959S}. Specifically, we have assumed that most C atoms are in the form of C$^+$ at primordial-like [O/H], C at intermediate [O/H], and CO at high [O/H]. Similarly, for O, we have assumed that most of the O atoms exist in the form of neutral O in the low-[O/H] ISM, while roughly half of them are locked in CO at high [O/H]. While there are sound theoretical arguments \citep[e.g.,]{2005ApJ...626..627O,2009ApJ...693..216K,2010ApJ...709..308M,2012MNRAS.426..377G,2014ApJ...790...10S,2015MNRAS.446.2659C,2015MNRAS.450.4424B,2016MNRAS.456.3596G,2021arXiv210501681S,2021arXiv210303889H,2022MNRAS.510.5199C} as well as some observational support \citep{2009ASPC..417...71L,Rubio15a, Shi16a,2017ApJ...839..107P,2017ApJ...835..278S,2018ApJ...853..111J,2020A&A...643A.141M,2022arXiv220809327G} for our choice of C and O chemical states at different [O/H], it only statistically represents reality, and we expect considerable scatter or co-existence of different chemical states in individual low-metallicity ISM environments. 

Noteworthy in this context are the suite of simulations at very low metallicities (ranging from $-6 \leq \rm{[O/H]} \leq -3$) carried out by \cite{2016MNRAS.463.2781C}, \cite{2019MNRAS.482.3933C}, and \cite{2022MNRAS.510.5199C}. The authors start by assuming that the low metallicity ISM is enriched by a Population III supernova, which sets the initial abundances of the different elements and their chemical states. The authors follow the evolution of various chemical states of C and O, and find that carbon mostly exists as C or C$^+$ at densities less than $10^{4}-10^{5}\,\rm{cm^{-3}}$, beyond which it is locked in CO, irrespective of [O/H]. Similarly, regardless of [O/H], O is mostly present in the neutral form at densities less than $10^{4}\,\rm{cm^{-3}}$, OH and O$_2$ at densities around $10^{4}-10^{8}\,\rm{cm^{-3}}$, and H$_2$O at higher densities. In principle, we can adopt their results to construct the chemical makeup of C and O in our models as a function of the density. However, these authors do not consider dissociation of molecules from UV photons emitted by background stars or stars within the same cloud. One-zone calculations by \cite{2012PASJ...64..114O} that include this FUV feedback at [O/H] < -3 find that even a weak FUV field can significantly impact the chemical composition and subsequent thermal evolution of the clouds at densities lower than $10^8\,\rm{cm^{-3}}$. Since the densities at which we calculate $M_{\rm{ch}}$ are typically in the range $10^{3}-10^{7}\,\rm{cm^{-3}}$, this implies that the chemical composition of C and O as presented in the above simulations is very sensitive to FUV feedback.

As a simpler alternative, we create a grid of models with four distinct chemical compositions: (1). $\rm{H_2}$ + CO + \ion{O}{I}, (2). $\rm{H_2}$ + \ion{C}{I} + \ion{O}{I}, (3). $\rm{H_2}$ + \ion{C}{ii} + \ion{O}{i}, and (4). \ion{H}{i} + \ion{C}{ii} + \ion{O}{i} for all [O/H]. The first composition is such that all C is locked in CO, and the remaining O atoms not locked in CO exist as \ion{O}{i}. Note that not all of these combinations are realistic, at least not for all [O/H]. We deliberately sweep across a larger-than-reasonable parameter space to explore possible variations in $M_{\rm{ch}}$ at non-zero [C/O] for different chemical states.

\begin{figure}
\includegraphics[width=1.0\columnwidth]{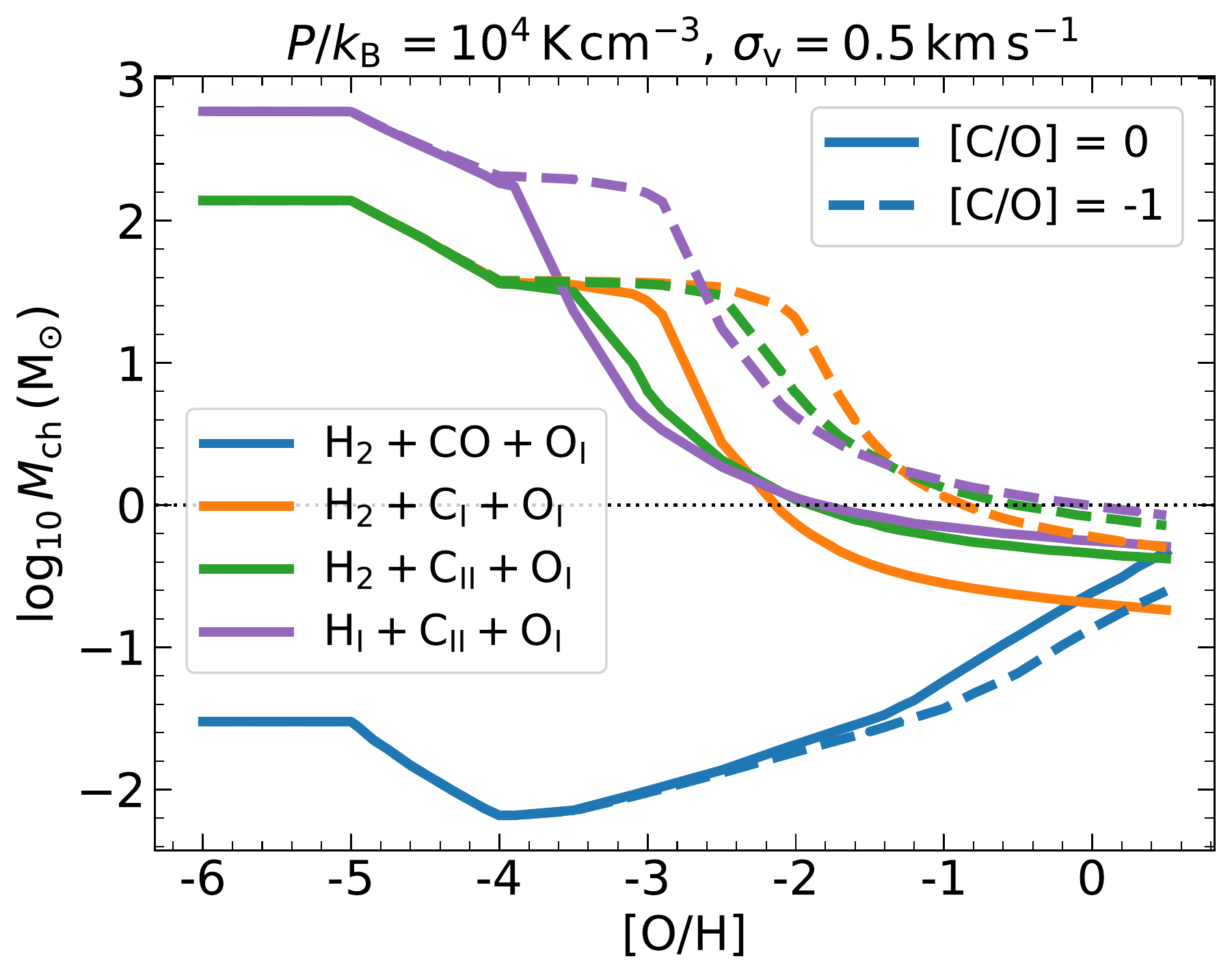}
\caption{Same as \autoref{fig:Mch_pres} but for different chemical states across [O/H] (as shown in the legend) for a fixed pressure $P/k_{\rm{B}} = 10^4\,\rm{K\,cm^{-3}}$ and effective velocity dispersion $\sigma_{\rm{v}}=0.5\,\rm{km\,s^{-1}}$. Variations in $M_{\rm{ch}}$ for non-Solar-scaled [C/O] are robust to \textit{realistic} chemical states of C and O at [O/H] < $-$1 (orange, green and purple curves).}
\label{fig:chemicalcomposition_Mch}
\end{figure}

In the previous sections, we have demonstrated that the effects of a non-zero [C/O] for different chemical compositions are only significant for the case where the pressure and velocity dispersion are low. Therefore, we only present results for this case in \autoref{fig:chemicalcomposition_Mch}. The trends we find corroborate our findings above. The case where the gas is almost fully molecular (H$_2$ + CO + \ion{O}{I}) exhibits a bottom-heavy IMF even at very low [O/H]. This is because low $J$ lines of CO provide more cooling than the fine structure lines of \ion{C}{i}, \ion{C}{ii}, and \ion{O}{i} for the densities we are concerned with, and the CO cooling rate is largely insensitive to the CO abundance. Since the low $J$ lines of CO are very optically thick, so a reduction in CO abundance is compensated by a reduction in optical depth, leaving the net cooling rate almost unchanged \citep{2007A&A...468..627V,2014MNRAS.437.1662K}. However, as we discuss above, this case is not realistic, since we do not expect a CO-dominated chemical state at low C and O abundance for the densities we are concerned with. 

We see that for the other two models where hydrogen is molecular (orange, and green curves), the transition from a top- to bottom-heavy IMF occurs at higher [O/H] when [C/O] = $-$1, compared to when [C/O] = 0. Moreover, for all the models where CO is absent (orange, green and purple curves, all having different chemical states of H and C), the resulting $M_{\rm{ch}}$ increases by a factor $\sim 2$ at [O/H] = 0 when [C/O] = $-$1, compared to when [C/O] = 0. While the resulting $M_{\rm{ch}}$ is not realistic since we know C and O primarily exist in the form of CO at high [O/H], it shows the impacts a non-zero [C/O] can have based on the chemical state of C and O.

To summarize, we find that the trends in $M_{\rm{ch}}$ for non-zero [C/O] are robust to variations in the chemical state of C and O. However, the exact [O/H] where the transition from a top- to bottom-heavy IMF occurs is sensitive to the chemical makeup of C and O in the gas.

\subsection{Effects of a varying dust-to-metal ratio}
\label{s:discussions_metalscaling}
It has been long known that there is a correlation between the ISM gas-phase C abundance and the dust-to-gas ratio \citep{1989ApJ...341..246C,1990ARA&A..28...37M}. The amount of C available in the gas-phase depends on the fraction of C locked in dust grains, which is different at different metallicities \citep[e.g.,][]{2022arXiv220708804K}. Thus, the dust-to-metal ratio can play a role in setting $M_{\rm{ch}}$. So far, we have used a simple scaling of the dust-to-metal ratio, $\delta \propto \rm{[O/H]}$. It is therefore worth analyzing how our results change if we use a different scaling of the dust-to-metal ratio with [O/H]. For this purpose, following \cite{2014A&A...563A..31R}, we set $\delta \propto 10^{\rm{[O/H]}}$ for $\rm{[O/H]} \geq -0.7$, and $\delta \propto 10^{3.1\rm{[O/H]}}$ for lower [O/H].

\begin{figure}
\includegraphics[width=1.0\columnwidth]{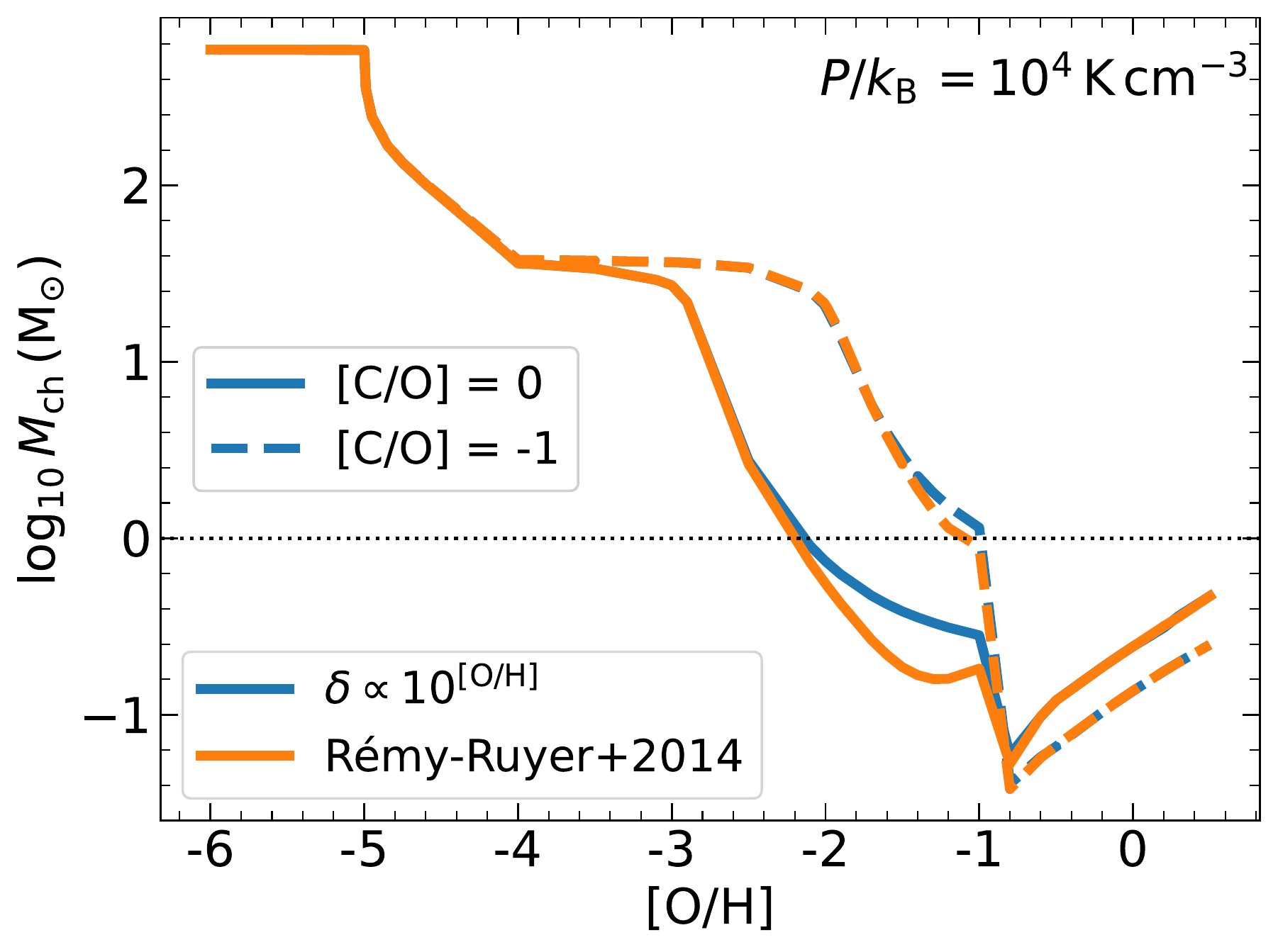}
\caption{Blue curves are the same as \autoref{fig:Mch_pres} for pressure $P/k_{\rm{B}} = 10^4\,\rm{K\,cm^{-3}}$ and effective velocity dispersion $\sigma_{\rm{v}} = 0.5\,\rm{km\,s^{-1}}$, using a linear scaling of the dust-to-metal ratio $\delta$ with [O/H]. Orange curves are for models using the dust-to-metal ratio scaling with [O/H] from \citet{2014A&A...563A..31R}. The \citet{2014A&A...563A..31R} scaling of $\delta$ with [O/H] makes little difference to the resulting trends in $M_{\rm{ch}}$.}
\label{fig:Mch_remyruyer_lowpres}
\end{figure}

\autoref{fig:Mch_remyruyer_lowpres} plots the trends in $M_{\rm{ch}}$ at low pressure and high density (or, low effective velocity dispersion) for the modified scaling of dust-to-metal ratio with [O/H], for the two cases where [C/O] = 0 and [C/O] = $-$1. While a different dust-to-metal ratio scaling produces slightly different trends in $M_{\rm{ch}}$ due to $\Gamma_{\rm{gd}}$ and $\Lambda_{\rm{gd}}$, the variations in $M_{\rm{ch}}$ due to a non-zero [C/O] are similar to those we have noticed in \autoref{s:results_IMF} and \autoref{s:results_chemicalcompositions}. We also find that the trends in $M_{\rm{ch}}$ at high pressure for a non-zero [C/O] do not vary if we change the scaling of the dust-to-metal ratio with [O/H].

Note that there are several other models/simulations that describe the evolution of $\delta$ with [O/H] \citep{2017MNRAS.471.3152P,2019MNRAS.485.1727H,2019MNRAS.489.4072V,2019MNRAS.490.1425L,2020MNRAS.493.2490T}. However, none of them seem to reproduce the observed data across a wide range of metallicities and redshifts \cite[see the discussion in][]{2022MNRAS.513.1531P}. We therefore do not attempt to test variations in $M_{\rm{ch}}$ due to the $\delta-\rm{[O/H]}$ relations predicted by these models, but note that this is an important avenue with significant potential for future exploration.

\section{Impact of variation in [C/O] on the metal-poor IMF}
\label{s:impact_co}
Having obtained a basic intuition of variations in $M_{\rm{ch}}$ as a function of [C/O], we now turn to available measurements of [C/O] at different [O/H], and use these data to systematically predict variations in $M_{\rm{ch}}$. We focus on two sets of data: one from high-resolution spectroscopic observations of metal-poor stars in the Milky Way, and other from gas-phase abundance measurements in \ion{H}{ii} regions of metal-poor dwarf galaxies. 

\subsection{Insights from observations of metal-poor stars}
\label{s:metalpoor}
The last decade has seen an immense progress in abundance measurements of metal-poor stars in the Milky Way \citep{2013ApJ...762...28N,2020MNRAS.496.4964A,2020MNRAS.492.4986Y,2021arXiv210706430Y,2022arXiv220608299L} as well as in dwarf galaxies \citep[e.g.,][]{2010ApJ...708..560F,2016ApJ...826..110F,2016ApJ...817...41J,2014A&A...562A.146I,2018ApJS..238...36A,2018ApJ...852...99N,2018ApJ...856..142C,2020A&A...636A.111A}. The abundance patterns observed in metal-poor stars carry signatures of the primordial (or, primordial-like) ISM they were born in, and thus provide much needed observational constraints on the metal-poor ISM \citep[e.g.,][]{2011Natur.477...67C,2015MNRAS.453.2771J,2019MNRAS.488L.109N,2019ApJ...871..146F,2019ApJ...876...97E,2021ApJ...915L..30S,2022arXiv220803891M}. Carbon, in particular, has received special emphasis in these studies because early results discovered high [C/Fe] in stars with [Fe/H] < $-$2, now known as carbon-enhanced metal-poor stars \citep[CEMP,][]{2005ARA&A..43..531B,2015ARA&A..53..631F}, with [C/Fe] > $+$0.7 \citep{2007ApJ...655..492A}.\footnote{What matters for gas cooling by metals in the ISM is the [C/O] ratio, and not the [C/Fe] or [O/Fe] ratios, since the abundance of Fe in the gas-phase is tiny. Thus, for the purpose of our work, a high [C/Fe] or [O/Fe] has similar implications on the IMF as \cite{2022MNRAS.509.1959S} if [C/O] = 0.} While there are still discrepancies that exist between measurements from different surveys \citep{2022MNRAS.515.4082A}, the general consensus is that the evolution of C is highly non-linear in metal-poor environments. Thus, the measured trends in [C/O] in metal-poor stars are promising as a diagnostic of the metal-poor IMF.

\begin{figure*}
\includegraphics[width=1.0\textwidth]{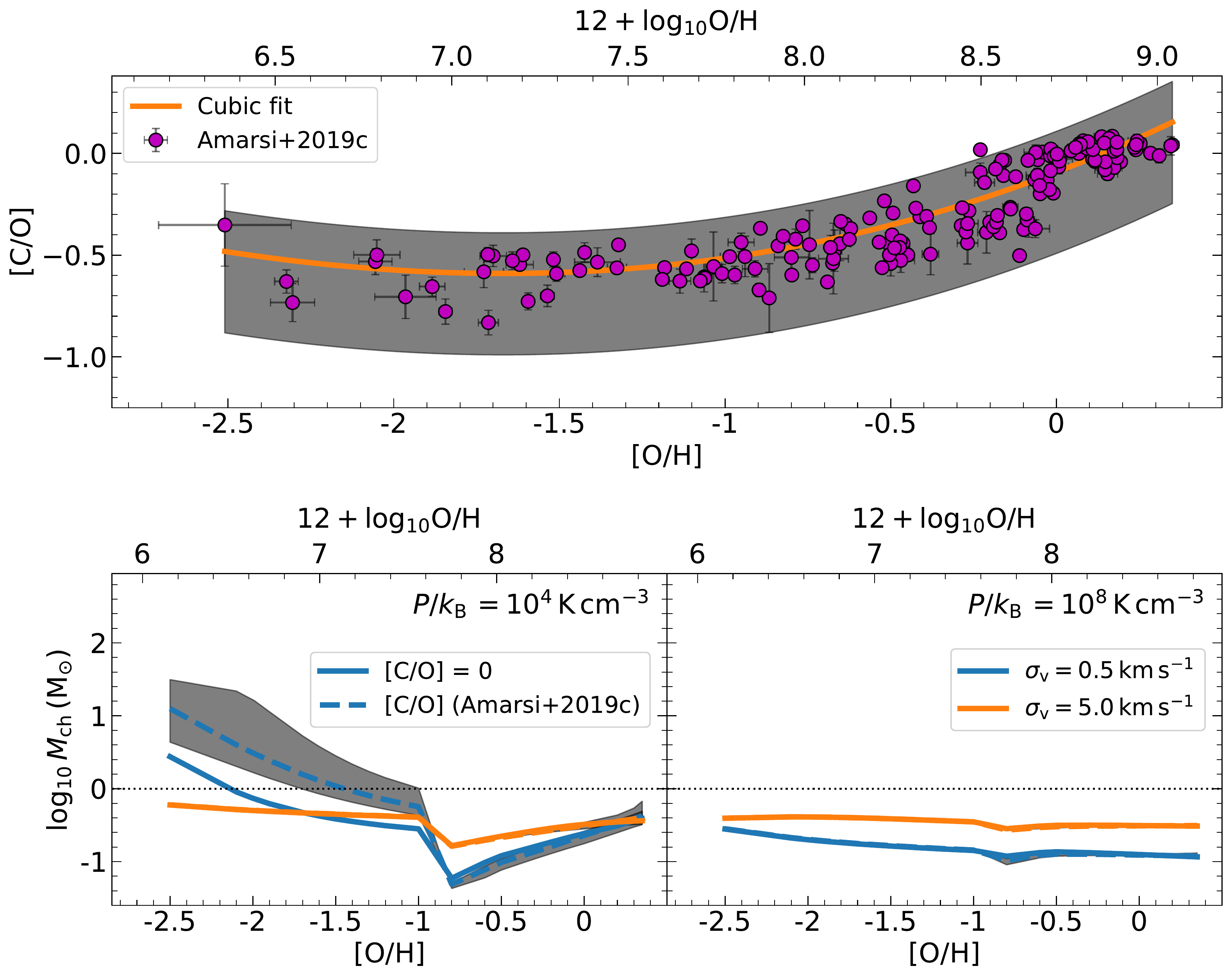}
\caption{\textit{Top panel:} Observed trends in [C/O] as a function of [O/H] from 3D-NLTE abundance modeling of metal-poor Milky Way stars \protect\citep{2019A&A...630A.104A}. The orange curve represents a cubic polynomial fit to the data, and the grey-shaded region encompasses [C/O] higher (lower) than the best-fit by 0.2 (0.4) dex. \textit{Bottom panel:} Trends in the characteristic IMF mass, $M_{\rm{ch}}$, as a function of [O/H] from the models. The solid curves represent the models from \protect\cite{2022MNRAS.509.1959S} that use [C/O] = 0. The dashed curves represent models where [C/O] varies with [O/H] following the cubic fit to the data plotted in the top panel. The grey-shaded region corresponds to $M_{\rm{ch}}$ values when [C/O] is higher (lower) than the best-fit by 0.2 (0.4) dex. The conversion from [O/H] to $12 + \log_{10}\,\rm{O/H}$ is different by 0.22 dex between the top ands the bottom panels due to different normalizations for Solar O abundance in stars versus in the gas-phase because of depletion ihn the latter.}
\label{fig:amarsi}
\end{figure*}

Direct measurements of O in metal-poor stars are challenging as the atomic lines in the optical become vanishingly weak at [O/H] < $-$2; at such low metallicities only the near-UV molecular OH lines can be used, which requires significant investment in time on large telescopes with highly efficient UV spectrographs \citep{1984PASA....5..547B,1991ApJ...383L..71B,2011ApJ...743..140B}. Moreover, molecular lines are expected to be strongly susceptible to assumptions about the model atmosphere at low metallicity \citep{2006A&A...451..621G,2006ApJ...644L.121C,2007A&A...469..687C,2008ApJ...684..588F,2010A&A...519A..46G,2016A&A...593A..48G,2017A&A...598L..10G,2019ApJ...879...37N}. Thus, measured [C/O] ratios in these stars are hard to come by. Moreover, the abundances, if measured using approximate 1D or local thermodynamic equilibrium (LTE) stellar atmosphere models \citep[e.g.,][]{2013ApJ...762...28N,2014A&A...568A..25N}, need complex corrections \citep[e.g.,][]{2016MNRAS.455.3735A,2018A&ARv..26....6N}. In fact, C abundances are severely impacted by 3D-NLTE corrections, which has led to several CEMP stars being re-classified as C-normal stars \citep{2019ApJ...879...37N,2019A&A...622L...4A}. 

To avoid these caveats, we use the C and O abundances presented by \cite{2019A&A...630A.104A} for a sample of 187 FG-type main sequence stars in the Galactic thin and thick discs as well as in the metal-poor halo. The abundances were inferred via a re-analysis of the data presented in \cite{2007A&A...469..319N,2014A&A...568A..25N}, by using 3D radiative-hydrodynamic stellar atmosphere models from the Stagger-grid \citep{2011JPhCS.328a2003C,2013A&A...557A..26M} combined with 3D radiative transfer taking NLTE effects into account. The C and O abundances are based on \ion{C}{i} and \ion{O}{i} lines of high excitation potential \citep{2018A&A...616A..89A,2019A&A...624A.111A}. The advantage of this is that they have similar sensitivities to stellar parameters, and so uncertainties in the measured stellar effective temperature and surface gravity should cancel out to first order in the [C/O] ratio \citep{2019A&A...622L...4A}.

The top panel of \autoref{fig:amarsi} shows the trends in [C/O] as a function of [O/H] in the observed sample.\footnote{The [C/H] and [O/H] measured by \cite{2019A&A...630A.104A} are normalized via a similar analysis for the Sun as for their stars, corresponding to Solar carbon and oxygen abundances of $12 + \log_{10}\rm{C/H} = 8.43$ and $12 + \log_{10}\rm{O/H} = 8.68$ respectively. The reader might notice that the Solar abundance for C and O atoms per H nucleus we use ($x_{\rm{C,MW}} = 1.4\times10^{-4}$ and $x_{\rm{O,MW}} = 3\times10^{-4}$) for the models is lower than the above values by $\sim 40$ per cent. The reason for this discrepancy is that a fraction of the C and O atoms in the ISM are locked in dust grains (e.g., SiO, FeO, etc.). This difference does not matter for this work since we normalize the model results by the gas-phase Solar O abundance, and the data by the O abundance measured in the Sun assuming no depletion. We refer the reader to \cite{2011piim.book.....D} for a detailed discussion of C and O abundances in different components of the Galaxy.\label{foot:solarO}} We fit the observed trend with a simple cubic polynomial (orange curve) weighted by the measured $1\sigma$ uncertainty in [C/O]
\begin{equation}
\mathrm{[C/O]} = a_{\mathrm{C}}\mathrm{[O/H]^3} + b_{\mathrm{C}}\mathrm{[O/H]^2} + c_{\mathrm{C}}\mathrm{[O/H]} + d_{\mathrm{C}}   \,,
\end{equation}
where the best-fit $a_{\rm{C}} = -0.02, b_{\rm{C}} = 0.14, c_{\rm{C}} = 0.60$, and $d_{\rm{C}} = -0.09$. To cover the scatter present in the data, we simply increase (decrease) the best-fit [C/O] by $+$0.2 ($-$0.4) dex. We notice that the mean [C/O] $\propto$ [O/H] at [O/H] > $-$1 however it remains roughly constant for $-2 \leq \rm{[O/H]} \leq -1$ \citep[see also,][]{2019MNRAS.490.2838R,2020A&A...639A..37R}. Below [O/H] = $-$2, the data are too scarce to draw any meaningful conclusions, although the cubic polynomial fit shows an upturn in [C/O] at the lowest [O/H] due to the most metal-poor star. This upturn has also been noticed in other works; for example, in damped $\rm{Ly}-\alpha$ absorbers \citep{2005A&A...440..499A,2015ApJ...800...12C,2017MNRAS.467..802C} as well as EAGLE cosmological simulations \citep{2018MNRAS.473..984S}. Some authors attribute it to enrichment from Population III supernovae \citep{2009A&A...500.1143F,2012MNRAS.421L..29S}, while others argue for enrichment from low-metallicity AGB stars \citep{2018MNRAS.473..984S}. If this upturn were real, it would imply that [C/O] tends to approach zero again at low [O/H], so the resulting $M_{\rm{ch}}$ at low [O/H] would be similar to that reported in \cite{2022MNRAS.509.1959S}; for example, $M_{\rm{ch}} \approx 20\,\rm{M_{\odot}}$ at [O/H] = -3.

Before we present the results from the model, a key question to address here is whether the [C/O] ratios that \cite{2019A&A...630A.104A} find are representative of the [C/O] ratio of the ISM out of which these stars formed \citep[e.g.,][]{2015A&A...579A..28B}, because our IMF models use ISM abundances. Given that all the stars in this sample are C-normal ([C/Fe] < $+$0.7, \citealt{2019A&A...630A.104A}, figure 11), it is not likely that C was accreted later on due to mass transfer by a companion (as is the case for \textit{s}-process rich CEMP-s stars; see \citealt{2015A&A...576A.118A} and \citealt{2019A&A...621A.108A}). Moreover, these main-sequence stars have not undergone any helium burning, precluding any intrinsic enhancement of C on the stellar surface. Thus, we can safely assume that the C abundances measured in these stars reflects the intrinsic ISM C abundance out of which they were born. Another category of CEMP stars is called CEMP-no, which do not show an enhancement in \textit{s}-process elements \citep[e.g.,][]{2016ApJ...833...20Y,2019A&A...621A.108A}. A fraction of these stars are proposed to have evolved in isolation \citep{2019ApJ...878...97Y}, so their [C/O] ratios should also reflect the ISM [C/O] that they were born with \citep{2015A&A...579A..28B,2016A&A...586A.160H,2019ApJ...879...37N,2022arXiv220608299L,2022arXiv220912224Z}. While the available O measurements in CEMP-no stars suggest [C/O] $\sim$ 0 \citep[figure 8]{2015ARA&A..53..631F}, the scatter is quite large, and for a large fraction of stars only O abundance limits are available due to the reasons we mention above. Future observations that focus on obtaining O abundances in CEMP-no stars is therefore crucial to constraining their IMF.\footnote{In the absence of O measurements, Mg is often used as a tracer for oxygen (or, more broadly speaking, for $\alpha$ elements). However, at low [O/H], available data suggests that Mg does not trace O very well. In fact, there is considerable discrepancy in [O/Mg] at low [O/H] within different Galactic surveys \citep{2021MNRAS.506..150B,2022ApJS..259...35A,2022arXiv220800071H}.}

Using the best-fit [C/O] -- [O/H] relation, we produce the trends in $M_{\rm{ch}}$ -- [O/H] at low and high pressures. The bottom panels of \autoref{fig:amarsi} shows the results. Not surprisingly, the trends in $M_{\rm{ch}}$ at high pressure remain unaffected due to varying [C/O] as suggested by the data. The only appreciable difference in $M_{\rm{ch}}$ for a varying [C/O] occurs in models with low pressure and high density (blue curves in the bottom-left panel). At low metallicity ([O/H] < $-$1), as compared to the fiducial model with [C/O] = 0, we find that a varying [C/O] produces $M_{\rm{ch}}$ higher by a factor $\lesssim 5$. Additionally, the metallicity ([O/H]) where the IMF transitions from top- to bottom-heavy shifts by somewhere between $+$0.4 dex and $+$1.1 dex; the exact location of the transition is uncertain owing to the scatter present in the data, which corresponds to the grey-shaded region around the dashed blue curve in the bottom panel of \autoref{fig:amarsi}. Interestingly, at high metallicity ([O/H] > $-$1), we also find that the scatter in the data introduces uncertainty in $M_{\rm{ch}}$ by a factor $\lesssim 2$. While a factor of 2 variation in $M_{\rm{ch}}$ is not substantial when $M_{\rm{ch}}$ is large, it can have measurable consequences when $M_{\rm{ch}}$ is sub-Solar, because sub-Solar stars live for much longer.

Overall, the non-constant [C/O] as measured in metal-poor stars would have influenced the IMF for the next generation of stars in a low pressure and high density ISM. Additionally, in such an environment, the data suggests that the transition to a bottom-heavy IMF would occur later on, after the ISM was enriched to 1 dex higher in O. However, a non-constant [C/O] would have had no significant impact on the characteristic stellar mass if the ISM was highly pressurized.  

\begin{figure*}
\includegraphics[width=1.0\textwidth]{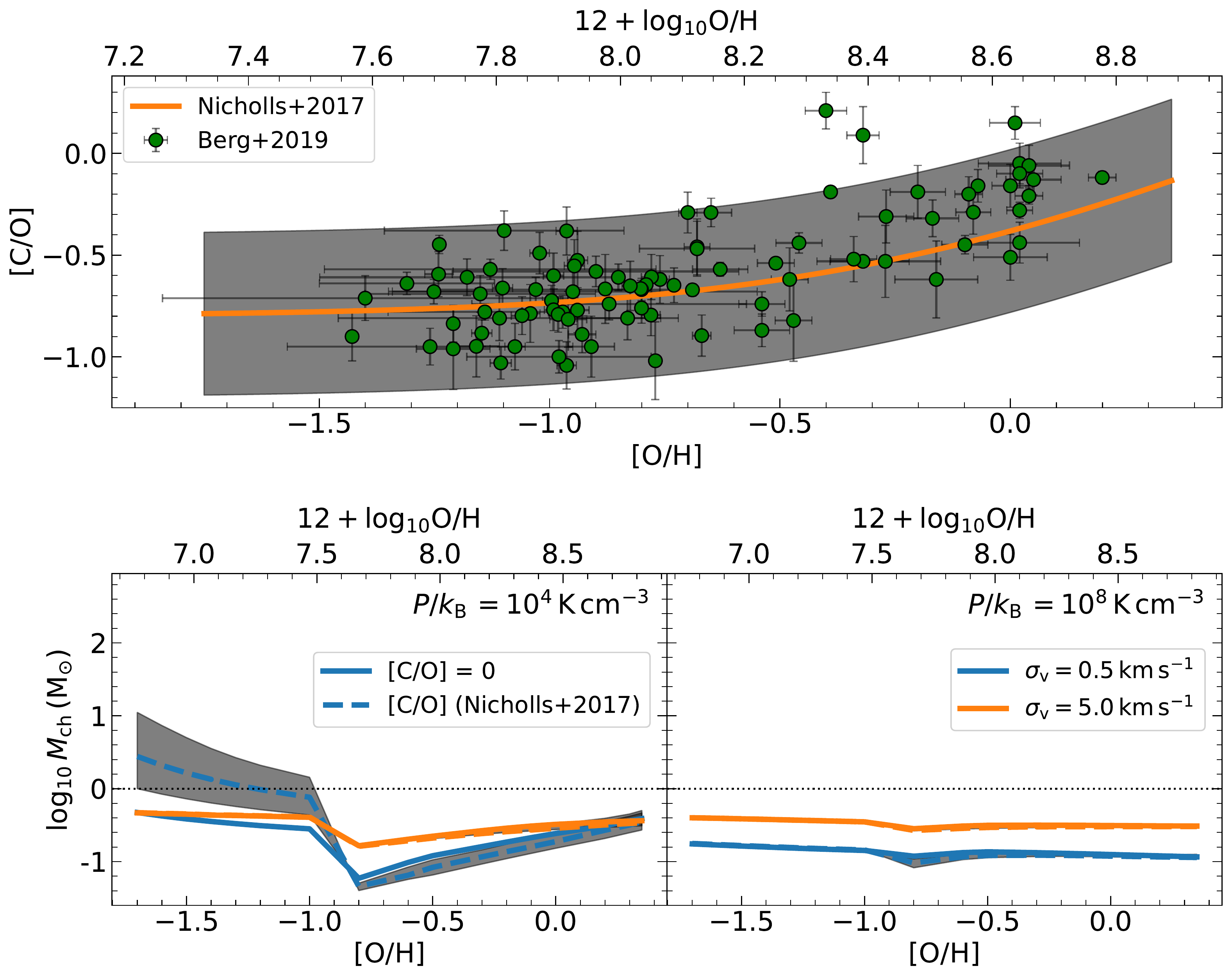}
\caption{Same as \autoref{fig:amarsi} but for gas-phase (ISM) measurements from \ion{H}{ii} regions in metal-poor dwarf galaxies. The data plotted in the top panel is taken from \protect\cite{2019ApJ...874...93B}. The orange curve in the top panel represents the empirical [C/O]$-$[O/H] relation from \protect\cite{2017MNRAS.466.4403N}. The grey-shaded region encompasses [C/O] different than the best-fit by $\pm$0.4 dex.}
\label{fig:berg}
\end{figure*}

\subsection{Insights from observations of metal-poor dwarf galaxies}
\label{s:galaxies}
Metal-poor dwarf galaxies are ideal laboratories to test variations in the IMF since the physical conditions therein are quite different from those in the Milky Way \citep[e.g.,][]{2018ApJ...855...20G}. The abundances of nebular C and O have been measured in several metal-poor dwarf galaxies using high resolution ultra-violet (UV) spectroscopy \citep{1995ApJ...443...64G,1999ApJ...513..168G,2016ApJ...827..126B,2019ApJ...874...93B,2022A&A...659A..16L}. The collisionally excited UV lines of \ion{C}{iii} and \ion{O}{iii} that are used to measure the [C/O] ratio do not suffer from reddening since the interstellar extinction is invariant over 1600 -- 2000 \AA \,\citep{2018ApJ...863...14B,2022arXiv220909047M}. These lines are also largely insensitive to variations in the physical conditions of the \ion{H}{ii} regions. Additional measurements have also been obtained using the optical recombination lines of \ion{C}{ii} and \ion{O}{ii} \citep{2007ApJ...656..168L,2009ApJ...700..654E,2014MNRAS.443..624E}. The resulting trends from the optical data are similar to that obtained from UV lines \citep[e.g.,][]{2014MNRAS.443..624E,2017MNRAS.467.3759T}, however the abundances can differ by as high as 0.3 dex (the so-called abundance discrepancy problem; see \citealt{2007ApJ...670..457G,2007ApJ...656..168L}). Additional uncertainty can arise from ionization correction factors that are applied to take into account the existence of other ionized states of C and O which are not directly measured \citep{1978A&A....66..257S,1999ApJ...513..168G,2014MNRAS.443..624E,2016ApJ...827..126B}.

The top panel of \autoref{fig:berg} shows measurements of the [C/O] ratio as a function of [O/H] in 93 metal-poor dwarf galaxies that we adopt from \cite{2019ApJ...874...93B}, which includes data compiled from various studies, including at redshifts $2-3$ \citep{2005ApJ...631..231P,2007ApJ...670..457G,2007ApJ...656..168L,2009ApJ...700..654E,2014MNRAS.443..624E,2010ApJ...719.1168E,2012MNRAS.427.1973C,2014ApJ...790..144B,2014MNRAS.445.3200S,2014MNRAS.440.1794J,2016ApJ...821L..27V,2016ApJ...826..159S,2017NatAs...1E..52A,2018ApJ...859..164B,2018ApJ...853...87R}.\footnote{Similar to the stellar data we use in \autoref{s:metalpoor}, the \cite{2019ApJ...874...93B} data is normalized to a Solar C abundance of $12 + \log_{10}\rm{C/H} = 8.43$ and Solar O abundance of $12 + \log_{10}\rm{O/H} = 8.69$ \citep{2010Ap&SS.328..179G}.} Note that these data are not corrected for the depletion of C and O onto dust, largely due to uncertainties in the chemical evolution of C in the diffuse and the ionized ISM at low metallicities \citep{2011AJ....141...22S,2011piim.book.....D,2014arXiv1402.4765J}. Some authors argue for a correction factor due to dust depletion of $\sim 0.1$ dex for both C and O at high metallicities \citep{2010ApJ...724..791P,2014MNRAS.443..624E}, which would leave the [C/O] ratio invariant to dust depletion. However, this effect is proposed to be smaller at low metallicities due to small extinctions \citep{2019ApJ...874...93B}.

In the top panel of \autoref{fig:berg}, we also overplot the empirical relation developed by \cite{2017MNRAS.466.4403N} based on scaling relations using stellar and gas-phase abundances data
\begin{equation}
\mathrm{[C/O]} = \log_{10} \left[10^{e_\mathrm{C}} + 10^{\log_{10}\rm{(O/H)} + f_{\mathrm{C}}}\right]\,,
\end{equation}
where $e_{\rm{C}} = -0.8$ and $f_{\rm{C}} = 2.72$. The data show a trend where [C/O] is roughly constant for [O/H] < -0.7, and linearly increases with [O/H] otherwise \citep{2017MNRAS.466.4403N}. However, given the uncertainty and the scatter in the measurements, the trend can be equally well described by a monotonically increasing function in [O/H] \citep{1995ApJ...443...64G}. The former has been interpreted to reflect the primary production of C at low [O/H] from intermediate-mass stars, and a pseudo-secondary production at high [O/H], possibly from winds of $M_{\star} > 10\,\rm{M_{\odot}}$ stars, while the latter reflects the possibility of pseudo-secondary C production down to the lowest measured [O/H]. Nonetheless, it is clear that the average [C/O] < 0 for the range of [O/H] covered by the data. A negative [C/O] could also be shaped by dwarf galaxies preferentially losing oxygen in galactic outflows, as predicted by both the theory of ISM metallicity \citep{1999ApJ...513..142M,2007ApJ...658..941D,2017ApJ...835..136R,2018ApJ...869...94E,2019A&A...630A.140R,2021MNRAS.502.5935S,2021MNRAS.504...53S} and observations of galactic outflows in dwarf galaxies \citep{2018MNRAS.481.1690C,Lopez20a,Cameron21a}.

We use the empirical relation of \cite{2017MNRAS.466.4403N} to predict variations in $M_{\rm{ch}}$ for the four canonical models with low/high pressures and effective velocity dispersions (or, densities). To encapsulate the scatter present in the data, we simply change the resulting [C/O] by $\pm$0.4 dex at all [O/H], as shown with the grey-shaded region in the top panel of \autoref{fig:berg}. The bottom panel of \autoref{fig:berg} presents the results, which are qualitatively similar to what we find in \autoref{s:metalpoor} (\autoref{fig:amarsi}) from observations of metal-poor stars. As in \autoref{s:results}, we see that a varying [C/O] only impacts the model with low pressure and high density (blue curves in the bottom-left panel). The consequence of a varying [C/O] ratio according to the \citeauthor{2017MNRAS.466.4403N} relation is that $M_{\rm{ch}}$ can be larger by a factor $\lesssim 7$ compared to the default case where C and O scale together. Additionally, we also see that the IMF does not transition to bottom-heavy until the gas-phase oxygen abundance exceeds 7.5 on the absolute scale (or, [O/H] > $-$1.2). This is considerably different from the case where C and O scale together, and the IMF transitions to bottom-heavy at an oxygen abundance that is smaller by an order of magnitude. This is further complicated by the scatter present in the data, as the grey-shaded band in the bottom panel of \autoref{fig:berg}. The impact of the observed trends in [C/O] as a function of [O/H] remains negligible at high pressures.

Thus, we find that the non-zero [C/O] ratio in metal-poor dwarf galaxies can have a strong impact on the characteristic stellar mass of IMF in a low-pressure but dense ISM. However, it does not have an appreciable impact on the IMF if the ISM is highly pressurized. Measurements of the [C/O] ratio in galaxies with metallicities $7 < 12+\log_{10}\,\rm{O/H} < 7.7$ are highly desirable to constrain the transition of the IMF from top- to bottom-heavy.

\section{Conclusions and future outlook}
\label{s:conclusions}
There is compelling evidence for a non-Solar scaling of the abundance of carbon with respect to oxygen in metal-poor environments. Specifically, observations find that the [C/O] ratio linearly changes with [O/H] at high metallicities, but flattens out at low [O/H], possibly also displaying an upturn below [O/H] = $-$3. At [O/H] < $-$1, the [C/O] ratio can vary by almost an order of magnitude \citep{2019A&A...630A.104A,2019ApJ...874...93B}. While there has been immense work carried out to reproduce the observed [C/O] -- [O/H] relation in metal-poor stars as well as metal-poor dwarf galaxies using IMF variations, the impact of the C and O abundances on the IMF has been largely unexplored.

In this work, we extend the calculations of collapsing dusty molecular clouds undergoing protostellar radiation feedback across a wide range of pressures, densities, and metallicities from primordial to super-Solar \cite{2022MNRAS.509.1959S}. We particularly emphasize the impacts of non-zero (or, non-Solar-scaled) [C/O] ratio on the characteristic stellar mass (or, the peak mass, $M_{\rm{ch}}$) of the IMF at low [O/H]. We find that as long as the ISM pressure is high $P/k_{\rm{B}}=10^8\,\rm{K\,cm^{-3}}$, typical of super-star clusters and starburst galaxies, a non-constant [C/O] has no impact on $M_{\rm{ch}}$, because dust completely dominates the thermodynamics of collapsing gas even at metallicities as low as [O/H] = $-$3, beyond which molecular $\rm{H_2}$ takes over as the dominant gas coolant. A varying [C/O] ratio also does not seem to matter for the case where the ISM pressure is low $P/k_{\rm{B}}=10^4\,\rm{K\,cm^{-3}}$, typical of main sequence star-forming galaxies, so long as the density is also low, or equivalently the cloud velocity dispersion is high.

Nonetheless, in the regime of pressure and gas density / velocity dispersion characteristic of giant molecular clouds in local galaxies, a varying [C/O] ratio significantly impacts $M_{\rm{ch}}$. Using the observed trends in [C/O] as a function of [O/H] from two distinct sources -- metal-poor stars in the Milky Way \citep{2019A&A...630A.104A}, and metal-poor dwarf galaxies \citep{2019ApJ...874...93B} -- we show that the resulting $M_{\rm{ch}}$ implied by these trends can be different by a factor $\lesssim 7$ at $-3 \leq \rm{[O/H]} \leq -1$. This is due to the reduction in the overall gas cooling rate due to low C abundances as compared to O. Additionally, the metallicity at which the IMF transitions from top-heavy to bottom-heavy (\textit{i.e.,} where $M_{\rm{ch}}$ decreases from $ >1\,\rm{M_{\odot}}$ to $ < 1\,\rm{M_{\odot}}$) also shifts to higher [O/H] by $\sim$ 1 dex. These results are rather insensitive to the choice of the chemical state of C and O in the ISM, or the adopted scaling of the dust-to-metal ratio with [O/H]. 

Thus, we find that the abundances of C and O significantly impact the IMF at low metallicities where cooling provided by fine structure metal lines dominates gas thermodynamics, at least in the case where the ISM pressure is low and density is high. Our results support the hypothesis that fine structure cooling by C and O lines plays a key role in the transition from Population III to Population II star formation \citep{2003Natur.425..812B,2007MNRAS.380L..40F}. This finding also has potential consequences for the origin of CEMP-no stars; however, measurements of O abundance in these stars are needed. A characteristic IMF mass of $1-10\,\rm{M_{\odot}}$ at $-2.5 \leq \rm{[O/H]} \leq -1.5$ implies an overabundance of AGB stars that can re-enrich the metal-poor ISM with C on timescales of the order of a few 100 Myr \citep{2014ApJ...797...44F,2015ApJS..219...40C,2022arXiv220905587G}, potentially even driving [C/O] > 0 for the subsequent generation of star formation (e.g., CEMP-no stars). Note, however, that our models are restricted to predicting changes in the characteristic mass of the IMF, and not its entire functional form. It is possible that the slope at the high-mass end of the IMF is also impacted by a varying [C/O] ratio at low [O/H], an avenue we plan to investigate in future work.

Finding the IMF and the elemental abundances is an iterative process, since both impact each other in numerous ways. Given the new era of discoveries of very high-redshift galaxies by JWST with measurements of ISM abundances \citep[e.g.,][]{Adams_Conselice_2022,2022arXiv220713693K,2022arXiv220803281T}, it is now more important than ever to self-consistently model the IMF and the ISM abundances to correctly predict the yields of different elements \citep{2015MNRAS.452.1447K}, figure out dominant nucleosynthesis and feedback channels at low metallicities \citep{2020ApJ...900..179K}, interpret metal distribution and ionization budget in the ISM \citep{2021MNRAS.502.5935S}, fit spectral energy distribution (SED) models to integrated galaxy spectra \citep{2020MNRAS.498.5581B}, and more broadly, investigate star formation and metal enrichment in diverse, metal-poor environments.

\section*{Acknowledgements}
We thank Sven Buder for discussions on oxygen and magnesium abundances in stars, Gary Da Costa for discussions on [C/O] in metal-poor stars, and Shyam Menon for discussions on super star clusters. We also thank an anonymous reviewer for their comments that helped improve the manuscript. PS is supported by the Australian Government Research Training Program (RTP) Scholarship. PS also acknowledges support in the form of Oort Fellowship at Leiden Observatory, and the International Astronomical Union -- Gruber Foundation Fellowship. AMA acknowledges support from the Swedish Research Council (VR 2020-03940). KG is supported by the Australian Research Council (ARC) through the Discovery Early Career Researcher Award (DECRA) Fellowship DE220100766 funded by the Australian Government. MRK and PS acknowledge support from the ARC Future Fellowship funding scheme, award FT180100375. GC is supported by Research Fellowships of the Japan Society for the Promotion of Science. AR gratefully acknowledges support from the Italian funding scheme ``The quest for the first stars'' (Code 2017T4ARJ5\_001). TN is supported by the Stromlo Fellowship at Australian National University. Parts of this research were supported by the ARC Centre of Excellence for All Sky Astrophysics in 3 Dimensions (ASTRO 3D), through project number CE170100013. Analysis was performed using \texttt{NUMPY} \citep{oliphant2006guide,2020arXiv200610256H} and \texttt{SCIPY} \citep{2020NatMe..17..261V}; plots were created using \texttt{MATPLOTLIB} \citep{Hunter:2007}. This research has made extensive use of NASA's Astrophysics Data System Bibliographic Services, image to data software \texttt{WEBPLOT} \texttt{DIGITIZER}, and the Leiden Atomic and Molecular Database \citep[LAMDA,][]{2005A&A...432..369S,2020Atoms...8...15V}. This research has also made extensive use of \texttt{MATHEMATICA} for numerical analyses.

\section*{Data Availability}
The data we use in this work are available in \cite{2019A&A...630A.104A} and \cite{2019ApJ...874...93B}.
 


\bibliographystyle{mnras}
\bibliography{references}

\begin{thebibliography}{}
\makeatletter
\relax
\def\mn@urlcharsother{\let\do\@makeother \do\$\do\&\do\#\do\^\do\_\do\%\do\~}
\def\mn@doi{\begingroup\mn@urlcharsother \@ifnextchar [ {\mn@doi@}
  {\mn@doi@[]}}
\def\mn@doi@[#1]#2{\def\@tempa{#1}\ifx\@tempa\@empty \href
  {http://dx.doi.org/#2} {doi:#2}\else \href {http://dx.doi.org/#2} {#1}\fi
  \endgroup}
\def\mn@eprint#1#2{\mn@eprint@#1:#2::\@nil}
\def\mn@eprint@arXiv#1{\href {http://arxiv.org/abs/#1} {{\tt arXiv:#1}}}
\def\mn@eprint@dblp#1{\href {http://dblp.uni-trier.de/rec/bibtex/#1.xml}
  {dblp:#1}}
\def\mn@eprint@#1:#2:#3:#4\@nil{\def\@tempa {#1}\def\@tempb {#2}\def\@tempc
  {#3}\ifx \@tempc \@empty \let \@tempc \@tempb \let \@tempb \@tempa \fi \ifx
  \@tempb \@empty \def\@tempb {arXiv}\fi \@ifundefined
  {mn@eprint@\@tempb}{\@tempb:\@tempc}{\expandafter \expandafter \csname
  mn@eprint@\@tempb\endcsname \expandafter{\@tempc}}}

\bibitem[\protect\citeauthoryear{{Abate}, {Pols}, {Karakas}  \&
  {Izzard}}{{Abate} et~al.}{2015}]{2015A&A...576A.118A}
{Abate} C.,  {Pols} O.~R.,  {Karakas} A.~I.,   {Izzard} R.~G.,  2015, \mn@doi
  [\aap] {10.1051/0004-6361/201424739}, \href
  {https://ui.adsabs.harvard.edu/abs/2015A&A...576A.118A} {576, A118}

\bibitem[\protect\citeauthoryear{{Abdurro'uf} et~al.,}{{Abdurro'uf}
  et~al.}{2022}]{2022ApJS..259...35A}
{Abdurro'uf} et~al., 2022, \mn@doi [\apjs] {10.3847/1538-4365/ac4414}, \href
  {https://ui.adsabs.harvard.edu/abs/2022ApJS..259...35A} {259, 35}

\bibitem[\protect\citeauthoryear{{Abohalima} \& {Frebel}}{{Abohalima} \&
  {Frebel}}{2018}]{2018ApJS..238...36A}
{Abohalima} A.,  {Frebel} A.,  2018, \mn@doi [\apjs]
  {10.3847/1538-4365/aadfe9}, \href
  {https://ui.adsabs.harvard.edu/abs/2018ApJS..238...36A} {238, 36}

\bibitem[\protect\citeauthoryear{{Adams} et~al.,}{{Adams}
  et~al.}{2022}]{Adams_Conselice_2022}
{Adams} N.~J.,  et~al., 2022, arXiv e-prints, \href
  {https://ui.adsabs.harvard.edu/abs/2022arXiv220711217A} {p. arXiv:2207.11217}

\bibitem[\protect\citeauthoryear{{Akerman}, {Carigi}, {Nissen}, {Pettini}  \&
  {Asplund}}{{Akerman} et~al.}{2004}]{2004A&A...414..931A}
{Akerman} C.~J.,  {Carigi} L.,  {Nissen} P.~E.,  {Pettini} M.,   {Asplund} M.,
  2004, \mn@doi [\aap] {10.1051/0004-6361:20034188}, \href
  {https://ui.adsabs.harvard.edu/abs/2004A&A...414..931A} {414, 931}

\bibitem[\protect\citeauthoryear{{Akerman}, {Ellison}, {Pettini}  \&
  {Steidel}}{{Akerman} et~al.}{2005}]{2005A&A...440..499A}
{Akerman} C.~J.,  {Ellison} S.~L.,  {Pettini} M.,   {Steidel} C.~C.,  2005,
  \mn@doi [\aap] {10.1051/0004-6361:20052947}, \href
  {https://ui.adsabs.harvard.edu/abs/2005A&A...440..499A} {440, 499}

\bibitem[\protect\citeauthoryear{{Amarsi}, {Asplund}, {Collet}  \&
  {Leenaarts}}{{Amarsi} et~al.}{2016}]{2016MNRAS.455.3735A}
{Amarsi} A.~M.,  {Asplund} M.,  {Collet} R.,   {Leenaarts} J.,  2016, \mn@doi
  [\mnras] {10.1093/mnras/stv2608}, \href
  {https://ui.adsabs.harvard.edu/abs/2016MNRAS.455.3735A} {455, 3735}

\bibitem[\protect\citeauthoryear{{Amarsi}, {Barklem}, {Asplund}, {Collet}  \&
  {Zatsarinny}}{{Amarsi} et~al.}{2018}]{2018A&A...616A..89A}
{Amarsi} A.~M.,  {Barklem} P.~S.,  {Asplund} M.,  {Collet} R.,   {Zatsarinny}
  O.,  2018, \mn@doi [\aap] {10.1051/0004-6361/201832770}, \href
  {https://ui.adsabs.harvard.edu/abs/2018A&A...616A..89A} {616, A89}

\bibitem[\protect\citeauthoryear{{Amarsi}, {Nissen}, {Asplund}, {Lind}  \&
  {Barklem}}{{Amarsi} et~al.}{2019a}]{2019A&A...622L...4A}
{Amarsi} A.~M.,  {Nissen} P.~E.,  {Asplund} M.,  {Lind} K.,   {Barklem} P.~S.,
  2019a, \mn@doi [\aap] {10.1051/0004-6361/201834480}, \href
  {https://ui.adsabs.harvard.edu/abs/2019A&A...622L...4A} {622, L4}

\bibitem[\protect\citeauthoryear{{Amarsi}, {Barklem}, {Collet}, {Grevesse}  \&
  {Asplund}}{{Amarsi} et~al.}{2019b}]{2019A&A...624A.111A}
{Amarsi} A.~M.,  {Barklem} P.~S.,  {Collet} R.,  {Grevesse} N.,   {Asplund} M.,
   2019b, \mn@doi [\aap] {10.1051/0004-6361/201833603}, \href
  {https://ui.adsabs.harvard.edu/abs/2019A&A...624A.111A} {624, A111}

\bibitem[\protect\citeauthoryear{{Amarsi}, {Nissen}  \&
  {Sk{\'u}lad{\'o}ttir}}{{Amarsi} et~al.}{2019c}]{2019A&A...630A.104A}
{Amarsi} A.~M.,  {Nissen} P.~E.,   {Sk{\'u}lad{\'o}ttir} {\'A}.,  2019c,
  \mn@doi [\aap] {10.1051/0004-6361/201936265}, \href
  {https://ui.adsabs.harvard.edu/abs/2019A&A...630A.104A} {630, A104}

\bibitem[\protect\citeauthoryear{{Amor{\'\i}n} et~al.,}{{Amor{\'\i}n}
  et~al.}{2017}]{2017NatAs...1E..52A}
{Amor{\'\i}n} R.,  et~al., 2017, \mn@doi [Nature Astronomy]
  {10.1038/s41550-017-0052}, \href
  {https://ui.adsabs.harvard.edu/abs/2017NatAs...1E..52A} {1, 0052}

\bibitem[\protect\citeauthoryear{{Aoki}, {Beers}, {Christlieb}, {Norris},
  {Ryan}  \& {Tsangarides}}{{Aoki} et~al.}{2007}]{2007ApJ...655..492A}
{Aoki} W.,  {Beers} T.~C.,  {Christlieb} N.,  {Norris} J.~E.,  {Ryan} S.~G.,
  {Tsangarides} S.,  2007, \mn@doi [\apj] {10.1086/509817}, \href
  {https://ui.adsabs.harvard.edu/abs/2007ApJ...655..492A} {655, 492}

\bibitem[\protect\citeauthoryear{{Aoki}, {Aoki}  \& {Fran{\c{c}}ois}}{{Aoki}
  et~al.}{2020}]{2020A&A...636A.111A}
{Aoki} M.,  {Aoki} W.,   {Fran{\c{c}}ois} P.,  2020, \mn@doi [\aap]
  {10.1051/0004-6361/201936535}, \href
  {https://ui.adsabs.harvard.edu/abs/2020A&A...636A.111A} {636, A111}

\bibitem[\protect\citeauthoryear{{Arellano-C{\'o}rdova}
  et~al.,}{{Arellano-C{\'o}rdova} et~al.}{2022}]{2022arXiv220802562A}
{Arellano-C{\'o}rdova} K.~Z.,  et~al., 2022, arXiv e-prints, \href
  {https://ui.adsabs.harvard.edu/abs/2022arXiv220802562A} {p. arXiv:2208.02562}

\bibitem[\protect\citeauthoryear{{Arentsen}, {Starkenburg}, {Shetrone}, {Venn},
  {Depagne}  \& {McConnachie}}{{Arentsen} et~al.}{2019}]{2019A&A...621A.108A}
{Arentsen} A.,  {Starkenburg} E.,  {Shetrone} M.~D.,  {Venn} K.~A.,  {Depagne}
  {\'E}.,   {McConnachie} A.~W.,  2019, \mn@doi [\aap]
  {10.1051/0004-6361/201834146}, \href
  {https://ui.adsabs.harvard.edu/abs/2019A&A...621A.108A} {621, A108}

\bibitem[\protect\citeauthoryear{{Arentsen} et~al.,}{{Arentsen}
  et~al.}{2020}]{2020MNRAS.496.4964A}
{Arentsen} A.,  et~al., 2020, \mn@doi [\mnras] {10.1093/mnras/staa1661}, \href
  {https://ui.adsabs.harvard.edu/abs/2020MNRAS.496.4964A} {496, 4964}

\bibitem[\protect\citeauthoryear{{Arentsen}, {Placco}, {Lee}, {Aguado},
  {Martin}, {Starkenburg}  \& {Yoon}}{{Arentsen}
  et~al.}{2022}]{2022MNRAS.515.4082A}
{Arentsen} A.,  {Placco} V.~M.,  {Lee} Y.~S.,  {Aguado} D.~S.,  {Martin} N.~F.,
   {Starkenburg} E.,   {Yoon} J.,  2022, \mn@doi [\mnras]
  {10.1093/mnras/stac2062}, \href
  {https://ui.adsabs.harvard.edu/abs/2022MNRAS.515.4082A} {515, 4082}

\bibitem[\protect\citeauthoryear{{Asplund}, {Grevesse}, {Sauval}  \&
  {Scott}}{{Asplund} et~al.}{2009}]{2009ARA&A..47..481A}
{Asplund} M.,  {Grevesse} N.,  {Sauval} A.~J.,   {Scott} P.,  2009, \mn@doi
  [\araa] {10.1146/annurev.astro.46.060407.145222}, \href
  {https://ui.adsabs.harvard.edu/abs/2009ARA&A..47..481A} {47, 481}

\bibitem[\protect\citeauthoryear{{Bastian}, {Saglia}, {Goudfrooij},
  {Kissler-Patig}, {Maraston}, {Schweizer}  \& {Zoccali}}{{Bastian}
  et~al.}{2006}]{2006A&A...448..881B}
{Bastian} N.,  {Saglia} R.~P.,  {Goudfrooij} P.,  {Kissler-Patig} M.,
  {Maraston} C.,  {Schweizer} F.,   {Zoccali} M.,  2006, \mn@doi [\aap]
  {10.1051/0004-6361:20054177}, \href
  {https://ui.adsabs.harvard.edu/abs/2006A&A...448..881B} {448, 881}

\bibitem[\protect\citeauthoryear{{Bastian}, {Covey}  \& {Meyer}}{{Bastian}
  et~al.}{2010}]{2010ARA&A..48..339B}
{Bastian} N.,  {Covey} K.~R.,   {Meyer} M.~R.,  2010, \mn@doi [\araa]
  {10.1146/annurev-astro-082708-101642}, \href
  {https://ui.adsabs.harvard.edu/abs/2010ARA&A..48..339B} {48, 339}

\bibitem[\protect\citeauthoryear{{Bate}}{{Bate}}{2014}]{2014MNRAS.442..285B}
{Bate} M.~R.,  2014, \mn@doi [\mnras] {10.1093/mnras/stu795}, \href
  {https://ui.adsabs.harvard.edu/abs/2014MNRAS.442..285B} {442, 285}

\bibitem[\protect\citeauthoryear{{Bate}}{{Bate}}{2019}]{2019MNRAS.484.2341B}
{Bate} M.~R.,  2019, \mn@doi [\mnras] {10.1093/mnras/stz103}, \href
  {https://ui.adsabs.harvard.edu/abs/2019MNRAS.484.2341B} {484, 2341}

\bibitem[\protect\citeauthoryear{{Bate} \& {Keto}}{{Bate} \&
  {Keto}}{2015}]{2015MNRAS.449.2643B}
{Bate} M.~R.,  {Keto} E.~R.,  2015, \mn@doi [\mnras] {10.1093/mnras/stv451},
  \href {https://ui.adsabs.harvard.edu/abs/2015MNRAS.449.2643B} {449, 2643}

\bibitem[\protect\citeauthoryear{{Bayliss}, {Rigby}, {Sharon}, {Wuyts},
  {Florian}, {Gladders}, {Johnson}  \& {Oguri}}{{Bayliss}
  et~al.}{2014}]{2014ApJ...790..144B}
{Bayliss} M.~B.,  {Rigby} J.~R.,  {Sharon} K.,  {Wuyts} E.,  {Florian} M.,
  {Gladders} M.~D.,  {Johnson} T.,   {Oguri} M.,  2014, \mn@doi [\apj]
  {10.1088/0004-637X/790/2/144}, \href
  {https://ui.adsabs.harvard.edu/abs/2014ApJ...790..144B} {790, 144}

\bibitem[\protect\citeauthoryear{{Beers} \& {Christlieb}}{{Beers} \&
  {Christlieb}}{2005}]{2005ARA&A..43..531B}
{Beers} T.~C.,  {Christlieb} N.,  2005, \mn@doi [\araa]
  {10.1146/annurev.astro.42.053102.134057}, \href
  {https://ui.adsabs.harvard.edu/abs/2005ARA&A..43..531B} {43, 531}

\bibitem[\protect\citeauthoryear{{Bellstedt} et~al.,}{{Bellstedt}
  et~al.}{2020}]{2020MNRAS.498.5581B}
{Bellstedt} S.,  et~al., 2020, \mn@doi [\mnras] {10.1093/mnras/staa2620}, \href
  {https://ui.adsabs.harvard.edu/abs/2020MNRAS.498.5581B} {498, 5581}

\bibitem[\protect\citeauthoryear{{Berg}, {Skillman}, {Henry}, {Erb}  \&
  {Carigi}}{{Berg} et~al.}{2016}]{2016ApJ...827..126B}
{Berg} D.~A.,  {Skillman} E.~D.,  {Henry} R. B.~C.,  {Erb} D.~K.,   {Carigi}
  L.,  2016, \mn@doi [\apj] {10.3847/0004-637X/827/2/126}, \href
  {https://ui.adsabs.harvard.edu/abs/2016ApJ...827..126B} {827, 126}

\bibitem[\protect\citeauthoryear{{Berg}, {Erb}, {Auger}, {Pettini}  \&
  {Brammer}}{{Berg} et~al.}{2018}]{2018ApJ...859..164B}
{Berg} D.~A.,  {Erb} D.~K.,  {Auger} M.~W.,  {Pettini} M.,   {Brammer} G.~B.,
  2018, \mn@doi [\apj] {10.3847/1538-4357/aab7fa}, \href
  {https://ui.adsabs.harvard.edu/abs/2018ApJ...859..164B} {859, 164}

\bibitem[\protect\citeauthoryear{{Berg}, {Erb}, {Henry}, {Skillman}  \&
  {McQuinn}}{{Berg} et~al.}{2019}]{2019ApJ...874...93B}
{Berg} D.~A.,  {Erb} D.~K.,  {Henry} R. B.~C.,  {Skillman} E.~D.,   {McQuinn}
  K. B.~W.,  2019, \mn@doi [\apj] {10.3847/1538-4357/ab020a}, \href
  {https://ui.adsabs.harvard.edu/abs/2019ApJ...874...93B} {874, 93}

\bibitem[\protect\citeauthoryear{{Bessell}, {Hughes}  \& {Cottrell}}{{Bessell}
  et~al.}{1984}]{1984PASA....5..547B}
{Bessell} M.~S.,  {Hughes} S.~M.~G.,   {Cottrell} P.~L.,  1984, \mn@doi [\pasa]
  {10.1017/S1323358000017574}, \href
  {https://ui.adsabs.harvard.edu/abs/1984PASA....5..547B} {5, 547}

\bibitem[\protect\citeauthoryear{{Bessell}, {Sutherland}  \& {Ruan}}{{Bessell}
  et~al.}{1991}]{1991ApJ...383L..71B}
{Bessell} M.~S.,  {Sutherland} R.~S.,   {Ruan} K.,  1991, \mn@doi [\apjl]
  {10.1086/186244}, \href
  {https://ui.adsabs.harvard.edu/abs/1991ApJ...383L..71B} {383, L71}

\bibitem[\protect\citeauthoryear{{Bialy} \& {Sternberg}}{{Bialy} \&
  {Sternberg}}{2015}]{2015MNRAS.450.4424B}
{Bialy} S.,  {Sternberg} A.,  2015, \mn@doi [\mnras] {10.1093/mnras/stv851},
  \href {https://ui.adsabs.harvard.edu/abs/2015MNRAS.450.4424B} {450, 4424}

\bibitem[\protect\citeauthoryear{{Boesgaard}, {Rich}, {Levesque}  \&
  {Bowler}}{{Boesgaard} et~al.}{2011}]{2011ApJ...743..140B}
{Boesgaard} A.~M.,  {Rich} J.~A.,  {Levesque} E.~M.,   {Bowler} B.~P.,  2011,
  \mn@doi [\apj] {10.1088/0004-637X/743/2/140}, \href
  {https://ui.adsabs.harvard.edu/abs/2011ApJ...743..140B} {743, 140}

\bibitem[\protect\citeauthoryear{{Bolatto}, {Leroy}, {Rosolowsky}, {Walter}  \&
  {Blitz}}{{Bolatto} et~al.}{2008}]{2008ApJ...686..948B}
{Bolatto} A.~D.,  {Leroy} A.~K.,  {Rosolowsky} E.,  {Walter} F.,   {Blitz} L.,
  2008, \mn@doi [\apj] {10.1086/591513}, \href
  {https://ui.adsabs.harvard.edu/abs/2008ApJ...686..948B} {686, 948}

\bibitem[\protect\citeauthoryear{{Bonifacio} et~al.,}{{Bonifacio}
  et~al.}{2015}]{2015A&A...579A..28B}
{Bonifacio} P.,  et~al., 2015, \mn@doi [\aap] {10.1051/0004-6361/201425266},
  \href {https://ui.adsabs.harvard.edu/abs/2015A&A...579A..28B} {579, A28}

\bibitem[\protect\citeauthoryear{{Bonnor}}{{Bonnor}}{1957}]{1957MNRAS.117..104B}
{Bonnor} W.~B.,  1957, \mn@doi [\mnras] {10.1093/mnras/117.1.104}, \href
  {https://ui.adsabs.harvard.edu/abs/1957MNRAS.117..104B} {117, 104}

\bibitem[\protect\citeauthoryear{{Bromm} \& {Loeb}}{{Bromm} \&
  {Loeb}}{2003}]{2003Natur.425..812B}
{Bromm} V.,  {Loeb} A.,  2003, \mn@doi [\nat] {10.1038/nature02071}, \href
  {https://ui.adsabs.harvard.edu/abs/2003Natur.425..812B} {425, 812}

\bibitem[\protect\citeauthoryear{{Bromm}, {Ferrara}, {Coppi}  \&
  {Larson}}{{Bromm} et~al.}{2001}]{2001MNRAS.328..969B}
{Bromm} V.,  {Ferrara} A.,  {Coppi} P.~S.,   {Larson} R.~B.,  2001, \mn@doi
  [\mnras] {10.1046/j.1365-8711.2001.04915.x}, \href
  {https://ui.adsabs.harvard.edu/abs/2001MNRAS.328..969B} {328, 969}

\bibitem[\protect\citeauthoryear{{Buder} et~al.,}{{Buder}
  et~al.}{2021}]{2021MNRAS.506..150B}
{Buder} S.,  et~al., 2021, \mn@doi [\mnras] {10.1093/mnras/stab1242}, \href
  {https://ui.adsabs.harvard.edu/abs/2021MNRAS.506..150B} {506, 150}

\bibitem[\protect\citeauthoryear{{Byler}, {Dalcanton}, {Conroy}, {Johnson},
  {Levesque}  \& {Berg}}{{Byler} et~al.}{2018}]{2018ApJ...863...14B}
{Byler} N.,  {Dalcanton} J.~J.,  {Conroy} C.,  {Johnson} B.~D.,  {Levesque}
  E.~M.,   {Berg} D.~A.,  2018, \mn@doi [\apj] {10.3847/1538-4357/aacd50},
  \href {https://ui.adsabs.harvard.edu/abs/2018ApJ...863...14B} {863, 14}

\bibitem[\protect\citeauthoryear{{Caffau} et~al.,}{{Caffau}
  et~al.}{2011}]{2011Natur.477...67C}
{Caffau} E.,  et~al., 2011, \mn@doi [\nat] {10.1038/nature10377}, \href
  {https://ui.adsabs.harvard.edu/abs/2011Natur.477...67C} {477, 67}

\bibitem[\protect\citeauthoryear{{Cameron} et~al.,}{{Cameron}
  et~al.}{2021}]{Cameron21a}
{Cameron} A.~J.,  et~al., 2021, \apjl, \href
  {https://ui.adsabs.harvard.edu/abs/2021arXiv210813211C} {918, L16}

\bibitem[\protect\citeauthoryear{{Carigi} \& {Hernandez}}{{Carigi} \&
  {Hernandez}}{2008}]{2008MNRAS.390..582C}
{Carigi} L.,  {Hernandez} X.,  2008, \mn@doi [\mnras]
  {10.1111/j.1365-2966.2008.13743.x}, \href
  {https://ui.adsabs.harvard.edu/abs/2008MNRAS.390..582C} {390, 582}

\bibitem[\protect\citeauthoryear{{Carigi}, {Peimbert}, {Esteban}  \&
  {Garc{\'\i}a-Rojas}}{{Carigi} et~al.}{2005}]{2005ApJ...623..213C}
{Carigi} L.,  {Peimbert} M.,  {Esteban} C.,   {Garc{\'\i}a-Rojas} J.,  2005,
  \mn@doi [\apj] {10.1086/428491}, \href
  {https://ui.adsabs.harvard.edu/abs/2005ApJ...623..213C} {623, 213}

\bibitem[\protect\citeauthoryear{{Caselli} \& {Myers}}{{Caselli} \&
  {Myers}}{1995}]{1995ApJ...446..665C}
{Caselli} P.,  {Myers} P.~C.,  1995, \mn@doi [\apj] {10.1086/175825}, \href
  {https://ui.adsabs.harvard.edu/abs/1995ApJ...446..665C} {446, 665}

\bibitem[\protect\citeauthoryear{{Cazaux} \& {Spaans}}{{Cazaux} \&
  {Spaans}}{2009}]{2009A&A...496..365C}
{Cazaux} S.,  {Spaans} M.,  2009, \mn@doi [\aap] {10.1051/0004-6361:200811302},
  \href {https://ui.adsabs.harvard.edu/abs/2009A&A...496..365C} {496, 365}

\bibitem[\protect\citeauthoryear{{Chabrier}}{{Chabrier}}{2003}]{2003PASP..115..763C}
{Chabrier} G.,  2003, \mn@doi [\pasp] {10.1086/376392}, \href
  {https://ui.adsabs.harvard.edu/abs/2003PASP..115..763C} {115, 763}

\bibitem[\protect\citeauthoryear{{Chabrier}, {Hennebelle}  \&
  {Charlot}}{{Chabrier} et~al.}{2014}]{2014ApJ...796...75C}
{Chabrier} G.,  {Hennebelle} P.,   {Charlot} S.,  2014, \mn@doi [\apj]
  {10.1088/0004-637X/796/2/75}, \href
  {https://ui.adsabs.harvard.edu/abs/2014ApJ...796...75C} {796, 75}

\bibitem[\protect\citeauthoryear{{Chakrabarti} \& {McKee}}{{Chakrabarti} \&
  {McKee}}{2005}]{2005ApJ...631..792C}
{Chakrabarti} S.,  {McKee} C.~F.,  2005, \mn@doi [\apj] {10.1086/432659}, \href
  {https://ui.adsabs.harvard.edu/abs/2005ApJ...631..792C} {631, 792}

\bibitem[\protect\citeauthoryear{{Chakrabarti}, {Magnelli}, {McKee}, {Lutz},
  {Berta}, {Popesso}  \& {Pozzi}}{{Chakrabarti}
  et~al.}{2013}]{2013ApJ...773..113C}
{Chakrabarti} S.,  {Magnelli} B.,  {McKee} C.~F.,  {Lutz} D.,  {Berta} S.,
  {Popesso} P.,   {Pozzi} F.,  2013, \mn@doi [\apj]
  {10.1088/0004-637X/773/2/113}, \href
  {https://ui.adsabs.harvard.edu/abs/2013ApJ...773..113C} {773, 113}

\bibitem[\protect\citeauthoryear{{Chiaki} \& {Wise}}{{Chiaki} \&
  {Wise}}{2019}]{2019MNRAS.482.3933C}
{Chiaki} G.,  {Wise} J.~H.,  2019, \mn@doi [\mnras] {10.1093/mnras/sty2984},
  \href {https://ui.adsabs.harvard.edu/abs/2019MNRAS.482.3933C} {482, 3933}

\bibitem[\protect\citeauthoryear{{Chiaki} \& {Yoshida}}{{Chiaki} \&
  {Yoshida}}{2022}]{2022MNRAS.510.5199C}
{Chiaki} G.,  {Yoshida} N.,  2022, \mn@doi [\mnras] {10.1093/mnras/stab2799},
  \href {https://ui.adsabs.harvard.edu/abs/2022MNRAS.510.5199C} {510, 5199}

\bibitem[\protect\citeauthoryear{{Chiaki}, {Marassi}, {Nozawa}, {Yoshida},
  {Schneider}, {Omukai}, {Limongi}  \& {Chieffi}}{{Chiaki}
  et~al.}{2015}]{2015MNRAS.446.2659C}
{Chiaki} G.,  {Marassi} S.,  {Nozawa} T.,  {Yoshida} N.,  {Schneider} R.,
  {Omukai} K.,  {Limongi} M.,   {Chieffi} A.,  2015, \mn@doi [\mnras]
  {10.1093/mnras/stu2298}, \href
  {https://ui.adsabs.harvard.edu/abs/2015MNRAS.446.2659C} {446, 2659}

\bibitem[\protect\citeauthoryear{{Chiaki}, {Yoshida}  \& {Hirano}}{{Chiaki}
  et~al.}{2016}]{2016MNRAS.463.2781C}
{Chiaki} G.,  {Yoshida} N.,   {Hirano} S.,  2016, \mn@doi [\mnras]
  {10.1093/mnras/stw2120}, \href
  {https://ui.adsabs.harvard.edu/abs/2016MNRAS.463.2781C} {463, 2781}

\bibitem[\protect\citeauthoryear{{Chisholm}, {Tremonti}  \&
  {Leitherer}}{{Chisholm} et~al.}{2018}]{2018MNRAS.481.1690C}
{Chisholm} J.,  {Tremonti} C.,   {Leitherer} C.,  2018, \mn@doi [\mnras]
  {10.1093/mnras/sty2380}, \href
  {https://ui.adsabs.harvard.edu/abs/2018MNRAS.481.1690C} {481, 1690}

\bibitem[\protect\citeauthoryear{{Chiti} et~al.,}{{Chiti}
  et~al.}{2018}]{2018ApJ...856..142C}
{Chiti} A.,  et~al., 2018, \mn@doi [\apj] {10.3847/1538-4357/aab663}, \href
  {https://ui.adsabs.harvard.edu/abs/2018ApJ...856..142C} {856, 142}

\bibitem[\protect\citeauthoryear{{Chon}, {Omukai}  \& {Schneider}}{{Chon}
  et~al.}{2021}]{2021MNRAS.508.4175C}
{Chon} S.,  {Omukai} K.,   {Schneider} R.,  2021, \mn@doi [\mnras]
  {10.1093/mnras/stab2497}, \href
  {https://ui.adsabs.harvard.edu/abs/2021MNRAS.508.4175C} {508, 4175}

\bibitem[\protect\citeauthoryear{{Chon}, {Ono}, {Omukai}  \&
  {Schneider}}{{Chon} et~al.}{2022}]{2022MNRAS.514.4639C}
{Chon} S.,  {Ono} H.,  {Omukai} K.,   {Schneider} R.,  2022, \mn@doi [\mnras]
  {10.1093/mnras/stac1549}, \href
  {https://ui.adsabs.harvard.edu/abs/2022MNRAS.514.4639C} {514, 4639}

\bibitem[\protect\citeauthoryear{{Christensen} et~al.,}{{Christensen}
  et~al.}{2012}]{2012MNRAS.427.1973C}
{Christensen} L.,  et~al., 2012, \mn@doi [\mnras]
  {10.1111/j.1365-2966.2012.22007.x}, \href
  {https://ui.adsabs.harvard.edu/abs/2012MNRAS.427.1973C} {427, 1973}

\bibitem[\protect\citeauthoryear{{Cohen}, {Tielens}, {Bregman}, {Witteborn},
  {Rank}, {Allamandola}, {Wooden}  \& {Jourdain de Muizon}}{{Cohen}
  et~al.}{1989}]{1989ApJ...341..246C}
{Cohen} M.,  {Tielens} A.~G.~G.~M.,  {Bregman} J.,  {Witteborn} F.~C.,  {Rank}
  D.~M.,  {Allamandola} L.~J.,  {Wooden} D.,   {Jourdain de Muizon} M.,  1989,
  \mn@doi [\apj] {10.1086/167489}, \href
  {https://ui.adsabs.harvard.edu/abs/1989ApJ...341..246C} {341, 246}

\bibitem[\protect\citeauthoryear{{Collet}, {Asplund}  \& {Trampedach}}{{Collet}
  et~al.}{2006}]{2006ApJ...644L.121C}
{Collet} R.,  {Asplund} M.,   {Trampedach} R.,  2006, \mn@doi [\apjl]
  {10.1086/505643}, \href
  {https://ui.adsabs.harvard.edu/abs/2006ApJ...644L.121C} {644, L121}

\bibitem[\protect\citeauthoryear{{Collet}, {Asplund}  \& {Trampedach}}{{Collet}
  et~al.}{2007}]{2007A&A...469..687C}
{Collet} R.,  {Asplund} M.,   {Trampedach} R.,  2007, \mn@doi [\aap]
  {10.1051/0004-6361:20066321}, \href
  {https://ui.adsabs.harvard.edu/abs/2007A&A...469..687C} {469, 687}

\bibitem[\protect\citeauthoryear{{Collet}, {Magic}  \& {Asplund}}{{Collet}
  et~al.}{2011}]{2011JPhCS.328a2003C}
{Collet} R.,  {Magic} Z.,   {Asplund} M.,  2011, in Journal of Physics
  Conference Series. p. 012003 (\mn@eprint {arXiv} {1110.5475}),
  \mn@doi{10.1088/1742-6596/328/1/012003}

\bibitem[\protect\citeauthoryear{{Conroy} \& {van Dokkum}}{{Conroy} \& {van
  Dokkum}}{2012}]{2012ApJ...760...71C}
{Conroy} C.,  {van Dokkum} P.~G.,  2012, \mn@doi [\apj]
  {10.1088/0004-637X/760/1/71}, \href
  {https://ui.adsabs.harvard.edu/abs/2012ApJ...760...71C} {760, 71}

\bibitem[\protect\citeauthoryear{{Cooke}, {Pettini}  \& {Jorgenson}}{{Cooke}
  et~al.}{2015}]{2015ApJ...800...12C}
{Cooke} R.~J.,  {Pettini} M.,   {Jorgenson} R.~A.,  2015, \mn@doi [\apj]
  {10.1088/0004-637X/800/1/12}, \href
  {https://ui.adsabs.harvard.edu/abs/2015ApJ...800...12C} {800, 12}

\bibitem[\protect\citeauthoryear{{Cooke}, {Pettini}  \& {Steidel}}{{Cooke}
  et~al.}{2017}]{2017MNRAS.467..802C}
{Cooke} R.~J.,  {Pettini} M.,   {Steidel} C.~C.,  2017, \mn@doi [\mnras]
  {10.1093/mnras/stx037}, \href
  {https://ui.adsabs.harvard.edu/abs/2017MNRAS.467..802C} {467, 802}

\bibitem[\protect\citeauthoryear{{Cristallo}, {Straniero}, {Piersanti}  \&
  {Gobrecht}}{{Cristallo} et~al.}{2015}]{2015ApJS..219...40C}
{Cristallo} S.,  {Straniero} O.,  {Piersanti} L.,   {Gobrecht} D.,  2015,
  \mn@doi [\apjs] {10.1088/0067-0049/219/2/40}, \href
  {https://ui.adsabs.harvard.edu/abs/2015ApJS..219...40C} {219, 40}

\bibitem[\protect\citeauthoryear{{Dalcanton}}{{Dalcanton}}{2007}]{2007ApJ...658..941D}
{Dalcanton} J.~J.,  2007, \mn@doi [\apj] {10.1086/508913}, \href
  {https://ui.adsabs.harvard.edu/abs/2007ApJ...658..941D} {658, 941}

\bibitem[\protect\citeauthoryear{{Dame}, {Hartmann}  \& {Thaddeus}}{{Dame}
  et~al.}{2001}]{2001ApJ...547..792D}
{Dame} T.~M.,  {Hartmann} D.,   {Thaddeus} P.,  2001, \mn@doi [\apj]
  {10.1086/318388}, \href
  {https://ui.adsabs.harvard.edu/abs/2001ApJ...547..792D} {547, 792}

\bibitem[\protect\citeauthoryear{{Draine}}{{Draine}}{2011}]{2011piim.book.....D}
{Draine} B.~T.,  2011, {Physics of the Interstellar and Intergalactic Medium}.
Princeton University Press

\bibitem[\protect\citeauthoryear{{Ebert}}{{Ebert}}{1955}]{1955ZA.....37..217E}
{Ebert} R.,  1955, \zap, \href
  {https://ui.adsabs.harvard.edu/abs/1955ZA.....37..217E} {37, 217}

\bibitem[\protect\citeauthoryear{{Edmunds} \& {Pagel}}{{Edmunds} \&
  {Pagel}}{1978}]{1978MNRAS.185P..77E}
{Edmunds} M.~G.,  {Pagel} B.~E.~J.,  1978, \mn@doi [\mnras]
  {10.1093/mnras/185.1.77P}, \href
  {https://ui.adsabs.harvard.edu/abs/1978MNRAS.185P..77E} {185, 77P}

\bibitem[\protect\citeauthoryear{{Elmegreen} \& {Efremov}}{{Elmegreen} \&
  {Efremov}}{1997}]{1997ApJ...480..235E}
{Elmegreen} B.~G.,  {Efremov} Y.~N.,  1997, \mn@doi [\apj] {10.1086/303966},
  \href {https://ui.adsabs.harvard.edu/abs/1997ApJ...480..235E} {480, 235}

\bibitem[\protect\citeauthoryear{{Emerick}, {Bryan}, {Mac Low}, {C{\^o}t{\'e}},
  {Johnston}  \& {O'Shea}}{{Emerick} et~al.}{2018}]{2018ApJ...869...94E}
{Emerick} A.,  {Bryan} G.~L.,  {Mac Low} M.-M.,  {C{\^o}t{\'e}} B.,  {Johnston}
  K.~V.,   {O'Shea} B.~W.,  2018, \mn@doi [\apj] {10.3847/1538-4357/aaec7d},
  \href {https://ui.adsabs.harvard.edu/abs/2018ApJ...869...94E} {869, 94}

\bibitem[\protect\citeauthoryear{{Erb}, {Pettini}, {Shapley}, {Steidel}, {Law}
  \& {Reddy}}{{Erb} et~al.}{2010}]{2010ApJ...719.1168E}
{Erb} D.~K.,  {Pettini} M.,  {Shapley} A.~E.,  {Steidel} C.~C.,  {Law} D.~R.,
  {Reddy} N.~A.,  2010, \mn@doi [\apj] {10.1088/0004-637X/719/2/1168}, \href
  {https://ui.adsabs.harvard.edu/abs/2010ApJ...719.1168E} {719, 1168}

\bibitem[\protect\citeauthoryear{{Esteban}, {Bresolin}, {Peimbert},
  {Garc{\'\i}a-Rojas}, {Peimbert}  \& {Mesa-Delgado}}{{Esteban}
  et~al.}{2009}]{2009ApJ...700..654E}
{Esteban} C.,  {Bresolin} F.,  {Peimbert} M.,  {Garc{\'\i}a-Rojas} J.,
  {Peimbert} A.,   {Mesa-Delgado} A.,  2009, \mn@doi [\apj]
  {10.1088/0004-637X/700/1/654}, \href
  {https://ui.adsabs.harvard.edu/abs/2009ApJ...700..654E} {700, 654}

\bibitem[\protect\citeauthoryear{{Esteban}, {Garc{\'\i}a-Rojas}, {Carigi},
  {Peimbert}, {Bresolin}, {L{\'o}pez-S{\'a}nchez}  \& {Mesa-Delgado}}{{Esteban}
  et~al.}{2014}]{2014MNRAS.443..624E}
{Esteban} C.,  {Garc{\'\i}a-Rojas} J.,  {Carigi} L.,  {Peimbert} M.,
  {Bresolin} F.,  {L{\'o}pez-S{\'a}nchez} A.~R.,   {Mesa-Delgado} A.,  2014,
  \mn@doi [\mnras] {10.1093/mnras/stu1177}, \href
  {https://ui.adsabs.harvard.edu/abs/2014MNRAS.443..624E} {443, 624}

\bibitem[\protect\citeauthoryear{{Ezzeddine} et~al.,}{{Ezzeddine}
  et~al.}{2019}]{2019ApJ...876...97E}
{Ezzeddine} R.,  et~al., 2019, \mn@doi [\apj] {10.3847/1538-4357/ab14e7}, \href
  {https://ui.adsabs.harvard.edu/abs/2019ApJ...876...97E} {876, 97}

\bibitem[\protect\citeauthoryear{{Fabbian}, {Nissen}, {Asplund}, {Pettini}  \&
  {Akerman}}{{Fabbian} et~al.}{2009}]{2009A&A...500.1143F}
{Fabbian} D.,  {Nissen} P.~E.,  {Asplund} M.,  {Pettini} M.,   {Akerman} C.,
  2009, \mn@doi [\aap] {10.1051/0004-6361/200810095}, \href
  {https://ui.adsabs.harvard.edu/abs/2009A&A...500.1143F} {500, 1143}

\bibitem[\protect\citeauthoryear{{Federrath} \& {Klessen}}{{Federrath} \&
  {Klessen}}{2013}]{2013ApJ...763...51F}
{Federrath} C.,  {Klessen} R.~S.,  2013, \mn@doi [\apj]
  {10.1088/0004-637X/763/1/51}, \href
  {https://ui.adsabs.harvard.edu/abs/2013ApJ...763...51F} {763, 51}

\bibitem[\protect\citeauthoryear{{Filion}, {Platais}, {Wyse}  \&
  {Kozhurina-Platais}}{{Filion} et~al.}{2022}]{2022arXiv220910461F}
{Filion} C.,  {Platais} I.,  {Wyse} R. F.~G.,   {Kozhurina-Platais} V.,  2022,
  arXiv e-prints, \href {https://ui.adsabs.harvard.edu/abs/2022arXiv220910461F}
  {p. arXiv:2209.10461}

\bibitem[\protect\citeauthoryear{{Finn}, {Johnson}, {Brogan}, {Wilson},
  {Indebetouw}, {Harris}, {Kamenetzky}  \& {Bemis}}{{Finn}
  et~al.}{2019}]{2019ApJ...874..120F}
{Finn} M.~K.,  {Johnson} K.~E.,  {Brogan} C.~L.,  {Wilson} C.~D.,  {Indebetouw}
  R.,  {Harris} W.~E.,  {Kamenetzky} J.,   {Bemis} A.,  2019, \mn@doi [\apj]
  {10.3847/1538-4357/ab0d1e}, \href
  {https://ui.adsabs.harvard.edu/abs/2019ApJ...874..120F} {874, 120}

\bibitem[\protect\citeauthoryear{{Fishlock}, {Karakas}, {Lugaro}  \&
  {Yong}}{{Fishlock} et~al.}{2014}]{2014ApJ...797...44F}
{Fishlock} C.~K.,  {Karakas} A.~I.,  {Lugaro} M.,   {Yong} D.,  2014, \mn@doi
  [\apj] {10.1088/0004-637X/797/1/44}, \href
  {https://ui.adsabs.harvard.edu/abs/2014ApJ...797...44F} {797, 44}

\bibitem[\protect\citeauthoryear{{Frebel} \& {Norris}}{{Frebel} \&
  {Norris}}{2015}]{2015ARA&A..53..631F}
{Frebel} A.,  {Norris} J.~E.,  2015, \mn@doi [\araa]
  {10.1146/annurev-astro-082214-122423}, \href
  {https://ui.adsabs.harvard.edu/abs/2015ARA&A..53..631F} {53, 631}

\bibitem[\protect\citeauthoryear{{Frebel}, {Johnson}  \& {Bromm}}{{Frebel}
  et~al.}{2007}]{2007MNRAS.380L..40F}
{Frebel} A.,  {Johnson} J.~L.,   {Bromm} V.,  2007, \mn@doi [\mnras]
  {10.1111/j.1745-3933.2007.00344.x}, \href
  {https://ui.adsabs.harvard.edu/abs/2007MNRAS.380L..40F} {380, L40}

\bibitem[\protect\citeauthoryear{{Frebel}, {Collet}, {Eriksson}, {Christlieb}
  \& {Aoki}}{{Frebel} et~al.}{2008}]{2008ApJ...684..588F}
{Frebel} A.,  {Collet} R.,  {Eriksson} K.,  {Christlieb} N.,   {Aoki} W.,
  2008, \mn@doi [\apj] {10.1086/590327}, \href
  {https://ui.adsabs.harvard.edu/abs/2008ApJ...684..588F} {684, 588}

\bibitem[\protect\citeauthoryear{{Frebel}, {Simon}, {Geha}  \&
  {Willman}}{{Frebel} et~al.}{2010}]{2010ApJ...708..560F}
{Frebel} A.,  {Simon} J.~D.,  {Geha} M.,   {Willman} B.,  2010, \mn@doi [\apj]
  {10.1088/0004-637X/708/1/560}, \href
  {https://ui.adsabs.harvard.edu/abs/2010ApJ...708..560F} {708, 560}

\bibitem[\protect\citeauthoryear{{Frebel}, {Norris}, {Gilmore}  \&
  {Wyse}}{{Frebel} et~al.}{2016}]{2016ApJ...826..110F}
{Frebel} A.,  {Norris} J.~E.,  {Gilmore} G.,   {Wyse} R. F.~G.,  2016, \mn@doi
  [\apj] {10.3847/0004-637X/826/2/110}, \href
  {https://ui.adsabs.harvard.edu/abs/2016ApJ...826..110F} {826, 110}

\bibitem[\protect\citeauthoryear{{Frebel}, {Ji}, {Ezzeddine}, {Hansen},
  {Chiti}, {Thompson}  \& {Merle}}{{Frebel} et~al.}{2019}]{2019ApJ...871..146F}
{Frebel} A.,  {Ji} A.~P.,  {Ezzeddine} R.,  {Hansen} T.~T.,  {Chiti} A.,
  {Thompson} I.~B.,   {Merle} T.,  2019, \mn@doi [\apj]
  {10.3847/1538-4357/aae848}, \href
  {https://ui.adsabs.harvard.edu/abs/2019ApJ...871..146F} {871, 146}

\bibitem[\protect\citeauthoryear{{Gallagher}, {Caffau}, {Bonifacio}, {Ludwig},
  {Steffen}  \& {Spite}}{{Gallagher} et~al.}{2016}]{2016A&A...593A..48G}
{Gallagher} A.~J.,  {Caffau} E.,  {Bonifacio} P.,  {Ludwig} H.~G.,  {Steffen}
  M.,   {Spite} M.,  2016, \mn@doi [\aap] {10.1051/0004-6361/201628602}, \href
  {https://ui.adsabs.harvard.edu/abs/2016A&A...593A..48G} {593, A48}

\bibitem[\protect\citeauthoryear{{Gallagher}, {Caffau}, {Bonifacio}, {Ludwig},
  {Steffen}, {Homeier}  \& {Plez}}{{Gallagher}
  et~al.}{2017}]{2017A&A...598L..10G}
{Gallagher} A.~J.,  {Caffau} E.,  {Bonifacio} P.,  {Ludwig} H.~G.,  {Steffen}
  M.,  {Homeier} D.,   {Plez} B.,  2017, \mn@doi [\aap]
  {10.1051/0004-6361/201630272}, \href
  {https://ui.adsabs.harvard.edu/abs/2017A&A...598L..10G} {598, L10}

\bibitem[\protect\citeauthoryear{{Galli} \& {Palla}}{{Galli} \&
  {Palla}}{1998}]{1998A&A...335..403G}
{Galli} D.,  {Palla} F.,  1998, \aap, \href
  {https://ui.adsabs.harvard.edu/abs/1998A&A...335..403G} {335, 403}

\bibitem[\protect\citeauthoryear{{Garc{\'\i}a P{\'e}rez}, {Asplund}, {Primas},
  {Nissen}  \& {Gustafsson}}{{Garc{\'\i}a P{\'e}rez}
  et~al.}{2006}]{2006A&A...451..621G}
{Garc{\'\i}a P{\'e}rez} A.~E.,  {Asplund} M.,  {Primas} F.,  {Nissen} P.~E.,
  {Gustafsson} B.,  2006, \mn@doi [\aap] {10.1051/0004-6361:20053181}, \href
  {https://ui.adsabs.harvard.edu/abs/2006A&A...451..621G} {451, 621}

\bibitem[\protect\citeauthoryear{{Garc{\'\i}a-Rojas} \&
  {Esteban}}{{Garc{\'\i}a-Rojas} \& {Esteban}}{2007}]{2007ApJ...670..457G}
{Garc{\'\i}a-Rojas} J.,  {Esteban} C.,  2007, \mn@doi [\apj] {10.1086/521871},
  \href {https://ui.adsabs.harvard.edu/abs/2007ApJ...670..457G} {670, 457}

\bibitem[\protect\citeauthoryear{{Garnett}, {Skillman}, {Dufour}, {Peimbert},
  {Torres-Peimbert}, {Terlevich}, {Terlevich}  \& {Shields}}{{Garnett}
  et~al.}{1995}]{1995ApJ...443...64G}
{Garnett} D.~R.,  {Skillman} E.~D.,  {Dufour} R.~J.,  {Peimbert} M.,
  {Torres-Peimbert} S.,  {Terlevich} R.,  {Terlevich} E.,   {Shields} G.~A.,
  1995, \mn@doi [\apj] {10.1086/175503}, \href
  {https://ui.adsabs.harvard.edu/abs/1995ApJ...443...64G} {443, 64}

\bibitem[\protect\citeauthoryear{{Garnett}, {Shields}, {Peimbert},
  {Torres-Peimbert}, {Skillman}, {Dufour}, {Terlevich}  \&
  {Terlevich}}{{Garnett} et~al.}{1999}]{1999ApJ...513..168G}
{Garnett} D.~R.,  {Shields} G.~A.,  {Peimbert} M.,  {Torres-Peimbert} S.,
  {Skillman} E.~D.,  {Dufour} R.~J.,  {Terlevich} E.,   {Terlevich} R.~J.,
  1999, \mn@doi [\apj] {10.1086/306860}, \href
  {https://ui.adsabs.harvard.edu/abs/1999ApJ...513..168G} {513, 168}

\bibitem[\protect\citeauthoryear{{Gennaro} et~al.,}{{Gennaro}
  et~al.}{2018}]{2018ApJ...855...20G}
{Gennaro} M.,  et~al., 2018, \mn@doi [\apj] {10.3847/1538-4357/aaa973}, \href
  {https://ui.adsabs.harvard.edu/abs/2018ApJ...855...20G} {855, 20}

\bibitem[\protect\citeauthoryear{{Gieser} et~al.,}{{Gieser}
  et~al.}{2021}]{2021A&A...648A..66G}
{Gieser} C.,  et~al., 2021, \mn@doi [\aap] {10.1051/0004-6361/202039670}, \href
  {https://ui.adsabs.harvard.edu/abs/2021A&A...648A..66G} {648, A66}

\bibitem[\protect\citeauthoryear{{Gil-Pons}, {Doherty}, {Campbell}  \&
  {Guti{\'e}rrez}}{{Gil-Pons} et~al.}{2022}]{2022arXiv220905587G}
{Gil-Pons} P.,  {Doherty} C.~L.,  {Campbell} S.~W.,   {Guti{\'e}rrez} J.,
  2022, arXiv e-prints, \href
  {https://ui.adsabs.harvard.edu/abs/2022arXiv220905587G} {p. arXiv:2209.05587}

\bibitem[\protect\citeauthoryear{{Glover} \& {Clark}}{{Glover} \&
  {Clark}}{2012}]{2012MNRAS.426..377G}
{Glover} S. C.~O.,  {Clark} P.~C.,  2012, \mn@doi [\mnras]
  {10.1111/j.1365-2966.2012.21737.x}, \href
  {https://ui.adsabs.harvard.edu/abs/2012MNRAS.426..377G} {426, 377}

\bibitem[\protect\citeauthoryear{{Glover} \& {Clark}}{{Glover} \&
  {Clark}}{2016}]{2016MNRAS.456.3596G}
{Glover} S. C.~O.,  {Clark} P.~C.,  2016, \mn@doi [\mnras]
  {10.1093/mnras/stv2863}, \href
  {https://ui.adsabs.harvard.edu/abs/2016MNRAS.456.3596G} {456, 3596}

\bibitem[\protect\citeauthoryear{{Gonz{\'a}lez Hern{\'a}ndez}, {Bonifacio},
  {Ludwig}, {Caffau}, {Behara}  \& {Freytag}}{{Gonz{\'a}lez Hern{\'a}ndez}
  et~al.}{2010}]{2010A&A...519A..46G}
{Gonz{\'a}lez Hern{\'a}ndez} J.~I.,  {Bonifacio} P.,  {Ludwig} H.~G.,  {Caffau}
  E.,  {Behara} N.~T.,   {Freytag} B.,  2010, \mn@doi [\aap]
  {10.1051/0004-6361/201014397}, \href
  {https://ui.adsabs.harvard.edu/abs/2010A&A...519A..46G} {519, A46}

\bibitem[\protect\citeauthoryear{{Grasha}, {Roy}, {Sutherland}  \&
  {Kewley}}{{Grasha} et~al.}{2021}]{2021ApJ...908..241G}
{Grasha} K.,  {Roy} A.,  {Sutherland} R.~S.,   {Kewley} L.~J.,  2021, \mn@doi
  [\apj] {10.3847/1538-4357/abd6bf}, \href
  {https://ui.adsabs.harvard.edu/abs/2021ApJ...908..241G} {908, 241}

\bibitem[\protect\citeauthoryear{{Grasha} et~al.,}{{Grasha}
  et~al.}{2022}]{2022ApJ...929..118G}
{Grasha} K.,  et~al., 2022, \mn@doi [\apj] {10.3847/1538-4357/ac5ab2}, \href
  {https://ui.adsabs.harvard.edu/abs/2022ApJ...929..118G} {929, 118}

\bibitem[\protect\citeauthoryear{{Grevesse}, {Asplund}, {Sauval}  \&
  {Scott}}{{Grevesse} et~al.}{2010}]{2010Ap&SS.328..179G}
{Grevesse} N.,  {Asplund} M.,  {Sauval} A.~J.,   {Scott} P.,  2010, \mn@doi
  [\apss] {10.1007/s10509-010-0288-z}, \href
  {https://ui.adsabs.harvard.edu/abs/2010Ap&SS.328..179G} {328, 179}

\bibitem[\protect\citeauthoryear{{Gu}, {Greene}, {Newman}, {Kreisch},
  {Quenneville}, {Ma}  \& {Blakeslee}}{{Gu} et~al.}{2022}]{2022ApJ...932..103G}
{Gu} M.,  {Greene} J.~E.,  {Newman} A.~B.,  {Kreisch} C.,  {Quenneville} M.~E.,
   {Ma} C.-P.,   {Blakeslee} J.~P.,  2022, \mn@doi [\apj]
  {10.3847/1538-4357/ac69ea}, \href
  {https://ui.adsabs.harvard.edu/abs/2022ApJ...932..103G} {932, 103}

\bibitem[\protect\citeauthoryear{{Guadarrama}, {Vorobyov}, {Rab}  \&
  {G{\"u}del}}{{Guadarrama} et~al.}{2022}]{2022arXiv220809327G}
{Guadarrama} R.,  {Vorobyov} E.~I.,  {Rab} C.,   {G{\"u}del} M.,  2022, \mn@doi
  [\aap] {10.1051/0004-6361/202140995}, \href
  {https://ui.adsabs.harvard.edu/abs/2022A&A...667A..28G} {667, A28}

\bibitem[\protect\citeauthoryear{{Gustafsson}, {Karlsson}, {Olsson},
  {Edvardsson}  \& {Ryde}}{{Gustafsson} et~al.}{1999}]{1999A&A...342..426G}
{Gustafsson} B.,  {Karlsson} T.,  {Olsson} E.,  {Edvardsson} B.,   {Ryde} N.,
  1999, \aap, \href {https://ui.adsabs.harvard.edu/abs/1999A&A...342..426G}
  {342, 426}

\bibitem[\protect\citeauthoryear{{Guszejnov}, {Krumholz}  \&
  {Hopkins}}{{Guszejnov} et~al.}{2016}]{2016MNRAS.458..673G}
{Guszejnov} D.,  {Krumholz} M.~R.,   {Hopkins} P.~F.,  2016, \mn@doi [\mnras]
  {10.1093/mnras/stw315}, \href
  {https://ui.adsabs.harvard.edu/abs/2016MNRAS.458..673G} {458, 673}

\bibitem[\protect\citeauthoryear{{Hansen}, {Andersen}, {Nordstr{\"o}m},
  {Beers}, {Placco}, {Yoon}  \& {Buchhave}}{{Hansen}
  et~al.}{2016}]{2016A&A...586A.160H}
{Hansen} T.~T.,  {Andersen} J.,  {Nordstr{\"o}m} B.,  {Beers} T.~C.,  {Placco}
  V.~M.,  {Yoon} J.,   {Buchhave} L.~A.,  2016, \mn@doi [\aap]
  {10.1051/0004-6361/201527235}, \href
  {https://ui.adsabs.harvard.edu/abs/2016A&A...586A.160H} {586, A160}

\bibitem[\protect\citeauthoryear{{Harris} et~al.,}{{Harris}
  et~al.}{2020}]{2020arXiv200610256H}
{Harris} C.~R.,  et~al., 2020, \mn@doi [\nat] {10.1038/s41586-020-2649-2},
  \href {https://ui.adsabs.harvard.edu/abs/2020arXiv200610256H} {585, 357}

\bibitem[\protect\citeauthoryear{{Hayes} et~al.,}{{Hayes}
  et~al.}{2022}]{2022arXiv220800071H}
{Hayes} C.~R.,  et~al., 2022, \mn@doi [\apjs] {10.3847/1538-4365/ac839f}, \href
  {https://ui.adsabs.harvard.edu/abs/2022ApJS..262...34H} {262, 34}

\bibitem[\protect\citeauthoryear{{Henry}, {Edmunds}  \& {K{\"o}ppen}}{{Henry}
  et~al.}{2000}]{2000ApJ...541..660H}
{Henry} R.~B.~C.,  {Edmunds} M.~G.,   {K{\"o}ppen} J.,  2000, \mn@doi [\apj]
  {10.1086/309471}, \href
  {https://ui.adsabs.harvard.edu/abs/2000ApJ...541..660H} {541, 660}

\bibitem[\protect\citeauthoryear{{Hopkins}}{{Hopkins}}{2018}]{2018PASA...35...39H}
{Hopkins} A.~M.,  2018, \mn@doi [\pasa] {10.1017/pasa.2018.29}, \href
  {https://ui.adsabs.harvard.edu/abs/2018PASA...35...39H} {35, e039}

\bibitem[\protect\citeauthoryear{{Hou}, {Aoyama}, {Hirashita}, {Nagamine}  \&
  {Shimizu}}{{Hou} et~al.}{2019}]{2019MNRAS.485.1727H}
{Hou} K.-C.,  {Aoyama} S.,  {Hirashita} H.,  {Nagamine} K.,   {Shimizu} I.,
  2019, \mn@doi [\mnras] {10.1093/mnras/stz121}, \href
  {https://ui.adsabs.harvard.edu/abs/2019MNRAS.485.1727H} {485, 1727}

\bibitem[\protect\citeauthoryear{{Hu}, {Sternberg}  \& {van Dishoeck}}{{Hu}
  et~al.}{2021}]{2021arXiv210303889H}
{Hu} C.-Y.,  {Sternberg} A.,   {van Dishoeck} E.~F.,  2021, \mn@doi [\apj]
  {10.3847/1538-4357/ac0dbd}, \href
  {https://ui.adsabs.harvard.edu/abs/2021ApJ...920...44H} {920, 44}

\bibitem[\protect\citeauthoryear{Hunter}{Hunter}{2007}]{Hunter:2007}
Hunter J.~D.,  2007, \mn@doi [Computing in Science \& Engineering]
  {10.1109/MCSE.2007.55}, 9, 90

\bibitem[\protect\citeauthoryear{{Ishigaki}, {Aoki}, {Arimoto}  \&
  {Okamoto}}{{Ishigaki} et~al.}{2014}]{2014A&A...562A.146I}
{Ishigaki} M.~N.,  {Aoki} W.,  {Arimoto} N.,   {Okamoto} S.,  2014, \mn@doi
  [\aap] {10.1051/0004-6361/201322796}, \href
  {https://ui.adsabs.harvard.edu/abs/2014A&A...562A.146I} {562, A146}

\bibitem[\protect\citeauthoryear{{Israelian}, {Ecuvillon}, {Rebolo},
  {Garc{\'\i}a-L{\'o}pez}, {Bonifacio}  \& {Molaro}}{{Israelian}
  et~al.}{2004}]{2004A&A...421..649I}
{Israelian} G.,  {Ecuvillon} A.,  {Rebolo} R.,  {Garc{\'\i}a-L{\'o}pez} R.,
  {Bonifacio} P.,   {Molaro} P.,  2004, \mn@doi [\aap]
  {10.1051/0004-6361:20047132}, \href
  {https://ui.adsabs.harvard.edu/abs/2004A&A...421..649I} {421, 649}

\bibitem[\protect\citeauthoryear{{Iwamoto}, {Umeda}, {Tominaga}, {Nomoto}  \&
  {Maeda}}{{Iwamoto} et~al.}{2005}]{2005Sci...309..451I}
{Iwamoto} N.,  {Umeda} H.,  {Tominaga} N.,  {Nomoto} K.,   {Maeda} K.,  2005,
  \mn@doi [Science] {10.1126/science.1112997}, \href
  {https://ui.adsabs.harvard.edu/abs/2005Sci...309..451I} {309, 451}

\bibitem[\protect\citeauthoryear{{James} et~al.,}{{James}
  et~al.}{2014}]{2014MNRAS.440.1794J}
{James} B.~L.,  et~al., 2014, \mn@doi [\mnras] {10.1093/mnras/stu287}, \href
  {https://ui.adsabs.harvard.edu/abs/2014MNRAS.440.1794J} {440, 1794}

\bibitem[\protect\citeauthoryear{{Jameson} et~al.,}{{Jameson}
  et~al.}{2018}]{2018ApJ...853..111J}
{Jameson} K.~E.,  et~al., 2018, \mn@doi [\apj] {10.3847/1538-4357/aaa4bb},
  \href {https://ui.adsabs.harvard.edu/abs/2018ApJ...853..111J} {853, 111}

\bibitem[\protect\citeauthoryear{{Jenkins}}{{Jenkins}}{2014}]{2014arXiv1402.4765J}
{Jenkins} E.~B.,  2014, arXiv e-prints, \href
  {https://ui.adsabs.harvard.edu/abs/2014arXiv1402.4765J} {p. arXiv:1402.4765}

\bibitem[\protect\citeauthoryear{{Ji}, {Frebel}, {Simon}  \& {Geha}}{{Ji}
  et~al.}{2016}]{2016ApJ...817...41J}
{Ji} A.~P.,  {Frebel} A.,  {Simon} J.~D.,   {Geha} M.,  2016, \mn@doi [\apj]
  {10.3847/0004-637X/817/1/41}, \href
  {https://ui.adsabs.harvard.edu/abs/2016ApJ...817...41J} {817, 41}

\bibitem[\protect\citeauthoryear{{Johnson}}{{Johnson}}{2015}]{2015MNRAS.453.2771J}
{Johnson} J.~L.,  2015, \mn@doi [\mnras] {10.1093/mnras/stv1815}, \href
  {https://ui.adsabs.harvard.edu/abs/2015MNRAS.453.2771J} {453, 2771}

\bibitem[\protect\citeauthoryear{{Johnson}, {Leroy}, {Indebetouw}, {Brogan},
  {Whitmore}, {Hibbard}, {Sheth}  \& {Evans}}{{Johnson}
  et~al.}{2015}]{2015ApJ...806...35J}
{Johnson} K.~E.,  {Leroy} A.~K.,  {Indebetouw} R.,  {Brogan} C.~L.,  {Whitmore}
  B.~C.,  {Hibbard} J.,  {Sheth} K.,   {Evans} A.~S.,  2015, \mn@doi [\apj]
  {10.1088/0004-637X/806/1/35}, \href
  {https://ui.adsabs.harvard.edu/abs/2015ApJ...806...35J} {806, 35}

\bibitem[\protect\citeauthoryear{{Johnson}, {Weinberg}, {Vincenzo}, {Bird}  \&
  {Griffith}}{{Johnson} et~al.}{2022}]{2022arXiv220204666J}
{Johnson} J.~W.,  {Weinberg} D.~H.,  {Vincenzo} F.,  {Bird} J.~C.,   {Griffith}
  E.~J.,  2022, arXiv e-prints, \href
  {https://ui.adsabs.harvard.edu/abs/2022arXiv220204666J} {p. arXiv:2202.04666}

\bibitem[\protect\citeauthoryear{{Katz} et~al.,}{{Katz}
  et~al.}{2022}]{2022arXiv220713693K}
{Katz} H.,  et~al., 2022, \mn@doi [\mnras] {10.1093/mnras/stac2657}, \href
  {https://ui.adsabs.harvard.edu/abs/2022MNRAS.tmp.2470K} {}

\bibitem[\protect\citeauthoryear{{Kepley}, {Leroy}, {Johnson}, {Sandstrom}  \&
  {Chen}}{{Kepley} et~al.}{2016}]{2016ApJ...828...50K}
{Kepley} A.~A.,  {Leroy} A.~K.,  {Johnson} K.~E.,  {Sandstrom} K.,   {Chen} C.
  H.~R.,  2016, \mn@doi [\apj] {10.3847/0004-637X/828/1/50}, \href
  {https://ui.adsabs.harvard.edu/abs/2016ApJ...828...50K} {828, 50}

\bibitem[\protect\citeauthoryear{{Kewley}, {Nicholls}, {Sutherland}, {Rigby},
  {Acharya}, {Dopita}  \& {Bayliss}}{{Kewley}
  et~al.}{2019}]{2019ApJ...880...16K}
{Kewley} L.~J.,  {Nicholls} D.~C.,  {Sutherland} R.,  {Rigby} J.~R.,  {Acharya}
  A.,  {Dopita} M.~A.,   {Bayliss} M.~B.,  2019, \mn@doi [\apj]
  {10.3847/1538-4357/ab16ed}, \href
  {https://ui.adsabs.harvard.edu/abs/2019ApJ...880...16K} {880, 16}

\bibitem[\protect\citeauthoryear{{Kobayashi}, {Karakas}  \&
  {Lugaro}}{{Kobayashi} et~al.}{2020}]{2020ApJ...900..179K}
{Kobayashi} C.,  {Karakas} A.~I.,   {Lugaro} M.,  2020, \mn@doi [\apj]
  {10.3847/1538-4357/abae65}, \href
  {https://ui.adsabs.harvard.edu/abs/2020ApJ...900..179K} {900, 179}

\bibitem[\protect\citeauthoryear{{Kobulnicky} \& {Skillman}}{{Kobulnicky} \&
  {Skillman}}{1996}]{1996ApJ...471..211K}
{Kobulnicky} H.~A.,  {Skillman} E.~D.,  1996, \mn@doi [\apj] {10.1086/177964},
  \href {https://ui.adsabs.harvard.edu/abs/1996ApJ...471..211K} {471, 211}

\bibitem[\protect\citeauthoryear{{Komiya}, {Suda}, {Minaguchi}, {Shigeyama},
  {Aoki}  \& {Fujimoto}}{{Komiya} et~al.}{2007}]{2007ApJ...658..367K}
{Komiya} Y.,  {Suda} T.,  {Minaguchi} H.,  {Shigeyama} T.,  {Aoki} W.,
  {Fujimoto} M.~Y.,  2007, \mn@doi [\apj] {10.1086/510826}, \href
  {https://ui.adsabs.harvard.edu/abs/2007ApJ...658..367K} {658, 367}

\bibitem[\protect\citeauthoryear{{Konstantopoulou} et~al.,}{{Konstantopoulou}
  et~al.}{2022}]{2022arXiv220708804K}
{Konstantopoulou} C.,  et~al., 2022, arXiv e-prints, \href
  {https://ui.adsabs.harvard.edu/abs/2022arXiv220708804K} {p. arXiv:2207.08804}

\bibitem[\protect\citeauthoryear{{Kroupa}}{{Kroupa}}{2001}]{2001MNRAS.322..231K}
{Kroupa} P.,  2001, \mn@doi [\mnras] {10.1046/j.1365-8711.2001.04022.x}, \href
  {https://ui.adsabs.harvard.edu/abs/2001MNRAS.322..231K} {322, 231}

\bibitem[\protect\citeauthoryear{{Krumholz}}{{Krumholz}}{2011}]{2011ApJ...743..110K}
{Krumholz} M.~R.,  2011, \mn@doi [\apj] {10.1088/0004-637X/743/2/110}, \href
  {https://ui.adsabs.harvard.edu/abs/2011ApJ...743..110K} {743, 110}

\bibitem[\protect\citeauthoryear{{Krumholz}}{{Krumholz}}{2014}]{2014MNRAS.437.1662K}
{Krumholz} M.~R.,  2014, \mn@doi [\mnras] {10.1093/mnras/stt2000}, \href
  {https://ui.adsabs.harvard.edu/abs/2014MNRAS.437.1662K} {437, 1662}

\bibitem[\protect\citeauthoryear{{Krumholz}, {McKee}  \&
  {Tumlinson}}{{Krumholz} et~al.}{2009}]{2009ApJ...693..216K}
{Krumholz} M.~R.,  {McKee} C.~F.,   {Tumlinson} J.,  2009, \mn@doi [\apj]
  {10.1088/0004-637X/693/1/216}, \href
  {https://ui.adsabs.harvard.edu/abs/2009ApJ...693..216K} {693, 216}

\bibitem[\protect\citeauthoryear{{Krumholz}, {Klein}  \& {McKee}}{{Krumholz}
  et~al.}{2011}]{2011ApJ...740...74K}
{Krumholz} M.~R.,  {Klein} R.~I.,   {McKee} C.~F.,  2011, \mn@doi [\apj]
  {10.1088/0004-637X/740/2/74}, \href
  {https://ui.adsabs.harvard.edu/abs/2011ApJ...740...74K} {740, 74}

\bibitem[\protect\citeauthoryear{{Krumholz}, {Fumagalli}, {da Silva}, {Rendahl}
   \& {Parra}}{{Krumholz} et~al.}{2015}]{2015MNRAS.452.1447K}
{Krumholz} M.~R.,  {Fumagalli} M.,  {da Silva} R.~L.,  {Rendahl} T.,   {Parra}
  J.,  2015, \mn@doi [\mnras] {10.1093/mnras/stv1374}, \href
  {https://ui.adsabs.harvard.edu/abs/2015MNRAS.452.1447K} {452, 1447}

\bibitem[\protect\citeauthoryear{{Krumholz}, {Myers}, {Klein}  \&
  {McKee}}{{Krumholz} et~al.}{2016}]{2016MNRAS.460.3272K}
{Krumholz} M.~R.,  {Myers} A.~T.,  {Klein} R.~I.,   {McKee} C.~F.,  2016,
  \mn@doi [\mnras] {10.1093/mnras/stw1236}, \href
  {https://ui.adsabs.harvard.edu/abs/2016MNRAS.460.3272K} {460, 3272}

\bibitem[\protect\citeauthoryear{{Lacchin}, {Matteucci}, {Vincenzo}  \&
  {Palla}}{{Lacchin} et~al.}{2020}]{2020MNRAS.495.3276L}
{Lacchin} E.,  {Matteucci} F.,  {Vincenzo} F.,   {Palla} M.,  2020, \mn@doi
  [\mnras] {10.1093/mnras/staa585}, \href
  {https://ui.adsabs.harvard.edu/abs/2020MNRAS.495.3276L} {495, 3276}

\bibitem[\protect\citeauthoryear{{Langer}}{{Langer}}{2009}]{2009ASPC..417...71L}
{Langer} W.~D.,  2009, in {Lis} D.~C.,  {Vaillancourt} J.~E.,  {Goldsmith}
  P.~F.,  {Bell} T.~A.,  {Scoville} N.~Z.,   {Zmuidzinas} J.,  eds,
  Astronomical Society of the Pacific Conference Series Vol. 417, Submillimeter
  Astrophysics and Technology: a Symposium Honoring Thomas G. Phillips. p.~71

\bibitem[\protect\citeauthoryear{{Li}, {Narayanan}  \& {Dav{\'e}}}{{Li}
  et~al.}{2019}]{2019MNRAS.490.1425L}
{Li} Q.,  {Narayanan} D.,   {Dav{\'e}} R.,  2019, \mn@doi [\mnras]
  {10.1093/mnras/stz2684}, \href
  {https://ui.adsabs.harvard.edu/abs/2019MNRAS.490.1425L} {490, 1425}

\bibitem[\protect\citeauthoryear{{Liang}, {Yin}, {Hammer}, {Deng}, {Flores}  \&
  {Zhang}}{{Liang} et~al.}{2006}]{2006ApJ...652..257L}
{Liang} Y.~C.,  {Yin} S.~Y.,  {Hammer} F.,  {Deng} L.~C.,  {Flores} H.,
  {Zhang} B.,  2006, \mn@doi [\apj] {10.1086/507592}, \href
  {https://ui.adsabs.harvard.edu/abs/2006ApJ...652..257L} {652, 257}

\bibitem[\protect\citeauthoryear{{Llerena} et~al.,}{{Llerena}
  et~al.}{2022}]{2022A&A...659A..16L}
{Llerena} M.,  et~al., 2022, \mn@doi [\aap] {10.1051/0004-6361/202141651},
  \href {https://ui.adsabs.harvard.edu/abs/2022A&A...659A..16L} {659, A16}

\bibitem[\protect\citeauthoryear{{L{\'o}pez-S{\'a}nchez}, {Esteban},
  {Garc{\'\i}a-Rojas}, {Peimbert}  \& {Rodr{\'\i}guez}}{{L{\'o}pez-S{\'a}nchez}
  et~al.}{2007}]{2007ApJ...656..168L}
{L{\'o}pez-S{\'a}nchez} {\'A}.~R.,  {Esteban} C.,  {Garc{\'\i}a-Rojas} J.,
  {Peimbert} M.,   {Rodr{\'\i}guez} M.,  2007, \mn@doi [\apj] {10.1086/510112},
  \href {https://ui.adsabs.harvard.edu/abs/2007ApJ...656..168L} {656, 168}

\bibitem[\protect\citeauthoryear{{Lopez}, {Mathur}, {Nguyen}, {Thompson}  \&
  {Olivier}}{{Lopez} et~al.}{2020}]{Lopez20a}
{Lopez} L.~A.,  {Mathur} S.,  {Nguyen} D.~D.,  {Thompson} T.~A.,   {Olivier}
  G.~M.,  2020, \mn@doi [\apj] {10.3847/1538-4357/abc010}, \href
  {https://ui.adsabs.harvard.edu/abs/2020ApJ...904..152L} {904, 152}

\bibitem[\protect\citeauthoryear{{Lucey} et~al.,}{{Lucey}
  et~al.}{2022}]{2022arXiv220608299L}
{Lucey} M.,  et~al., 2022, arXiv e-prints, \href
  {https://ui.adsabs.harvard.edu/abs/2022arXiv220608299L} {p. arXiv:2206.08299}

\bibitem[\protect\citeauthoryear{{Mac Low} \& {Ferrara}}{{Mac Low} \&
  {Ferrara}}{1999}]{1999ApJ...513..142M}
{Mac Low} M.-M.,  {Ferrara} A.,  1999, \mn@doi [\apj] {10.1086/306832}, \href
  {https://ui.adsabs.harvard.edu/abs/1999ApJ...513..142M} {513, 142}

\bibitem[\protect\citeauthoryear{{Madden} et~al.,}{{Madden}
  et~al.}{2020}]{2020A&A...643A.141M}
{Madden} S.~C.,  et~al., 2020, \mn@doi [\aap] {10.1051/0004-6361/202038860},
  \href {https://ui.adsabs.harvard.edu/abs/2020A&A...643A.141M} {643, A141}

\bibitem[\protect\citeauthoryear{{Maeder}}{{Maeder}}{1992}]{1992A&A...264..105M}
{Maeder} A.,  1992, \aap, \href
  {https://ui.adsabs.harvard.edu/abs/1992A&A...264..105M} {264, 105}

\bibitem[\protect\citeauthoryear{{Magic}, {Collet}, {Asplund}, {Trampedach},
  {Hayek}, {Chiavassa}, {Stein}  \& {Nordlund}}{{Magic}
  et~al.}{2013}]{2013A&A...557A..26M}
{Magic} Z.,  {Collet} R.,  {Asplund} M.,  {Trampedach} R.,  {Hayek} W.,
  {Chiavassa} A.,  {Stein} R.~F.,   {Nordlund} {\r{A}}.,  2013, \mn@doi [\aap]
  {10.1051/0004-6361/201321274}, \href
  {https://ui.adsabs.harvard.edu/abs/2013A&A...557A..26M} {557, A26}

\bibitem[\protect\citeauthoryear{{Mardini} et~al.,}{{Mardini}
  et~al.}{2022}]{2022arXiv220803891M}
{Mardini} M.~K.,  et~al., 2022, \mn@doi [\mnras] {10.1093/mnras/stac2783},
  \href {https://ui.adsabs.harvard.edu/abs/2022MNRAS.517.3993M} {517, 3993}

\bibitem[\protect\citeauthoryear{{Mart{\'\i}n-Navarro}
  et~al.,}{{Mart{\'\i}n-Navarro} et~al.}{2015}]{2015ApJ...806L..31M}
{Mart{\'\i}n-Navarro} I.,  et~al., 2015, \mn@doi [\apjl]
  {10.1088/2041-8205/806/2/L31}, \href
  {https://ui.adsabs.harvard.edu/abs/2015ApJ...806L..31M} {806, L31}

\bibitem[\protect\citeauthoryear{{Mathis}}{{Mathis}}{1990}]{1990ARA&A..28...37M}
{Mathis} J.~S.,  1990, \mn@doi [\araa] {10.1146/annurev.aa.28.090190.000345},
  \href {https://ui.adsabs.harvard.edu/abs/1990ARA&A..28...37M} {28, 37}

\bibitem[\protect\citeauthoryear{{Mattsson}}{{Mattsson}}{2010}]{2010A&A...515A..68M}
{Mattsson} L.,  2010, \mn@doi [\aap] {10.1051/0004-6361/200913315}, \href
  {https://ui.adsabs.harvard.edu/abs/2010A&A...515A..68M} {515, A68}

\bibitem[\protect\citeauthoryear{{McKee} \& {Krumholz}}{{McKee} \&
  {Krumholz}}{2010}]{2010ApJ...709..308M}
{McKee} C.~F.,  {Krumholz} M.~R.,  2010, \mn@doi [\apj]
  {10.1088/0004-637X/709/1/308}, \href
  {https://ui.adsabs.harvard.edu/abs/2010ApJ...709..308M} {709, 308}

\bibitem[\protect\citeauthoryear{{Meece}, {Smith}  \& {O'Shea}}{{Meece}
  et~al.}{2014}]{2014ApJ...783...75M}
{Meece} G.~R.,  {Smith} B.~D.,   {O'Shea} B.~W.,  2014, \mn@doi [\apj]
  {10.1088/0004-637X/783/2/75}, \href
  {https://ui.adsabs.harvard.edu/abs/2014ApJ...783...75M} {783, 75}

\bibitem[\protect\citeauthoryear{{Mingozzi} et~al.,}{{Mingozzi}
  et~al.}{2022}]{2022arXiv220909047M}
{Mingozzi} M.,  et~al., 2022, arXiv e-prints, \href
  {https://ui.adsabs.harvard.edu/abs/2022arXiv220909047M} {p. arXiv:2209.09047}

\bibitem[\protect\citeauthoryear{{Miville-Desch{\^e}nes}, {Murray}  \&
  {Lee}}{{Miville-Desch{\^e}nes} et~al.}{2017}]{2017ApJ...834...57M}
{Miville-Desch{\^e}nes} M.-A.,  {Murray} N.,   {Lee} E.~J.,  2017, \mn@doi
  [\apj] {10.3847/1538-4357/834/1/57}, \href
  {https://ui.adsabs.harvard.edu/abs/2017ApJ...834...57M} {834, 57}

\bibitem[\protect\citeauthoryear{{Munshi}, {Christensen}, {Quinn}, {Governato},
  {Wadsley}, {Loebman}  \& {Shen}}{{Munshi} et~al.}{2014}]{2014ApJ...781L..14M}
{Munshi} F.,  {Christensen} C.,  {Quinn} T.~R.,  {Governato} F.,  {Wadsley} J.,
   {Loebman} S.,   {Shen} S.,  2014, \mn@doi [\apjl]
  {10.1088/2041-8205/781/1/L14}, \href
  {https://ui.adsabs.harvard.edu/abs/2014ApJ...781L..14M} {781, L14}

\bibitem[\protect\citeauthoryear{{Myers}, {Krumholz}, {Klein}  \&
  {McKee}}{{Myers} et~al.}{2011}]{2011ApJ...735...49M}
{Myers} A.~T.,  {Krumholz} M.~R.,  {Klein} R.~I.,   {McKee} C.~F.,  2011,
  \mn@doi [\apj] {10.1088/0004-637X/735/1/49}, \href
  {https://ui.adsabs.harvard.edu/abs/2011ApJ...735...49M} {735, 49}

\bibitem[\protect\citeauthoryear{{Nagasawa} et~al.,}{{Nagasawa}
  et~al.}{2018}]{2018ApJ...852...99N}
{Nagasawa} D.~Q.,  et~al., 2018, \mn@doi [\apj] {10.3847/1538-4357/aaa01d},
  \href {https://ui.adsabs.harvard.edu/abs/2018ApJ...852...99N} {852, 99}

\bibitem[\protect\citeauthoryear{{Nicholls}, {Sutherland}, {Dopita}, {Kewley}
  \& {Groves}}{{Nicholls} et~al.}{2017}]{2017MNRAS.466.4403N}
{Nicholls} D.~C.,  {Sutherland} R.~S.,  {Dopita} M.~A.,  {Kewley} L.~J.,
  {Groves} B.~A.,  2017, \mn@doi [\mnras] {10.1093/mnras/stw3235}, \href
  {https://ui.adsabs.harvard.edu/abs/2017MNRAS.466.4403N} {466, 4403}

\bibitem[\protect\citeauthoryear{{Nissen} \& {Gustafsson}}{{Nissen} \&
  {Gustafsson}}{2018}]{2018A&ARv..26....6N}
{Nissen} P.~E.,  {Gustafsson} B.,  2018, \mn@doi [\aapr]
  {10.1007/s00159-018-0111-3}, \href
  {https://ui.adsabs.harvard.edu/abs/2018A&ARv..26....6N} {26, 6}

\bibitem[\protect\citeauthoryear{{Nissen}, {Akerman}, {Asplund}, {Fabbian},
  {Kerber}, {Kaufl}  \& {Pettini}}{{Nissen} et~al.}{2007}]{2007A&A...469..319N}
{Nissen} P.~E.,  {Akerman} C.,  {Asplund} M.,  {Fabbian} D.,  {Kerber} F.,
  {Kaufl} H.~U.,   {Pettini} M.,  2007, \mn@doi [\aap]
  {10.1051/0004-6361:20077344}, \href
  {https://ui.adsabs.harvard.edu/abs/2007A&A...469..319N} {469, 319}

\bibitem[\protect\citeauthoryear{{Nissen}, {Chen}, {Carigi}, {Schuster}  \&
  {Zhao}}{{Nissen} et~al.}{2014}]{2014A&A...568A..25N}
{Nissen} P.~E.,  {Chen} Y.~Q.,  {Carigi} L.,  {Schuster} W.~J.,   {Zhao} G.,
  2014, \mn@doi [\aap] {10.1051/0004-6361/201424184}, \href
  {https://ui.adsabs.harvard.edu/abs/2014A&A...568A..25N} {568, A25}

\bibitem[\protect\citeauthoryear{{Nordlander} et~al.,}{{Nordlander}
  et~al.}{2019}]{2019MNRAS.488L.109N}
{Nordlander} T.,  et~al., 2019, \mn@doi [\mnras] {10.1093/mnrasl/slz109}, \href
  {https://ui.adsabs.harvard.edu/abs/2019MNRAS.488L.109N} {488, L109}

\bibitem[\protect\citeauthoryear{{Norris} \& {Yong}}{{Norris} \&
  {Yong}}{2019}]{2019ApJ...879...37N}
{Norris} J.~E.,  {Yong} D.,  2019, \mn@doi [\apj] {10.3847/1538-4357/ab1f84},
  \href {https://ui.adsabs.harvard.edu/abs/2019ApJ...879...37N} {879, 37}

\bibitem[\protect\citeauthoryear{{Norris} et~al.,}{{Norris}
  et~al.}{2013}]{2013ApJ...762...28N}
{Norris} J.~E.,  et~al., 2013, \mn@doi [\apj] {10.1088/0004-637X/762/1/28},
  \href {https://ui.adsabs.harvard.edu/abs/2013ApJ...762...28N} {762, 28}

\bibitem[\protect\citeauthoryear{{Oliphant}}{{Oliphant}}{2006}]{oliphant2006guide}
{Oliphant} T.~E.,  2006, A guide to NumPy.
~ Vol. 1, Trelgol Publishing USA

\bibitem[\protect\citeauthoryear{{Omukai}}{{Omukai}}{2000}]{2000ApJ...534..809O}
{Omukai} K.,  2000, \mn@doi [\apj] {10.1086/308776}, \href
  {https://ui.adsabs.harvard.edu/abs/2000ApJ...534..809O} {534, 809}

\bibitem[\protect\citeauthoryear{{Omukai}}{{Omukai}}{2012}]{2012PASJ...64..114O}
{Omukai} K.,  2012, \mn@doi [\pasj] {10.1093/pasj/64.5.114}, \href
  {https://ui.adsabs.harvard.edu/abs/2012PASJ...64..114O} {64, 114}

\bibitem[\protect\citeauthoryear{{Omukai}, {Tsuribe}, {Schneider}  \&
  {Ferrara}}{{Omukai} et~al.}{2005}]{2005ApJ...626..627O}
{Omukai} K.,  {Tsuribe} T.,  {Schneider} R.,   {Ferrara} A.,  2005, \mn@doi
  [\apj] {10.1086/429955}, \href
  {https://ui.adsabs.harvard.edu/abs/2005ApJ...626..627O} {626, 627}

\bibitem[\protect\citeauthoryear{{Omukai}, {Hosokawa}  \& {Yoshida}}{{Omukai}
  et~al.}{2010}]{2010ApJ...722.1793O}
{Omukai} K.,  {Hosokawa} T.,   {Yoshida} N.,  2010, \mn@doi [\apj]
  {10.1088/0004-637X/722/2/1793}, \href
  {https://ui.adsabs.harvard.edu/abs/2010ApJ...722.1793O} {722, 1793}

\bibitem[\protect\citeauthoryear{{Padoan}, {Nordlund}  \& {Jones}}{{Padoan}
  et~al.}{1997}]{1997MNRAS.288..145P}
{Padoan} P.,  {Nordlund} A.,   {Jones} B. J.~T.,  1997, \mn@doi [\mnras]
  {10.1093/mnras/288.1.145}, \href
  {https://ui.adsabs.harvard.edu/abs/1997MNRAS.288..145P} {288, 145}

\bibitem[\protect\citeauthoryear{{Palla}, {Calura}, {Matteucci}, {Fan},
  {Vincenzo}  \& {Lacchin}}{{Palla} et~al.}{2020}]{2020MNRAS.494.2355P}
{Palla} M.,  {Calura} F.,  {Matteucci} F.,  {Fan} X.~L.,  {Vincenzo} F.,
  {Lacchin} E.,  2020, \mn@doi [\mnras] {10.1093/mnras/staa848}, \href
  {https://ui.adsabs.harvard.edu/abs/2020MNRAS.494.2355P} {494, 2355}

\bibitem[\protect\citeauthoryear{{Papadopoulos}}{{Papadopoulos}}{2010}]{2010ApJ...720..226P}
{Papadopoulos} P.~P.,  2010, \mn@doi [\apj] {10.1088/0004-637X/720/1/226},
  \href {https://ui.adsabs.harvard.edu/abs/2010ApJ...720..226P} {720, 226}

\bibitem[\protect\citeauthoryear{{Peimbert} \& {Peimbert}}{{Peimbert} \&
  {Peimbert}}{2010}]{2010ApJ...724..791P}
{Peimbert} A.,  {Peimbert} M.,  2010, \mn@doi [\apj]
  {10.1088/0004-637X/724/1/791}, \href
  {https://ui.adsabs.harvard.edu/abs/2010ApJ...724..791P} {724, 791}

\bibitem[\protect\citeauthoryear{{P{\'e}rez-Montero} \&
  {Contini}}{{P{\'e}rez-Montero} \& {Contini}}{2009}]{2009MNRAS.398..949P}
{P{\'e}rez-Montero} E.,  {Contini} T.,  2009, \mn@doi [\mnras]
  {10.1111/j.1365-2966.2009.15145.x}, \href
  {https://ui.adsabs.harvard.edu/abs/2009MNRAS.398..949P} {398, 949}

\bibitem[\protect\citeauthoryear{{Pilyugin} \& {Thuan}}{{Pilyugin} \&
  {Thuan}}{2005}]{2005ApJ...631..231P}
{Pilyugin} L.~S.,  {Thuan} T.~X.,  2005, \mn@doi [\apj] {10.1086/432408}, \href
  {https://ui.adsabs.harvard.edu/abs/2005ApJ...631..231P} {631, 231}

\bibitem[\protect\citeauthoryear{{Pineda} et~al.,}{{Pineda}
  et~al.}{2017}]{2017ApJ...839..107P}
{Pineda} J.~L.,  et~al., 2017, \mn@doi [\apj] {10.3847/1538-4357/aa683a}, \href
  {https://ui.adsabs.harvard.edu/abs/2017ApJ...839..107P} {839, 107}

\bibitem[\protect\citeauthoryear{{Popping} \& {P{\'e}roux}}{{Popping} \&
  {P{\'e}roux}}{2022}]{2022MNRAS.513.1531P}
{Popping} G.,  {P{\'e}roux} C.,  2022, \mn@doi [\mnras]
  {10.1093/mnras/stac695}, \href
  {https://ui.adsabs.harvard.edu/abs/2022MNRAS.513.1531P} {513, 1531}

\bibitem[\protect\citeauthoryear{{Popping}, {Somerville}  \&
  {Galametz}}{{Popping} et~al.}{2017}]{2017MNRAS.471.3152P}
{Popping} G.,  {Somerville} R.~S.,   {Galametz} M.,  2017, \mn@doi [\mnras]
  {10.1093/mnras/stx1545}, \href
  {https://ui.adsabs.harvard.edu/abs/2017MNRAS.471.3152P} {471, 3152}

\bibitem[\protect\citeauthoryear{{R{\'e}my-Ruyer} et~al.,}{{R{\'e}my-Ruyer}
  et~al.}{2014}]{2014A&A...563A..31R}
{R{\'e}my-Ruyer} A.,  et~al., 2014, \mn@doi [\aap]
  {10.1051/0004-6361/201322803}, \href
  {https://ui.adsabs.harvard.edu/abs/2014A&A...563A..31R} {563, A31}

\bibitem[\protect\citeauthoryear{{Rigby} et~al.,}{{Rigby}
  et~al.}{2018}]{2018ApJ...853...87R}
{Rigby} J.~R.,  et~al., 2018, \mn@doi [\apj] {10.3847/1538-4357/aaa2fc}, \href
  {https://ui.adsabs.harvard.edu/abs/2018ApJ...853...87R} {853, 87}

\bibitem[\protect\citeauthoryear{{Robles-Valdez},
  {Rodr{\'\i}guez-Gonz{\'a}lez}, {Hern{\'a}ndez-Mart{\'\i}nez}  \&
  {Esquivel}}{{Robles-Valdez} et~al.}{2017}]{2017ApJ...835..136R}
{Robles-Valdez} F.,  {Rodr{\'\i}guez-Gonz{\'a}lez} A.,
  {Hern{\'a}ndez-Mart{\'\i}nez} L.,   {Esquivel} A.,  2017, \mn@doi [\apj]
  {10.3847/1538-4357/835/2/136}, \href
  {https://ui.adsabs.harvard.edu/abs/2017ApJ...835..136R} {835, 136}

\bibitem[\protect\citeauthoryear{{Romano}}{{Romano}}{2022}]{2022arXiv221004350R}
{Romano} D.,  2022, arXiv e-prints, \href
  {https://ui.adsabs.harvard.edu/abs/2022arXiv221004350R} {p. arXiv:2210.04350}

\bibitem[\protect\citeauthoryear{{Romano}, {Matteucci}, {Zhang}, {Ivison}  \&
  {Ventura}}{{Romano} et~al.}{2019a}]{2019MNRAS.490.2838R}
{Romano} D.,  {Matteucci} F.,  {Zhang} Z.-Y.,  {Ivison} R.~J.,   {Ventura} P.,
  2019a, \mn@doi [\mnras] {10.1093/mnras/stz2741}, \href
  {https://ui.adsabs.harvard.edu/abs/2019MNRAS.490.2838R} {490, 2838}

\bibitem[\protect\citeauthoryear{{Romano}, {Calura}, {D'Ercole}  \&
  {Few}}{{Romano} et~al.}{2019b}]{2019A&A...630A.140R}
{Romano} D.,  {Calura} F.,  {D'Ercole} A.,   {Few} C.~G.,  2019b, \mn@doi
  [\aap] {10.1051/0004-6361/201935328}, \href
  {https://ui.adsabs.harvard.edu/abs/2019A&A...630A.140R} {630, A140}

\bibitem[\protect\citeauthoryear{{Romano}, {Franchini}, {Grisoni}, {Spitoni},
  {Matteucci}  \& {Morossi}}{{Romano} et~al.}{2020}]{2020A&A...639A..37R}
{Romano} D.,  {Franchini} M.,  {Grisoni} V.,  {Spitoni} E.,  {Matteucci} F.,
  {Morossi} C.,  2020, \mn@doi [\aap] {10.1051/0004-6361/202037972}, \href
  {https://ui.adsabs.harvard.edu/abs/2020A&A...639A..37R} {639, A37}

\bibitem[\protect\citeauthoryear{{Rossi}, {Salvadori}  \&
  {Sk{\'u}lad{\'o}ttir}}{{Rossi} et~al.}{2021}]{2021MNRAS.503.6026R}
{Rossi} M.,  {Salvadori} S.,   {Sk{\'u}lad{\'o}ttir} {\'A}.,  2021, \mn@doi
  [\mnras] {10.1093/mnras/stab821}, \href
  {https://ui.adsabs.harvard.edu/abs/2021MNRAS.503.6026R} {503, 6026}

\bibitem[\protect\citeauthoryear{{Roy}, {Dopita}, {Krumholz}, {Kewley},
  {Sutherland}  \& {Heger}}{{Roy} et~al.}{2021}]{2021MNRAS.502.4359R}
{Roy} A.,  {Dopita} M.~A.,  {Krumholz} M.~R.,  {Kewley} L.~J.,  {Sutherland}
  R.~S.,   {Heger} A.,  2021, \mn@doi [\mnras] {10.1093/mnras/stab376}, \href
  {https://ui.adsabs.harvard.edu/abs/2021MNRAS.502.4359R} {502, 4359}

\bibitem[\protect\citeauthoryear{{Rubio}, {Elmegreen}, {Hunter}, {Brinks},
  {Cort{\'e}s}  \& {Cigan}}{{Rubio} et~al.}{2015}]{Rubio15a}
{Rubio} M.,  {Elmegreen} B.~G.,  {Hunter} D.~A.,  {Brinks} E.,  {Cort{\'e}s}
  J.~R.,   {Cigan} P.,  2015, \mn@doi [\nat] {10.1038/nature14901}, \href
  {https://ui.adsabs.harvard.edu/abs/2015Natur.525..218R} {525, 218}

\bibitem[\protect\citeauthoryear{{Salvadori} \& {Ferrara}}{{Salvadori} \&
  {Ferrara}}{2012}]{2012MNRAS.421L..29S}
{Salvadori} S.,  {Ferrara} A.,  2012, \mn@doi [\mnras]
  {10.1111/j.1745-3933.2011.01200.x}, \href
  {https://ui.adsabs.harvard.edu/abs/2012MNRAS.421L..29S} {421, L29}

\bibitem[\protect\citeauthoryear{{Schneider} \& {Omukai}}{{Schneider} \&
  {Omukai}}{2010}]{2010MNRAS.402..429S}
{Schneider} R.,  {Omukai} K.,  2010, \mn@doi [\mnras]
  {10.1111/j.1365-2966.2009.15891.x}, \href
  {https://ui.adsabs.harvard.edu/abs/2010MNRAS.402..429S} {402, 429}

\bibitem[\protect\citeauthoryear{{Schneider}, {Omukai}, {Inoue}  \&
  {Ferrara}}{{Schneider} et~al.}{2006}]{2006MNRAS.369.1437S}
{Schneider} R.,  {Omukai} K.,  {Inoue} A.~K.,   {Ferrara} A.,  2006, \mn@doi
  [\mnras] {10.1111/j.1365-2966.2006.10391.x}, \href
  {https://ui.adsabs.harvard.edu/abs/2006MNRAS.369.1437S} {369, 1437}

\bibitem[\protect\citeauthoryear{{Schneider}, {Omukai}, {Bianchi}  \&
  {Valiante}}{{Schneider} et~al.}{2012}]{2012MNRAS.419.1566S}
{Schneider} R.,  {Omukai} K.,  {Bianchi} S.,   {Valiante} R.,  2012, \mn@doi
  [\mnras] {10.1111/j.1365-2966.2011.19818.x}, \href
  {https://ui.adsabs.harvard.edu/abs/2012MNRAS.419.1566S} {419, 1566}

\bibitem[\protect\citeauthoryear{{Schneider} et~al.,}{{Schneider}
  et~al.}{2015}]{2015A&A...578A..29S}
{Schneider} N.,  et~al., 2015, \mn@doi [\aap] {10.1051/0004-6361/201424375},
  \href {https://ui.adsabs.harvard.edu/abs/2015A&A...578A..29S} {578, A29}

\bibitem[\protect\citeauthoryear{{Sch{\"o}ier}, {van der Tak}, {van Dishoeck}
  \& {Black}}{{Sch{\"o}ier} et~al.}{2005}]{2005A&A...432..369S}
{Sch{\"o}ier} F.~L.,  {van der Tak} F.~F.~S.,  {van Dishoeck} E.~F.,   {Black}
  J.~H.,  2005, \mn@doi [\aap] {10.1051/0004-6361:20041729}, \href
  {https://ui.adsabs.harvard.edu/abs/2005A&A...432..369S} {432, 369}

\bibitem[\protect\citeauthoryear{{Schruba} et~al.,}{{Schruba}
  et~al.}{2017}]{2017ApJ...835..278S}
{Schruba} A.,  et~al., 2017, \mn@doi [\apj] {10.3847/1538-4357/835/2/278},
  \href {https://ui.adsabs.harvard.edu/abs/2017ApJ...835..278S} {835, 278}

\bibitem[\protect\citeauthoryear{{Sharda} \& {Krumholz}}{{Sharda} \&
  {Krumholz}}{2022}]{2022MNRAS.509.1959S}
{Sharda} P.,  {Krumholz} M.~R.,  2022, \mn@doi [\mnras]
  {10.1093/mnras/stab2921}, \href
  {https://ui.adsabs.harvard.edu/abs/2022MNRAS.509.1959S} {509, 1959}

\bibitem[\protect\citeauthoryear{{Sharda}, {Krumholz}  \& {Federrath}}{{Sharda}
  et~al.}{2019}]{2019MNRAS.490..513S}
{Sharda} P.,  {Krumholz} M.~R.,   {Federrath} C.,  2019, \mn@doi [\mnras]
  {10.1093/mnras/stz2618}, \href
  {https://ui.adsabs.harvard.edu/abs/2019MNRAS.490..513S} {490, 513}

\bibitem[\protect\citeauthoryear{{Sharda}, {Krumholz}, {Wisnioski}, {Forbes},
  {Federrath}  \& {Acharyya}}{{Sharda} et~al.}{2021a}]{2021MNRAS.502.5935S}
{Sharda} P.,  {Krumholz} M.~R.,  {Wisnioski} E.,  {Forbes} J.~C.,  {Federrath}
  C.,   {Acharyya} A.,  2021a, \mn@doi [\mnras] {10.1093/mnras/stab252}, \href
  {https://ui.adsabs.harvard.edu/abs/2021MNRAS.502.5935S} {502, 5935}

\bibitem[\protect\citeauthoryear{{Sharda}, {Federrath}, {Krumholz}  \&
  {Schleicher}}{{Sharda} et~al.}{2021b}]{2021MNRAS.503.2014S}
{Sharda} P.,  {Federrath} C.,  {Krumholz} M.~R.,   {Schleicher} D. R.~G.,
  2021b, \mn@doi [\mnras] {10.1093/mnras/stab531}, \href
  {https://ui.adsabs.harvard.edu/abs/2021MNRAS.503.2014S} {503, 2014}

\bibitem[\protect\citeauthoryear{{Sharda}, {Krumholz}, {Wisnioski}, {Acharyya},
  {Federrath}  \& {Forbes}}{{Sharda} et~al.}{2021c}]{2021MNRAS.504...53S}
{Sharda} P.,  {Krumholz} M.~R.,  {Wisnioski} E.,  {Acharyya} A.,  {Federrath}
  C.,   {Forbes} J.~C.,  2021c, \mn@doi [\mnras] {10.1093/mnras/stab868}, \href
  {https://ui.adsabs.harvard.edu/abs/2021MNRAS.504...53S} {504, 53}

\bibitem[\protect\citeauthoryear{{Sharda} et~al.,}{{Sharda}
  et~al.}{2022}]{2022MNRAS.509.2180S}
{Sharda} P.,  et~al., 2022, \mn@doi [\mnras] {10.1093/mnras/stab3048}, \href
  {https://ui.adsabs.harvard.edu/abs/2022MNRAS.509.2180S} {509, 2180}

\bibitem[\protect\citeauthoryear{{Sharma}, {Theuns}, {Frenk}  \&
  {Cooke}}{{Sharma} et~al.}{2018}]{2018MNRAS.473..984S}
{Sharma} M.,  {Theuns} T.,  {Frenk} C.~S.,   {Cooke} R.~J.,  2018, \mn@doi
  [\mnras] {10.1093/mnras/stx2392}, \href
  {https://ui.adsabs.harvard.edu/abs/2018MNRAS.473..984S} {473, 984}

\bibitem[\protect\citeauthoryear{{Shi}, {Wang}, {Zhang}, {Gao}, {Hao}, {Xia}
  \& {Gu}}{{Shi} et~al.}{2016}]{Shi16a}
{Shi} Y.,  {Wang} J.,  {Zhang} Z.-Y.,  {Gao} Y.,  {Hao} C.-N.,  {Xia} X.-Y.,
  {Gu} Q.,  2016, \mn@doi [Nature Communications] {10.1038/ncomms13789}, \href
  {https://ui.adsabs.harvard.edu/abs/2016NatCo...713789S} {7, 13789}

\bibitem[\protect\citeauthoryear{{Shima} \& {Hosokawa}}{{Shima} \&
  {Hosokawa}}{2021}]{2021arXiv210206312S}
{Shima} K.,  {Hosokawa} T.,  2021, \mn@doi [\mnras] {10.1093/mnras/stab2844},
  \href {https://ui.adsabs.harvard.edu/abs/2021MNRAS.508.4767S} {508, 4767}

\bibitem[\protect\citeauthoryear{{Sk{\'u}lad{\'o}ttir}
  et~al.,}{{Sk{\'u}lad{\'o}ttir} et~al.}{2021}]{2021ApJ...915L..30S}
{Sk{\'u}lad{\'o}ttir} {\'A}.,  et~al., 2021, \mn@doi [\apjl]
  {10.3847/2041-8213/ac0dc2}, \href
  {https://ui.adsabs.harvard.edu/abs/2021ApJ...915L..30S} {915, L30}

\bibitem[\protect\citeauthoryear{{Smith}}{{Smith}}{2020}]{2020ARA&A..58..577S}
{Smith} R.~J.,  2020, \mn@doi [\araa] {10.1146/annurev-astro-032620-020217},
  \href {https://ui.adsabs.harvard.edu/abs/2020ARA&A..58..577S} {58, 577}

\bibitem[\protect\citeauthoryear{{Sofia}, {Parvathi}, {Babu}  \&
  {Murthy}}{{Sofia} et~al.}{2011}]{2011AJ....141...22S}
{Sofia} U.~J.,  {Parvathi} V.~S.,  {Babu} B.~R.~S.,   {Murthy} J.,  2011,
  \mn@doi [\aj] {10.1088/0004-6256/141/1/22}, \href
  {https://ui.adsabs.harvard.edu/abs/2011AJ....141...22S} {141, 22}

\bibitem[\protect\citeauthoryear{{Spite} et~al.,}{{Spite}
  et~al.}{2005}]{2005A&A...430..655S}
{Spite} M.,  et~al., 2005, \mn@doi [\aap] {10.1051/0004-6361:20041274}, \href
  {https://ui.adsabs.harvard.edu/abs/2005A&A...430..655S} {430, 655}

\bibitem[\protect\citeauthoryear{{Stark} et~al.,}{{Stark}
  et~al.}{2014}]{2014MNRAS.445.3200S}
{Stark} D.~P.,  et~al., 2014, \mn@doi [\mnras] {10.1093/mnras/stu1618}, \href
  {https://ui.adsabs.harvard.edu/abs/2014MNRAS.445.3200S} {445, 3200}

\bibitem[\protect\citeauthoryear{{Stasi{\'n}ska}}{{Stasi{\'n}ska}}{1978}]{1978A&A....66..257S}
{Stasi{\'n}ska} G.,  1978, \aap, \href
  {https://ui.adsabs.harvard.edu/abs/1978A&A....66..257S} {66, 257}

\bibitem[\protect\citeauthoryear{{Steidel}, {Strom}, {Pettini}, {Rudie},
  {Reddy}  \& {Trainor}}{{Steidel} et~al.}{2016}]{2016ApJ...826..159S}
{Steidel} C.~C.,  {Strom} A.~L.,  {Pettini} M.,  {Rudie} G.~C.,  {Reddy} N.~A.,
    {Trainor} R.~F.,  2016, \mn@doi [\apj] {10.3847/0004-637X/826/2/159}, \href
  {https://ui.adsabs.harvard.edu/abs/2016ApJ...826..159S} {826, 159}

\bibitem[\protect\citeauthoryear{{Sternberg}, {Le Petit}, {Roueff}  \& {Le
  Bourlot}}{{Sternberg} et~al.}{2014}]{2014ApJ...790...10S}
{Sternberg} A.,  {Le Petit} F.,  {Roueff} E.,   {Le Bourlot} J.,  2014, \mn@doi
  [\apj] {10.1088/0004-637X/790/1/10}, \href
  {https://ui.adsabs.harvard.edu/abs/2014ApJ...790...10S} {790, 10}

\bibitem[\protect\citeauthoryear{{Sternberg}, {Gurman}  \& {Bialy}}{{Sternberg}
  et~al.}{2021}]{2021arXiv210501681S}
{Sternberg} A.,  {Gurman} A.,   {Bialy} S.,  2021, \mn@doi [\apj]
  {10.3847/1538-4357/ac167b}, \href
  {https://ui.adsabs.harvard.edu/abs/2021ApJ...920...83S} {920, 83}

\bibitem[\protect\citeauthoryear{{Suda}, {Yamada}, {Katsuta}, {Komiya},
  {Ishizuka}, {Aoki}  \& {Fujimoto}}{{Suda} et~al.}{2011}]{2011MNRAS.412..843S}
{Suda} T.,  {Yamada} S.,  {Katsuta} Y.,  {Komiya} Y.,  {Ishizuka} C.,  {Aoki}
  W.,   {Fujimoto} M.~Y.,  2011, \mn@doi [\mnras]
  {10.1111/j.1365-2966.2011.17943.x}, \href
  {https://ui.adsabs.harvard.edu/abs/2011MNRAS.412..843S} {412, 843}

\bibitem[\protect\citeauthoryear{{Suda} et~al.,}{{Suda}
  et~al.}{2013}]{2013MNRAS.432L..46S}
{Suda} T.,  et~al., 2013, \mn@doi [\mnras] {10.1093/mnrasl/slt033}, \href
  {https://ui.adsabs.harvard.edu/abs/2013MNRAS.432L..46S} {432, L46}

\bibitem[\protect\citeauthoryear{{Tacchella} et~al.,}{{Tacchella}
  et~al.}{2022}]{2022arXiv220803281T}
{Tacchella} S.,  et~al., 2022, arXiv e-prints, \href
  {https://ui.adsabs.harvard.edu/abs/2022arXiv220803281T} {p. arXiv:2208.03281}

\bibitem[\protect\citeauthoryear{{Tanvir}, {Krumholz}  \& {Federrath}}{{Tanvir}
  et~al.}{2022}]{2022arXiv220604999T}
{Tanvir} T.~S.,  {Krumholz} M.~R.,   {Federrath} C.,  2022, \mn@doi [\mnras]
  {10.1093/mnras/stac2642}, \href
  {https://ui.adsabs.harvard.edu/abs/2022MNRAS.516.5712T} {516, 5712}

\bibitem[\protect\citeauthoryear{{Toribio San Cipriano},
  {Dom{\'\i}nguez-Guzm{\'a}n}, {Esteban}, {Garc{\'\i}a-Rojas}, {Mesa-Delgado},
  {Bresolin}, {Rodr{\'\i}guez}  \& {Sim{\'o}n-D{\'\i}az}}{{Toribio San
  Cipriano} et~al.}{2017}]{2017MNRAS.467.3759T}
{Toribio San Cipriano} L.,  {Dom{\'\i}nguez-Guzm{\'a}n} G.,  {Esteban} C.,
  {Garc{\'\i}a-Rojas} J.,  {Mesa-Delgado} A.,  {Bresolin} F.,  {Rodr{\'\i}guez}
  M.,   {Sim{\'o}n-D{\'\i}az} S.,  2017, \mn@doi [\mnras]
  {10.1093/mnras/stx328}, \href
  {https://ui.adsabs.harvard.edu/abs/2017MNRAS.467.3759T} {467, 3759}

\bibitem[\protect\citeauthoryear{{Triani}, {Sinha}, {Croton}, {Pacifici}  \&
  {Dwek}}{{Triani} et~al.}{2020}]{2020MNRAS.493.2490T}
{Triani} D.~P.,  {Sinha} M.,  {Croton} D.~J.,  {Pacifici} C.,   {Dwek} E.,
  2020, \mn@doi [\mnras] {10.1093/mnras/staa446}, \href
  {https://ui.adsabs.harvard.edu/abs/2020MNRAS.493.2490T} {493, 2490}

\bibitem[\protect\citeauthoryear{{Tsujimoto} \& {Bekki}}{{Tsujimoto} \&
  {Bekki}}{2011}]{2011A&A...530A..78T}
{Tsujimoto} T.,  {Bekki} K.,  2011, \mn@doi [\aap]
  {10.1051/0004-6361/201016210}, \href
  {https://ui.adsabs.harvard.edu/abs/2011A&A...530A..78T} {530, A78}

\bibitem[\protect\citeauthoryear{{Turner}, {Beck}  \& {Ho}}{{Turner}
  et~al.}{2000}]{2000ApJ...532L.109T}
{Turner} J.~L.,  {Beck} S.~C.,   {Ho} P. T.~P.,  2000, \mn@doi [\apjl]
  {10.1086/312586}, \href
  {https://ui.adsabs.harvard.edu/abs/2000ApJ...532L.109T} {532, L109}

\bibitem[\protect\citeauthoryear{{Vanzella} et~al.,}{{Vanzella}
  et~al.}{2016}]{2016ApJ...821L..27V}
{Vanzella} E.,  et~al., 2016, \mn@doi [\apjl] {10.3847/2041-8205/821/2/L27},
  \href {https://ui.adsabs.harvard.edu/abs/2016ApJ...821L..27V} {821, L27}

\bibitem[\protect\citeauthoryear{{Vijayan}, {Clay}, {Thomas}, {Yates},
  {Wilkins}  \& {Henriques}}{{Vijayan} et~al.}{2019}]{2019MNRAS.489.4072V}
{Vijayan} A.~P.,  {Clay} S.~J.,  {Thomas} P.~A.,  {Yates} R.~M.,  {Wilkins}
  S.~M.,   {Henriques} B.~M.,  2019, \mn@doi [\mnras] {10.1093/mnras/stz1948},
  \href {https://ui.adsabs.harvard.edu/abs/2019MNRAS.489.4072V} {489, 4072}

\bibitem[\protect\citeauthoryear{{Virtanen} et~al.,}{{Virtanen}
  et~al.}{2020}]{2020NatMe..17..261V}
{Virtanen} P.,  et~al., 2020, \mn@doi [Nature Methods]
  {10.1038/s41592-019-0686-2}, \href
  {https://ui.adsabs.harvard.edu/abs/2020NatMe..17..261V} {17, 261}

\bibitem[\protect\citeauthoryear{{Welty}, {Lauroesch}, {Wong}  \&
  {York}}{{Welty} et~al.}{2016}]{2016ApJ...821..118W}
{Welty} D.~E.,  {Lauroesch} J.~T.,  {Wong} T.,   {York} D.~G.,  2016, \mn@doi
  [\apj] {10.3847/0004-637X/821/2/118}, \href
  {https://ui.adsabs.harvard.edu/abs/2016ApJ...821..118W} {821, 118}

\bibitem[\protect\citeauthoryear{{Yamaguchi}, {Furlanetto}  \&
  {Trapp}}{{Yamaguchi} et~al.}{2022}]{2022arXiv220909345Y}
{Yamaguchi} N.,  {Furlanetto} S.~R.,   {Trapp} A.~C.,  2022, arXiv e-prints,
  \href {https://ui.adsabs.harvard.edu/abs/2022arXiv220909345Y} {p.
  arXiv:2209.09345}

\bibitem[\protect\citeauthoryear{{Yin}, {Matteucci}  \& {Vladilo}}{{Yin}
  et~al.}{2011}]{2011A&A...531A.136Y}
{Yin} J.,  {Matteucci} F.,   {Vladilo} G.,  2011, \mn@doi [\aap]
  {10.1051/0004-6361/201015022}, \href
  {https://ui.adsabs.harvard.edu/abs/2011A&A...531A.136Y} {531, A136}

\bibitem[\protect\citeauthoryear{{Yong} et~al.,}{{Yong}
  et~al.}{2021}]{2021arXiv210706430Y}
{Yong} D.,  et~al., 2021, \mn@doi [\mnras] {10.1093/mnras/stab2001}, \href
  {https://ui.adsabs.harvard.edu/abs/2021MNRAS.507.4102Y} {507, 4102}

\bibitem[\protect\citeauthoryear{{Yoon} et~al.,}{{Yoon}
  et~al.}{2016}]{2016ApJ...833...20Y}
{Yoon} J.,  et~al., 2016, \mn@doi [\apj] {10.3847/0004-637X/833/1/20}, \href
  {https://ui.adsabs.harvard.edu/abs/2016ApJ...833...20Y} {833, 20}

\bibitem[\protect\citeauthoryear{{Yoon}, {Beers}, {Tian}  \& {Whitten}}{{Yoon}
  et~al.}{2019}]{2019ApJ...878...97Y}
{Yoon} J.,  {Beers} T.~C.,  {Tian} D.,   {Whitten} D.~D.,  2019, \mn@doi [\apj]
  {10.3847/1538-4357/ab1ead}, \href
  {https://ui.adsabs.harvard.edu/abs/2019ApJ...878...97Y} {878, 97}

\bibitem[\protect\citeauthoryear{{Youakim} et~al.,}{{Youakim}
  et~al.}{2020}]{2020MNRAS.492.4986Y}
{Youakim} K.,  et~al., 2020, \mn@doi [\mnras] {10.1093/mnras/stz3619}, \href
  {https://ui.adsabs.harvard.edu/abs/2020MNRAS.492.4986Y} {492, 4986}

\bibitem[\protect\citeauthoryear{{Zepeda} et~al.,}{{Zepeda}
  et~al.}{2022}]{2022arXiv220912224Z}
{Zepeda} J.,  et~al., 2022, arXiv e-prints, \href
  {https://ui.adsabs.harvard.edu/abs/2022arXiv220912224Z} {p. arXiv:2209.12224}

\bibitem[\protect\citeauthoryear{{Zubko}, {Dwek}  \& {Arendt}}{{Zubko}
  et~al.}{2004}]{2004ApJS..152..211Z}
{Zubko} V.,  {Dwek} E.,   {Arendt} R.~G.,  2004, \mn@doi [\apjs]
  {10.1086/382351}, \href
  {https://ui.adsabs.harvard.edu/abs/2004ApJS..152..211Z} {152, 211}

\bibitem[\protect\citeauthoryear{{van Dokkum} \& {Conroy}}{{van Dokkum} \&
  {Conroy}}{2010}]{2010Natur.468..940V}
{van Dokkum} P.~G.,  {Conroy} C.,  2010, \mn@doi [\nat] {10.1038/nature09578},
  \href {https://ui.adsabs.harvard.edu/abs/2010Natur.468..940V} {468, 940}

\bibitem[\protect\citeauthoryear{{van Dokkum} et~al.,}{{van Dokkum}
  et~al.}{2008}]{2008ApJ...677L...5V}
{van Dokkum} P.~G.,  et~al., 2008, \mn@doi [\apjl] {10.1086/587874}, \href
  {https://ui.adsabs.harvard.edu/abs/2008ApJ...677L...5V} {677, L5}

\bibitem[\protect\citeauthoryear{{van Zee} \& {Haynes}}{{van Zee} \&
  {Haynes}}{2006}]{2006ApJ...636..214V}
{van Zee} L.,  {Haynes} M.~P.,  2006, \mn@doi [\apj] {10.1086/498017}, \href
  {https://ui.adsabs.harvard.edu/abs/2006ApJ...636..214V} {636, 214}

\bibitem[\protect\citeauthoryear{{van der Tak}, {van Dishoeck}, {Evans}  \&
  {Blake}}{{van der Tak} et~al.}{2000}]{2000ApJ...537..283V}
{van der Tak} F. F.~S.,  {van Dishoeck} E.~F.,  {Evans} Neal~J. I.,   {Blake}
  G.~A.,  2000, \mn@doi [\apj] {10.1086/309011}, \href
  {https://ui.adsabs.harvard.edu/abs/2000ApJ...537..283V} {537, 283}

\bibitem[\protect\citeauthoryear{{van der Tak}, {Black}, {Sch{\"o}ier},
  {Jansen}  \& {van Dishoeck}}{{van der Tak}
  et~al.}{2007}]{2007A&A...468..627V}
{van der Tak} F.~F.~S.,  {Black} J.~H.,  {Sch{\"o}ier} F.~L.,  {Jansen} D.~J.,
   {van Dishoeck} E.~F.,  2007, \mn@doi [\aap] {10.1051/0004-6361:20066820},
  \href {https://ui.adsabs.harvard.edu/abs/2007A&A...468..627V} {468, 627}

\bibitem[\protect\citeauthoryear{{van der Tak}, {Lique}, {Faure}, {Black}  \&
  {van Dishoeck}}{{van der Tak} et~al.}{2020}]{2020Atoms...8...15V}
{van der Tak} F. F.~S.,  {Lique} F.,  {Faure} A.,  {Black} J.~H.,   {van
  Dishoeck} E.~F.,  2020, \mn@doi [Atoms] {10.3390/atoms8020015}, \href
  {https://ui.adsabs.harvard.edu/abs/2020Atoms...8...15V} {8, 15}

\makeatother
\end{thebibliography}

\bsp	
\label{lastpage}
\end{document}